\documentclass{aa} 

\usepackage{graphicx}
\usepackage{tabularx}
\usepackage{placeins}
\usepackage{abbriv}
\usepackage{acro}
\usepackage[english]{babel}
\usepackage{array}
\usepackage{capt-of}
\newcolumntype{L}[1]{>{\raggedright\let\newline\\\arraybackslash\hspace{0pt}}m{#1}}
\newcolumntype{C}[1]{>{\centering\let\newline\\\arraybackslash\hspace{0pt}}m{#1}}
\newcolumntype{R}[1]{>{\raggedleft\let\newline\\\arraybackslash\hspace{0pt}}m{#1}}

\usepackage{txfonts}
\usepackage[colorlinks=true, allcolors=blue]{hyperref}
\def\kms{km~s$^{-1}$}
\graphicspath{{folder/}{Figures/}}

\begin{document}

   \title{Complex organic molecules in low-mass protostars on solar system scales}

   \subtitle{I. Oxygen-bearing species}

   \author{
          M. L. van Gelder\inst{1}
          \and
          B. Tabone\inst{1}
          \and
          Ł. Tychoniec\inst{1}
          \and
          E. F. van Dishoeck\inst{1,2}
          \and
          H. Beuther\inst{3}
          \and
          A. C. A. Boogert\inst{4}
          \and
          A. Caratti o Garatti\inst{5}
          \and
          P. D. Klaassen\inst{6}
          \and
          H. Linnartz\inst{7}
          \and
          H. S. P. M\"uller\inst{8}
          \and
          V. Taquet\inst{9}
          }

\institute{
         Leiden Observatory, Leiden University, PO Box 9513, 2300RA Leiden, The Netherlands \\
         \email{vgelder@strw.leidenuniv.nl}
         \and
         Max Planck Institut f\"ur Extraterrestrische Physik (MPE), Giessenbachstrasse 1, 85748 Garching, Germany
         \and
          Max Planck Institute for Astronomy, K\"onigstuhl 17, 69117 Heidelberg, Germany
          \and
         Institute for Astronomy, University of Hawaii at Manoa, 2680 Woodlawn Drive, Honolulu, HI 96822, USA
         \and
         Dublin Institute for Advanced Studies, School of Cosmic Physics, Astronomy and Astrophysics Section, 31 Fitzwilliam Place, D04C932 Dublin 2, Ireland 
         \and
         UK Astronomy Technology Centre, Royal Observatory Edinburgh, Blackford Hill, Edinburgh EH9 3HJ, UK
		 \and
         Laboratory for Astrophysics, Leiden Observatory, Leiden University, PO Box 9531, 2300RA Leiden, The Netherlands
         \and       
		 I. Physikalisches Institut, Universit\"at zu K\"oln, Z\"ulpicher Str. 77, 50937 K\"oln, Germany   
         \and     
		 INAF, Osservatorio Astrofisico di Arcetri, Largo E. Fermi 5, 50125 Firenze, Italy
             }

   \date{Received XXX; accepted XXX}

 
  \abstract
   {Complex organic molecules (\acs{COM}s) are thought to form on icy dust grains in the earliest phase of star formation. The evolution of these \ac{COM}s from the youngest Class 0/I protostellar phases toward the more evolved Class II phase is still not fully understood. Since planet formation seems to start early, and mature disks are too cold for characteristic \ac{COM} emission lines, studying the inventory of \ac{COM}s on solar system scales in the Class~0/I stage is relevant. }
   {The aim is to determine the abundance ratios of oxygen-bearing \ac{COM}s in Class 0 protostellar systems on scales of $\sim$100~AU radius. These abundances will be inter-compared, and contrasted to other low-mass protostars such as IRAS16293-2422B and HH~212. Additionally, using both cold and hot \ac{COM} lines, the gas-phase abundances can be tracked from a cold to a hot component, and ultimately be compared with those in ices to be measured with the \ac{JWST}. The abundance of deuterated methanol allows us to probe the ambient temperature during the formation of this species.}
   {\acs{ALMA} Band~3 (3~mm) and Band~6 (1~mm) observations are obtained of seven Class 0 protostars in the Perseus and Serpens star-forming regions. By modeling the inner protostellar region using 'LTE' models, the excitation temperature and column densities are determined for several O-bearing \ac{COM}s including methanol (CH$_3$OH), acetaldehyde (CH$_3$CHO), methyl formate (CH$_3$OCHO), and dimethyl ether (CH$_3$OCH$_3$). Abundance ratios are taken with respect to CH$_3$OH. }
   {B1-c and B1-bS (both Perseus) and Serpens~S68N (Serpens) show \ac{COM} emission, {i.e.,} three out of the seven of the observed sources. No clear correlation seems to exist between the occurrence of \ac{COM}s and source luminosity.
   The abundances of several \ac{COM}s such as CH$_3$OCHO, CH$_3$OCH$_3$, acetone (CH$_3$COCH$_3$), and ethylene glycol ((CH$_2$OH)$_2$) are remarkably similar for the three \ac{COM}-rich sources, and to IRAS~16238-2422B and HH~212, even though these sources originate from four different star-forming regions (i.e., Perseus, Serpens, Ophiuchus, Orion). For other \ac{COM}s like CH$_3$CHO, ethanol (CH$_3$CH$_2$OH), and glycolaldehyde (CH$_2$OHCHO), the abundances differ by up to an order of magnitude, indicating that local source conditions are case determining.
   B1-c hosts a cold ($T_\mathrm{ex}\approx60$~K), more extended component of \ac{COM} emission with a column density of typically a few \% of the warm/hot ($T_\mathrm{ex}\sim 200$~K), central component.
   A D/H ratio of 1--{3} \% is derived for B1-c, S68N, and B1-bS based on the CH$_2$DOH/CH$_3$OH ratio (taking into account statistical weighting) suggesting a temperature of $\sim$15~K during the formation of methanol. This ratio is consistent with other low-mass protostars, but lower than for high-mass star-forming regions.
}
   {The abundance ratios of most O-bearing \ac{COM}s are roughly fixed between different star-forming regions, and are presumably set at an earlier cold prestellar phase. For several \ac{COM}s, local source properties become important. Future \acl{MIR} facilities such as \ac{JWST}/MIRI will be essential to directly observe \ac{COM} ices. Combining this with a larger sample of \ac{COM}-rich sources with \ac{ALMA} will allow for directly linking ice and gas-phase abundances in order to constrain the routes that produce and maintain chemical complexity during the star formation process.}


   \maketitle
%
\acresetall

\section{Introduction}
Complex organic molecules (\acs{COM}s) are molecules with six or more atoms of which at least one atom is carbon \citep[see review by][]{Herbst2009}. They have been observed toward high-mass and low-mass star-forming regions, both in young protostars \citep[e.g.,][]{Caselli2012,Jorgensen2016,McGuire2018,Bogelund2019}, and in protostellar outflows \citep[e.g.,][]{Arce2008,Oberg2010}. 
As a protostellar system develops from the embedded Class 0/I phase toward the more evolved Class II phase, the \ac{COM}s may eventually either be destroyed or become incorporated in the ice mantles of dust grains or planetesimals in disks \citep{Visser2009,Visser2011,Drozdovskaya2016}. Since planet formation may start already in the embedded Class 0/I phase \citep[e.g.,][]{Harsono2018,Manara2018}, studying the inventory and abundances of \ac{COM}s on solar system scales in the earliest stages of star formation is essential to probe the initial conditions of planet formation. With the launch of the \ac{JWST} in the near future, the ice abundances of \ac{COM}s can be directly linked to their gaseous counterpart. 

In cold prestellar envelopes, formation of \ac{COM}s happens on the ice mantles of dust grains through hydrogenation of smaller molecules such as CO \citep{Charnley1992,Watanabe2002,Fuchs2009}. This leads to the formation of methanol (CH$_3$OH), but these \ac{COM}s may even get as complex as, for example, glycolaldehyde (CH$_2$OHCHO), ethylene glycol ((CH$_2$OH)$_2$), and glycerol {(HOCH$_2$CH(OH)CH$_2$OH)} \citep{Oberg2009,Chuang2016,Chuang2017,Fedoseev2017}. When the dust grains move inwards toward protostars, the \ac{UV} radiation field increases \citep[e.g.,][]{Visser2009,Drozdovskaya2016}, leading to the dissociation of some \ac{COM}s into smaller radicals \citep{Gerakines1996,Oberg2016}. These radicals react with other \ac{COM}s in the ice mantles to form larger species \citep[][]{Garrod2006,Garrod2008,Oberg2009}, and at higher temperatures perhaps through thermal induced chemistry \citep{Theule2013}. When the temperature of the dust grains reaches $T\approx100-300$~K, all \ac{COM}s desorb from the grains into the gas phase where they can be transformed by high temperature gas-phase chemistry and \ac{UV} radiation \citep[e.g.,][]{Charnley1992,Balucani2015,Skouteris2018}. A fraction of the \ac{COM}s may be incorporated into larger bodies in the cold mid-plane of an accretion disk. 

Another reason for studying \ac{COM}s in the earliest (Class~0) phase of star formation is that gaseous \ac{COM} lines are hard to detect in the Class~I phase \citep[e.g.,][]{ArturdelaVillarmois2019}. An exception is when a (more evolved) source undergoes an accretion burst; the sudden increase in luminosity can release frozen \ac{COM}s from the grains \citep{vantHoff2018,Lee2019_V883Ori}. In the ice phase, they are difficult to observe, with only well constrained abundances for CH$_3$OH and some limits or tentative detections for formic acid (HCOOH), acetaldehyde (CH$_3$CHO), and ethanol (CH$_3$CH$_2$OH) \citep[see review by][]{Boogert2015,TerwisschavanScheltinga2018}. 

Most studies of gaseous \ac{COM}s have been carried out on high-mass hot cores in massive star-forming regions such as Sagittarius~B2 and Orion KL, where \ac{COM}s are typically present in high abundances \citep[e.g.,][]{Belloche2013,Neill2014,Crockett2014,Pagani2017}. 
In \acl{LYSO}s, single-dish observations with beams of a few thousand AU show that the abundances of oxygen-bearing \ac{COM}s with respect to CH$_3$OH seem to be narrowly distributed \citep{Bottinelli2004,Jorgensen2005,Bergner2017,Belloche2020}. 

With the \acf{ALMA}, it is now possible to spatially resolve individual protostellar systems on scales of $\sim$50~AU. Additionally, the spectral coverage and sensitivity of \ac{ALMA} allow a large variety of \ac{COM}s and corresponding isotopologues to be observed in optically thin lines, resulting in more accurate abundance determinations. Gaseous \ac{COM}s have been studied mostly in the Class~0 IRAS 16293-2422 binary system (hereafter IRAS~16293) as part of the \acs{ALMA} PILS survey \citep{Jorgensen2016}. This led to many detections of new species in the \ac{ISM} such as HONO \citep{Coutens2019}, {CH$_3$Cl \citep{Fayolle2017}}, and CHD$_2$OCHO \citep{Manigand2019}. Furthermore, IRAS 16293 has been studied extensively for both oxygen and nitrogen-bearing \ac{COM}s \citep{Jorgensen2018,Calcutt2018_nitriles,Ligterink2018_Nitrooxides, Ligterink2018_peptides,Manigand2020}. Other low-mass sources studied interferometrically on $\sim50-150$~AU scales include HH~212 \citep{Lee2019_HH212}, NGC~1333-IRAS~2A and IRAS~4A \citep[hereafter IRAS~2A and IRAS~4A;][]{Taquet2015,Taquet2019}{, L483 \citep{Jacobsen2019}, and three sources in the Serpens Main star forming region \citep{Bergner2019}}.

Single dish observations show hints of a cold, more extended component {of \ac{COM} emission} in Class 0 protostars where the excitation temperature may drop below 20--60~K. {This component is present additional to the 'typical' warm/hot, more compact \ac{COM} emission with $T\gtrsim100-300$~K} \citep{Bisschop2007_hotcores,Isokoski2013,Marcelino2018}. {Currently, the chemical relationship between the cold \ac{COM}s on large scales and warm/hot \ac{COM}s on small scales is still unknown}. The evolution of \ac{COM}s from a cold extended component to a hot central component can be traced by observing \ac{COM}s in multiple frequency bands. 

The Perseus and Serpens star-forming regions are known to host many embedded protostars \citep[e.g.,][]{Enoch2009,Tobin2016,Karska2018}. In this paper, four protostellar regions with strong ice features in infrared observations \citep{Boogert2008} are targeted with \ac{ALMA} in Band~3 and Band~6. These targets are also part part of the \ac{JWST}/MIRI \ac{GTO} program (project ID 1290). {\ac{JWST}/MIRI will, for instance, observe the 7.2~$\mu$m and 7.4~$\mu$m features for which CH$_3$CH$_2$OH and CH$_3$CHO, respectively, are strong candidates based on laboratory spectroscopy \citep[e.g.,][]{Schutte1999, TerwisschavanScheltinga2018}. Ultimately, this will allow for directly comparing gaseous and icy \ac{COM}s}. Some additional nearby sources falling within the \ac{ALMA} field of view are also included here, leading to a total of seven sources: B1-b, B1-bS, B1-bN and B1-c in the Perseus Barnard 1 cloud, and Serpens~S68N (hereafter S68N), Ser-emb~8~(N), and Serpens~SMM3 (hereafter SMM3) in the Serpens Main region. 

B1-b (sometimes also referred to as B1-bW) is a bright infrared source with strong ice features in CH$_3$OH and (potentially) CH$_3$CHO \citep{Oberg2011}. B1-bN is a suggested first hydrostatic core candidate \citep[e.g.,][]{Pezzuto2012,Hirano2014,Gerin2017}, and B1-bS is proposed as a hot corino \citep{Marcelino2018}. B1-c is a young Class 0 object showing both strong CH$_3$OH ice and gas \ac{COM} features in the hot corino \citep{Boogert2008,Bergner2017}, and a high velocity outflow \citep{Jorgensen2006,Hatchell2007}. Moreover, all Barnard~1 objects have been extensively studied in the \ac{VANDAM} survey \citep[][]{Tobin2016,Tychoniec2018}, showing compact disk-like structures at scales of several tens of AU. In Serpens, S68N and SMM3 are both Class 0 sources studied in detail by \cite{Kristensen2010}. \cite{Hull2017} and \citet[][]{TychoniecIAUS2018, Tychoniec2019} observed several protostars in Serpens with \ac{ALMA}, focusing on outflows. Interestingly, whereas Ser-emb~8~(N) hosts a high velocity outflow, S68N only has a slow velocity outflow. The characteristics of our targets are shown in Table~\ref{tab:obslog}. We focus here on the O-bearing \ac{COM}s in the data; the N-bearing species will be part of a future paper.

\begin{table*}[t]
\begin{center}
\caption{List of protostars discussed in this paper as well as their main astronomical properties.}
\begin{tabular}{@{\extracolsep{0pt}}llllccccccc@{}}
\hline
\hline
Object & Cloud & RA (J2000) & Dec (J2000) & $d$ & $L_\mathrm{bol}$ & $T_\mathrm{bol}$ & $M_\mathrm{env}$ & CH$_3$OH/H$_2$O$_\mathrm{ice}$ &\multicolumn{2}{c}{COMs detected?} \\ \cline{10-11}
 & & (hh:mm:ss.ss) & (dd:mm:ss.s) & pc & L$_\odot$ & K & M$_\odot$ & (\%) & Band~3 & Band~6 \\ 
\hline
\object{B1-b} & \object{Perseus} & 03:33:20.34 & 31:07:21.3 & 320\tablefootmark{1} & {0.32}\tablefootmark{2} & 151\tablefootmark{2} & 1.5\tablefootmark{2} & 11.2\tablefootmark{3} & n & n \\
\textit{\object{B1-bN}} & Perseus & 03:33:21.21 & 31:07:43.6 & 320\tablefootmark{1} & {0.28}\tablefootmark{4} & 17\tablefootmark{4} & 0.4\tablefootmark{4} & -- & n & -- \\
\textit{\object{B1-bS}} & Perseus & 03:33:21.36 & 31:07:26.3 & 320\tablefootmark{1} & {0.57}\tablefootmark{4} & 22\tablefootmark{4} & 0.4\tablefootmark{4} & -- & y & y \\
\object{B1-c} & Perseus & 03:33:17.88 & 31:09:31.8 & 320\tablefootmark{1} & {5.9}\tablefootmark{5} & 46\tablefootmark{5} & 3.8\tablefootmark{2} & $<$7.1\tablefootmark{3} & y & y \\
\hline
\object{S68N} & \object{Serpens} & 18:29:48.09 & 01:16:43.3 & 436\tablefootmark{6} & 5.4\tablefootmark{7} & 58\tablefootmark{6} & 9.4\tablefootmark{7} & --& y & y \\
\textit{\object{Ser-emb 8 (N)}} & Serpens & 18:29:48.73 & 01:16:55.6 & 436\tablefootmark{6} & -- & -- & -- & -- & n & y\tablefootmark{8} \\
\object{SMM3} & Serpens & 18:29:57.75 & 01:14:06.7 & 436\tablefootmark{6} & 27.5\tablefootmark{5} & 37\tablefootmark{5} & -- & -- & n & n\tablefootmark{9} \\
\hline
\end{tabular}
\label{tab:obslog}
\tablefoot{The objects in italic additionally fell within the field of view of the ALMA observations. In the last 2 columns, yes (y) indicates that \ac{COM}s were detected in this work, no (n) indicates that the spectra lacked any \ac{COM} detections, and -- indicates that the Band was not observed for the corresponding target. \\ 
\tablefoottext{1}{\cite{Ortiz-Leon2018}.}
\tablefoottext{2}{\cite{Enoch2009}.}
\tablefoottext{3}{\cite{Boogert2008}.}
\tablefoottext{4}{\cite{Hirano2014}.}
\tablefoottext{5}{\cite{Karska2018}.}
\tablefoottext{6}{\cite{Ortiz-Leon2017}.}
\tablefoottext{7}{\cite{Enoch2011}.}
\tablefoottext{8}{Only CH$_3$OH {emission related to the outflow} detected.}
\tablefoottext{9}{{Based on data from ALMA program 2017.1.01350.S}}
}
\end{center}
\end{table*}

\begin{figure*}[t]
\centering
\includegraphics[width=1\linewidth]{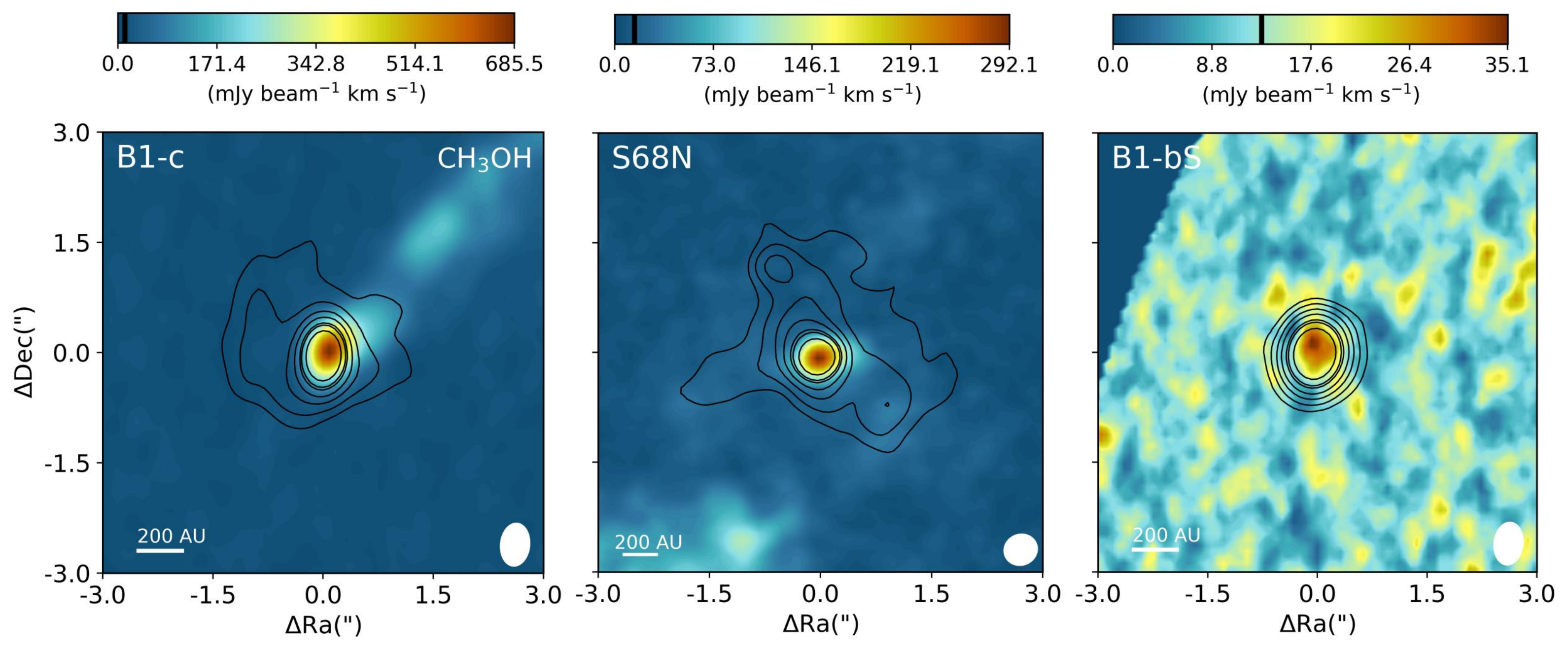}
\includegraphics[width=1\linewidth]{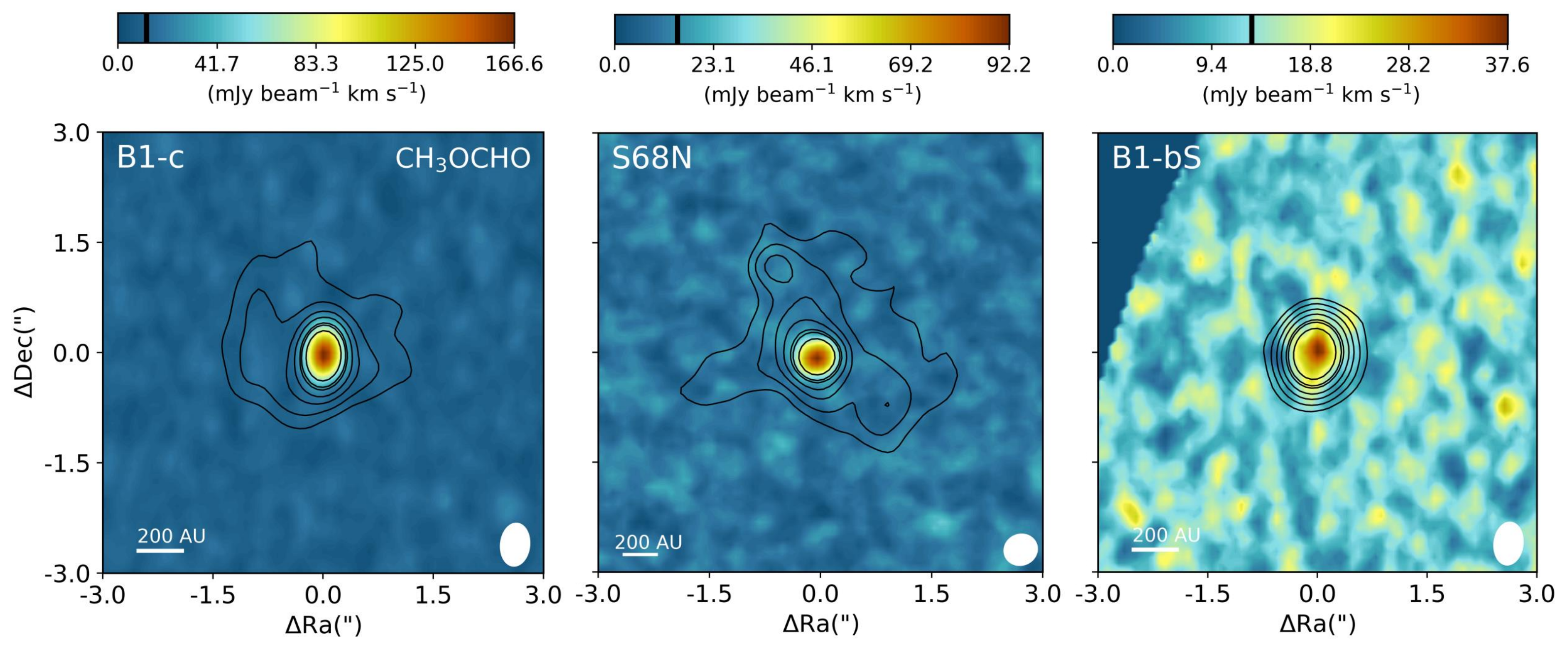}
\caption{ALMA Band~6 moment 0 images of the \ac{COM}-rich sources analyzed in this study: B1-c (left), S68N (middle), and B1-bS (right). In color the spatial distribution of CH$_3$OH 2$_{1,1}$--1$_{0,1}$ (top row, $E_\mathrm{up}=28$~K) and CH$_3$OCHO 21$_{7,14}$--20$_{7,13}$ emission (bottom row, $E_\mathrm{up}=170$~K) is shown, with the color scale shown on top of each image. {The images are integrated over [-10,10] \kms\ with respect to the $V_\mathrm{lsr}$}. In the colorbar, the 3$\sigma_\mathrm{line}$ level is indicated with the black bar, with $\sigma_\mathrm{line}=4.0$, $4.9$, and $4.4$~mJy~beam$^{-1}$~\kms\ for B1-c, S68N, and B1-bS, respectively. The continuum is shown with the black contours with increasingly [3,5,8,12,18,21,30]~$\sigma_\mathrm{cont}$, where $\sigma_\mathrm{cont}$ is $1.5$, $0.8$, and $1.0$ ~mJy~beam$^{-1}$~\kms\ for B1-c, S68N, and B1-bS, respectively. 
The size of the beam is shown in the lower right of each image, and in the lower left a scale bar is displayed. The image of B1-bS has a lower S/N since the source is located on the edge of the primary beam.
}
\label{fig:ALMAimages}
\end{figure*}

\begin{figure*}[t]
\centering
\includegraphics[width=1\linewidth]{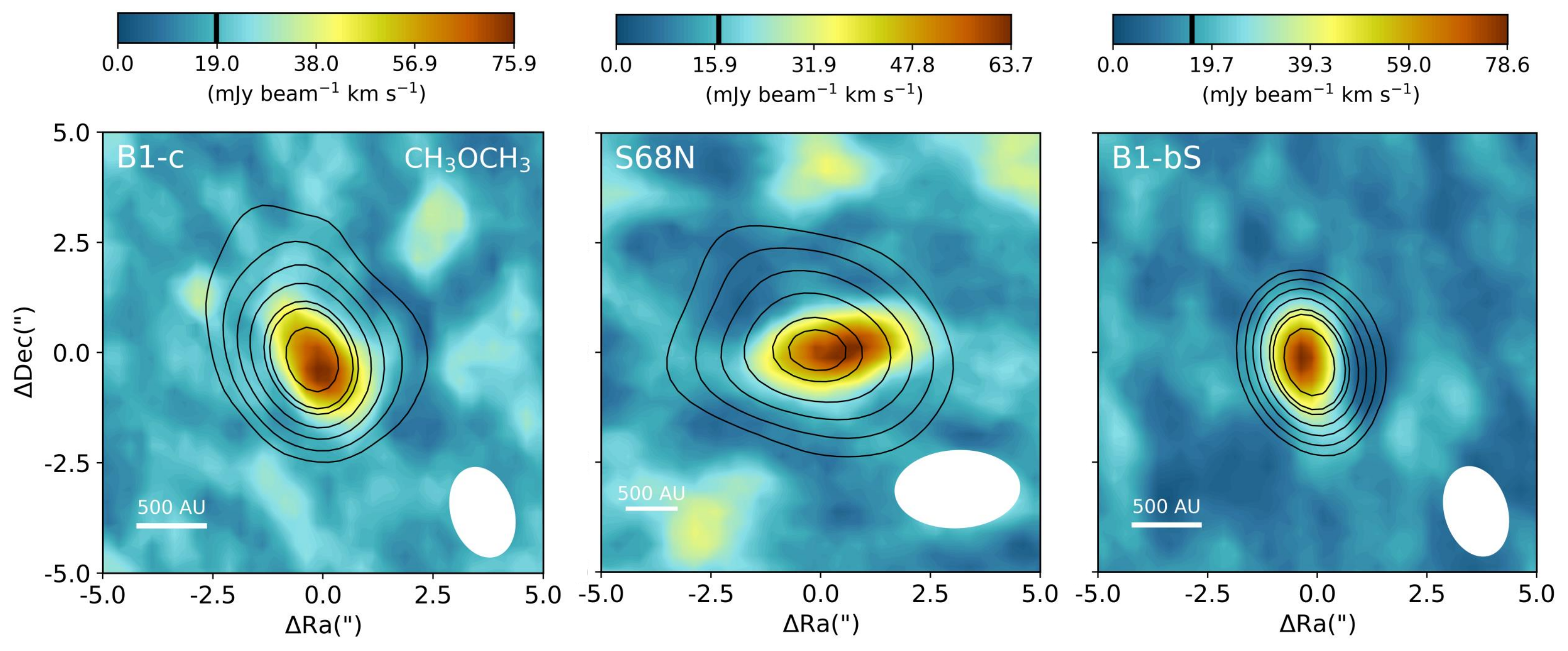}
\caption{Same as Fig.~\ref{fig:ALMAimages}, but now for CH$_3$OCH$_3$ 7$_{0,7}$--6$_{1,6}$ ($E_\mathrm{up}=25$~K) in Band~3.  Here, $\sigma_\mathrm{line}=6.3$, $5.5$, and $5.2$~mJy~beam$^{-1}$~\kms\ for B1-c, S68N, and B1-bS, respectively. The continuum contours are [3,5,8,12,18,21,30]~$\sigma_\mathrm{cont}$, where $\sigma_\mathrm{cont}$ is $0.46$, $0.54$, and $0.71$ ~mJy~beam$^{-1}$~\kms\ for B1-c, S68N, and B1-bS, respectively. 
}
\label{fig:ALMAB3images}
\end{figure*}

\section{Observations}
\label{sec:ALMAobs}
\ac{ALMA} Cycle 5 observations (project 2017.1.01174.S; PI: E.F. van Dishoeck) were taken in Band~3 (3~mm) and Band~6 (1~mm). The key targeted \ac{COM}s were CH$_3$OH, HCOOH, CH$_3$CHO, CH$_3$OCHO, CH$_3$CH$_2$OH, and NH$_2$CHO. 
We list the main technical properties of our \ac{ALMA} images in Table~\ref{tab:obsprop}.
For \mbox{B1-bN} and SMM3, we lack Band~6 observations {in our program}. {All data were taken using the 12m array, with a synthesized beam of} $\sim$1.5--2.5" ({$\sim$500-1000~AU at the distances of our sources;} C43-2 \& C43-3 configuration) and $\sim$0.45" ({$\sim$100-200~AU;} C43-4 configuration) for Band~3 and Band~6, respectively. The largest recoverable scales (LAS) were $\sim$20" {($\sim$6000--9000~AU)} and $\sim$6" {($\sim$2000-3000~AU)} for Band~3 and Band~6, respectively. The spectral resolution was $\sim$0.2~\kms\text{} for most spectral windows, with a few windows in Band~3 having $\sim$0.3--0.4~\kms. At this resolution, all lines are spectrally resolved. {The line sensitivity is $\sim$0.1~K.} The absolute uncertainty on the flux calibration is $\leq$15\% for all observations. {For SMM3, we use archival Band~6 data (project 2017.1.01350.S; PI: Ł. Tychoniec) at similar spatial resolution ($\sim$0.3") but lower spectral resolution (0.7~\kms). }

The images were primary beam corrected, with factors of $\sim$1 for sources in the center of the images (i.e., B1-b, B1-c, S68N, SMM3), and $\sim$3 and $\sim$4 for B1-bS and Ser-emb~8~(N) in Band~6, respectively. In Band~3 the primary beam correction for B1-bS, B1-bN, and Ser-emb~8~(N) was $\sim$1.2, $\sim$2, and $\sim$1.3, respectively.

Continuum images were made using line-free channels in all spectral windows. For the line-dense sources (e.g., B1-c) the line-free channels were carefully selected to exclude any line emission from the continuum images. Using the continuum solutions, all line data were continuum subtracted. From the pipeline product data, it was clear that all \ac{COM} emission (except CH$_3$OH) was unresolved and centrally peaked on all sources. Therefore, line images were made using a mask size of a few times the size of the synthesized beam with the CASA 5.1.1 {\it tclean} task using a Briggs weighting of 0.5. Figure~\ref{fig:ALMAimages} shows the continuum overlayed on the spatial distribution of \mbox{CH$_3$OH 2$_{1,1}$--1$_{0,1}$} and \mbox{CH$_3$OCHO 21$_{7,14}$--20$_{7,13}$} emission {in Band~6} from B1-c, S68N, and B1-bS. {Figure~\ref{fig:ALMAB3images} shows the emission of \mbox{CH$_3$OCH$_3$ 7$_{0,7}$--6$_{1,6}$} in Band~3}. Spectra were extracted from the central beam of the sources.

\section{Spectral modeling and results}
In the Band~3 and Band~6 spectra of B1-b, B1-bN, Ser-emb~8~(N), and SMM3, \ac{COM} emission features are absent at our detection sensitivity (see Table~\ref{tab:obsprop}), except for spatially extended CH$_3$OH emission in Ser-emb 8 (N) which is related to the outflow  \citep{TychoniecIAUS2018,Tychoniec2019}. These sources are therefore excluded in the further analysis and labeled as \ac{COM}-poor sources. The other sources, B1-c, S68N, and B1-bS, are labeled as \ac{COM}-rich sources, and analyzed further below. 

\subsection{Methodology}
\label{sec:specmod} 

\begin{figure*}[t]
\centering
\includegraphics[width=1.0\linewidth]{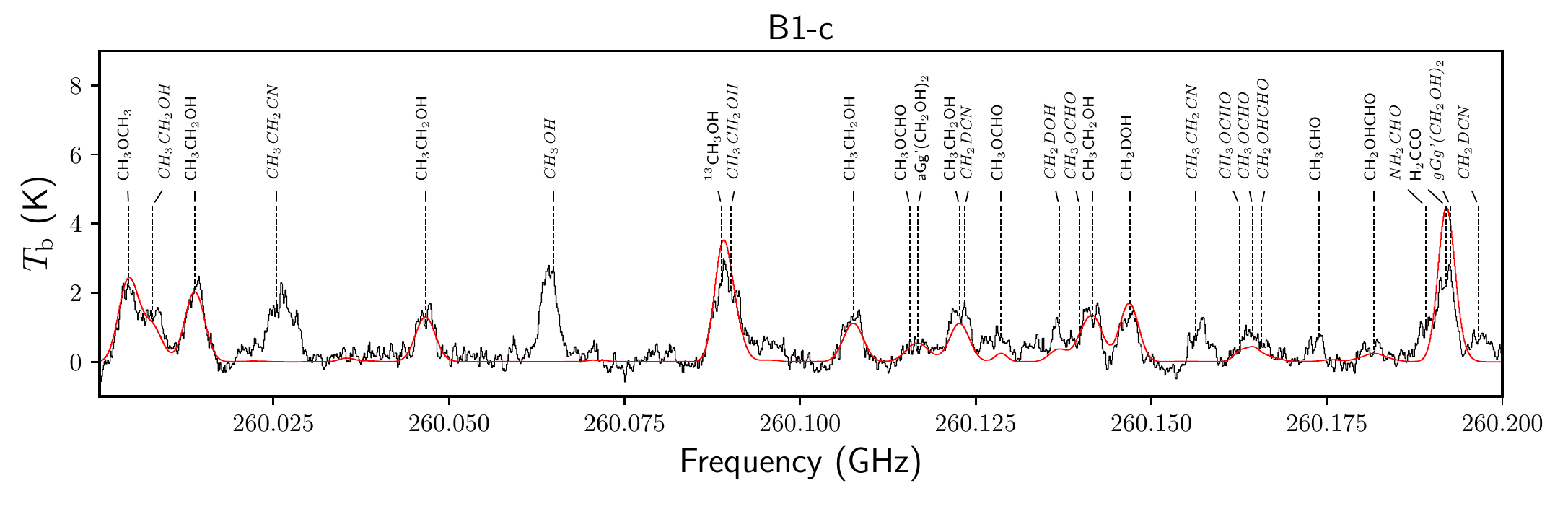}
\caption{Part of the Band~6 spectrum of B1-c shown in black with the best fitting model overplotted in red. The window is centered around a few CH$_3$CHO, CH$_3$OCHO, and CH$_3$CH$_2$OH lines. 
The focus here is on the O-bearing \ac{COM}s; for this reason several transitions clearly visible in the data (i.e., CH$_3$CH$_2$CN) do not show up in the fitting model. {Lines in italics were excluded during the fitting}. 
}
\label{fig:B1c_bestfitmodel}
\end{figure*}

The spectral analysis tool CASSIS\footnote{\url{http://cassis.irap.omp.eu/}} was used to determine column densities and excitation temperatures of detected species. The line lists of each molecule were taken from the CDMS {\citep{Muller2001,Muller2005,Endres2016}}\footnote{\url{https://cdms.astro.uni-koeln.de/}} and JPL {\citep{Pickett1998}}\footnote{\url{https://spec.jpl.nasa.gov/}} catalogs. {The laboratory spectroscopy of molecules discussed in this paper is introduced in Appendix~\ref{app:lab} and} a full list of lines can be found in Tables~\ref{tab:Band3lines} and \ref{tab:Band6lines} for Band~3 and Band~6, respectively. On average, the transition upper energies of the spectral lines are lower in Band~3 compared to Band~6. The Band~3 and Band~6 data are modeled separately. For a fixed $V_\mathrm{lsr}$, a grid of column densities, excitation temperatures, and \ac{FWHM} of the line was set with steps of 0.05 on a logarithmic scale, and 10~K and {0.1}~\kms\text{} on a linear scale, respectively. The total parameter space probed is $10^{13}$--$10^{17}$~cm$^{-2}$, 50--550~K, and 0.5--9.0~\kms, respectively; for CH$_2$DOH the explored parameter space of the column density is $10^{14}$--$5\times10^{17}$~cm$^{-2}$. For species with a small number of transitions in Band~6, the excitation temperature was fixed to $T_\mathrm{ex}=200$~K, and only the column density and \ac{FWHM} were varied. For most species in Band~3, we had to assume that $T_\mathrm{ex}$ is equal to the Band~6 temperature because the \ac{SN} is too low to detect multiple spectral lines. Since the \ac{FWHM} of the lines in B1-bS was close to the spectral resolution limit, we fixed the \ac{FWHM} to 1~\kms. 

A source size of 0.45" (equal to the Band~6 beam) is used for all sources. Given the equation for beam dilution:
\begin{align}
\mathrm{Dilution factor^{-1}} = \frac{\theta^2_\mathrm{s}}{\theta^2_\mathrm{b} + \theta^2_\mathrm{s}},
\label{eq:beamdil}
\end{align}
with $\theta_\mathrm{s}$ the (assumed) angular source size of 0.45" and $\theta_\mathrm{b}$ the beam size, the beam dilution factor is $\sim$20 and 2 for Band~3 and Band~6, respectively. {In reality, the size of the emitting region is smaller than the assumed 0.45" and any derived column densities are thus lower limits. The effect of a smaller source size is discussed in Section~\ref{subsec:sourcesize}}

Each species is modeled separately, and optically thick lines and blended lines are excluded from the fitting procedure.
Assuming that the populations of all levels can be described by a single excitation temperature $T_\mathrm{ex}$, the $\chi^2$ is computed for each grid point, after which the best fitting model is determined by the minimum reduced $\chi^2$. This method is often called '\ac{LTE}' as the high-density limit of molecular excitation in which the excitation temperature approaches the kinetic temperature $T_\mathrm{kin}$. Strictly speaking, thermodynamic equilibrium implies that not only the excitation but also the motions, ionization balance and radiation field are in equilibrium at the same temperature. The word 'local' refers to a specific region in an atmosphere with varying temperature and density, having its origin in stellar astrophysics. In radio astronomy of the interstellar medium, however, the word LTE often just implies a single excitation temperature to characterize the level populations of a molecule, which we will refer to here as 'LTE'.

The 2$\sigma$ uncertainty on the column density, excitation temperature, and \ac{FWHM} is computed from the grid.
The main contributors to the uncertainty are the assumption of 'LTE', the flux calibration error of \ac{ALMA}, and the assumption of Gaussian line profiles. For sources with a high line density, such as B1-c, many lines had to be excluded from the fit due to line blending, therefore increasing the uncertainty on the column densities and excitation temperatures.

\begin{table*}[t]
\begin{center}
\caption{Column densities of CH$_3$OH and abundance fractions (\%) with respect to CH$_3$OH for all \ac{COM} rich sources.}
\begin{tabular}{@{\extracolsep{3pt}}llcccccc@{}}
\hline \hline
& & \multicolumn{2}{c}{B1-c} & \multicolumn{2}{c}{S68N} & \multicolumn{2}{c}{B1-bS} \\
\cline{3-4} \cline{5-6} \cline{7-8} 
Species & Name & Band 3 & Band 6 & Band 3 & Band 6 & Band 3 & Band 6 \\
\hline
CH$_3$OH & Methanol & $18 \pm 4$ & $19 \pm 6$ & $3.6\pm 0.8$ & $14 \pm 6$ & $2.4 \pm 0.7$ & $5.0 \pm 0.6$ \\
\hline 
$^{13}$CH$_3$OH & & $\equiv 1.43$ & $1.0 \pm 0.3$ & $\equiv 1.43$ & $0.7 \pm 0.3$ & $\equiv 1.43$ & $0.3 \pm 0.1$ \\ 
CH$_3^{18}$OH & & $<1.0$ & $\equiv 0.18$ & $<2.8$ & $\equiv 0.18$ & $<4.2$ & $\equiv 0.18$ \\ 
CH$_2$DOH & & $5.7 \pm 2.1$ & $8.4 \pm 2.8$ & $<13$ & $4.3 \pm 1.9$ & $<9.9$ & $<4.7$ \\ 
CH$_3$CH$_2$OH & Ethanol & $1.6 \pm 0.9$ & $0.8 \pm 0.3$ & $<5.6$ & $0.22 \pm 0.09$ & $<8.3$ & $<0.6$ \\  
CH$_3$CHO & Acetaldehyde & $1.3 \pm 0.5$ & $0.2 \pm 0.1$ & $<1.4$ & $0.07 \pm 0.03$ & $<2.1$ & $0.13 \pm 0.02$ \\ 
CH$_3$OCHO & Methyl formate & $2.3 \pm 1.0$ & $1.0 \pm 0.3$ & $0.29 \pm 0.07$ & $1.1 \pm 0.5$ & $6.1 \pm 1.8$ & $0.75 \pm 0.10$ \\  
CH$_3$OCH$_3$ & Dimethyl ether & $1.0 \pm 0.2$ & $1.3 \pm 0.4$ & $1.8 \pm 0.4$ & $0.9 \pm 0.4$ & $4.3 \pm 1.2$ & $1.3 \pm 0.2$ \\ 
CH$_3$COCH$_3$ & Acetone & $<2.8$ & $0.4 \pm 0.1$ & $<14$ & $0.4 \pm 0.2$ & $<21$ & $0.10 \pm 0.03$ \\
aGg'(CH$_2$OH)$_2$ & Ethylene glycol & $<2.4$ & $0.3 \pm 0.1$ & $<12$ & $0.16 \pm 0.07$ & $<18$ & $<0.43$ \\  
gGg'(CH$_2$OH)$_2$ & & $<1.2$ & $0.4 \pm 0.1$ & $<6$ & $<0.81$ & $<8.9$ & $<0.43$ \\ 
CH$_2$OHCHO & Glycolaldehyde & $<0.89$ & $0.11 \pm 0.04$ & $<2.2$ & $<0.06$ & $<3.3$ & $<0.064$ \\ 
H$_2$CCO & Ketene & -- & $0.07 \pm 0.02$ & -- & $0.07 \pm 0.03$ & -- & $0.03 \pm 0.01$ \\ 
t-HCOOH & Formic acid & $0.5 \pm 0.1$ & $0.04 \pm 0.02$ & $<1.4$ & $0.07 \pm 0.03$ & $<4.2$ & $<0.1$ \\  
\hline
\end{tabular}
\label{tab:cassis_abundances}
\tablefoot{
The CH$_3$OH entry is the column density in $10^{17}$~cm$^{-2}$ in a 0.45" beam derived from $^{13}$CH$_3$OH for Band~3 and CH$_3^{18}$OH for Band~6 using $^{12}\mathrm{C}/^{13}\mathrm{C}=70$ and $^{16}\mathrm{O}/^{18}\mathrm{O}=560$, respectively. All other entries are abundances with respect to CH$_3$OH (in \%). 
The errors are derived from the 2$\sigma$ (95\%) uncertainties on the column densities. Abundance ratios which are set to their respective isotope ratio are indicated with a $\equiv$ symbol. 
}
\end{center}
\end{table*}

\subsection{Column densities and excitation temperatures}
\label{subsec:colTex}
The column densities and excitation temperatures of all \ac{COM}-rich sources are presented in Appendix~\ref{app:cassisresults}. We consider a derived column density an upper limit if the error is more than 80\% of the corresponding best fitting value. {The number of lines used for fitting each species is listed in Table~\ref{tab:Nlines}. Upper limits (3$\sigma$) to the column density are provided if no unblended lines are detected.} {For B1-c in Band~6}, unambiguous detections (i.e., $>5$ unblended lines identified) are made for $^{13}$CH$_3$OH, CH$_3^{18}$OH, CH$_2$DOH, CH$_3$CHO, CH$_3$OCHO, CH$_3$OCH$_3$, CH$_3$COCH$_3$, {and aGg'(CH$_2$OH)$_2$}. {For S68N and B1-bS, line blending and sensitivity hampered unambiguous detections for several of these COMs, but tentative detections can still be made for most species.} 
H$_2$CCO and \mbox{t-HCOOH} show $<3$ detected transitions, and their identifications are therefore considered tentative. 
{gGg'(CH$_2$OH)$_2$ and} CH$_2$OHCHO {are} only detected toward B1-c in Band~6; for all other sources (and B1-c Band~3) we put upper limits on the column density for an excitation temperature of 200~K. 
Due to the absence of any optically thin CH$_3$OH lines in our data, the column density of CH$_3$OH is determined by scaling it from $^{13}$CH$_3$OH for Band~3 and CH$_3^{18}$OH for Band~6 \citep[using \mbox{$^{12}$C/$^{13}$C~$\sim$~70} and \mbox{$^{16}$O/$^{18}$O~$\sim$~560} for the local \ac{ISM};][]{Milam2005,Wilson1994}. 
In Fig.~\ref{fig:B1c_bestfitmodel}, part of the Band~6 spectrum of B1-c is shown with the best fitting model overlayed. The full Band~6 spectra with best-fit model of all sources can be found in Appendix~\ref{app:full_spectra}. {The full Band~3 spectrum of B1-c is presented in Appendix~\ref{app:full_B3spectrum}}.

B1-c is the most \ac{COM} rich source of our sample. The measured column densities in Band~6 vary between \mbox{$1.9\times10^{18}$~cm$^{-2}$} for CH$_3$OH and \mbox{$7.0\times10^{14}$~cm$^{-2}$} for \mbox{t-HCOOH} in a $\sim$150~AU diameter region. Our derived column density for CH$_2$OHCHO is \mbox{$2.0\times10^{15}$~cm$^{-2}$}. Most \ac{COM}s show column densities on the order of \mbox{$\sim10^{15-16}$~cm$^{-2}$} in both Band~3 and Band~6.

The column densities toward S68N are a factor \mbox{$\sim1.5-2$} lower than toward B1-c. Similarly, the highest column density is derived for CH$_3$OH \mbox{($1.4\times10^{18}$~cm$^{-2}$)}, the lowest for \mbox{t-HCOOH} \mbox{($9.1\times10^{14}$~cm$^{-2}$)} in Band~6, and \mbox{$\sim10^{15-16}$~cm$^{-2}$} for most \ac{COM}s. 

B1-bS has the lowest column densities measured in this sample, up to an order of magnitude lower than toward B1-c with values of the order of $\sim10^{14-17}$~cm$^{-2}$. It was located in the edge of our ALMA field of view, resulting in a lower sensitivity and an increased level of noise. Our derived column densities (for a source size of 0.45") are up to a factor of a few lower than those derived by \cite{Marcelino2018} for a 0.35" source size, which is roughly the difference in beam dilution. 

For molecules with multiple lines for which a $T_\mathrm{ex}$ is derived, the average excitation temperature in Band~3 is $\sim$~30~K lower than in Band~6. 
{The average \ac{FWHM} for B1-c and S68N {in Band~6} is $3.3\pm 0.2$ and $5.0\pm 0.2$~\kms, respectively (the \ac{FWHM} for \mbox{B1-bS} was fixed to 1~\kms)}. {In Band~3 this is $3.1\pm 0.2$ and $4.5\pm 0.5$, respectively}.{The average \ac{FWHM} in Band~3 is thus slightly lower than in Band~6 for B1-c and S68N, but within the uncertainties.}
This matches the expectation that the Band~3 observations trace the colder more extended regions and Band~6 being more sensitive to warmer compact regions since the transition upper energies in Band~3 are on average lower than in Band~6 (see Tables~\ref{tab:Band3lines}~and~\ref{tab:Band6lines}) {and because the beam size and LAS are larger}. When the Band~3 excitation temperature of all species for which $T_\mathrm{ex}$ was fixed is decreased by 30~K, no significant changes in the column densities and abundances arise. {In Table~\ref{tab:B3100K}, the results of a fit with $T_\mathrm{ex}=100$~K to the Band~3 data are presented. The derived column densities are typically about a factor $\sim2$ lower than derived with $T_\mathrm{ex}$ as a free parameter or fixed to 200~K. }

\subsection{Relative abundances}
\label{subsec:relabun}
Clear isotopologue detections are only made for methanol. The $^{13}$CH$_3$OH/CH$_3^{18}$OH abundance ratio is $<6$ for all targets, which is lower than the elemental value of $\sim$~8. 
This hints at the possibility that even $^{13}$CH$_3$OH is optically thick for these sources.
The CH$_2$DOH/CH$_3$OH ratio is on the order of $\sim1-10$\%. Hints for features of isotopologues are seen for partially deuterated ethanol (CH$_2$DCH$_2$OH) and methyl formate (CH$_3$O$^{13}$CHO), but these lines are too weak to allow for an unambiguous confirmation and deriving physical properties.

In order to compare our derived column densities, we determine the abundance ratios of each species with respect to CH$_3$OH. Because the usual tracer of H$_2$, CO, comes from more extended regions, CH$_3$OH is chosen as the reference species. In Table~\ref{tab:cassis_abundances}, the derived abundances are presented for the three \ac{COM} rich sources. 
The abundances between B1-c and S68N vary within a factor 2-3. This may simply be an effect of the source properties; the lines in S68N are{, for example,} a factor $\sim1.5$ broader ($\sim$5.0~\kms\ and $\sim$3.3~\kms in Band~6, respectively) resulting in more blended lines. 
In S68N, our abundances of CH$_3$OCHO, CH$_3$OCH$_3$, and H$_2$CCO are an order of magnitude lower than those derived by \cite{Bergner2019}. 
The discrepancies are potentially due to different approaches to optical depth correction.
Our observations of B1-bS have a low \ac{SN}. Nevertheless our derived abundance ratios coincide within a factor of a few with the hot (200~K) component modeled by \cite{Marcelino2018}. 

The abundances in Band~3 for B1-c deviate within a factor of {$\sim$2} from the abundances derived in Band~6. Only CH$_3$CHO and \mbox{t-HCOOH} are more abundant based on the Band~3 data. In S68N, {CH$_3$OCHO} is less abundant in Band~3 whereas {CH$_3$OCH$_3$} is more abundant. In B1-bS, {both CH$_3$OCHO and CH$_3$OCH$_3$} are {significantly} more abundant in Band~3, possibly because the column density of CH$_3$OH is underestimated. 

\section{Discussion}
\label{sec:discussion}
\subsection{Occurrence of COMs in young protostars}
The emission of \ac{COM}s is assumed to originate from inside the $T_\mathrm{dust} \sim 100-300$~K radius. Using the relation of \cite{Bisschop2007_hotcores} for the 100~K radius in hot cores: 
\begin{align}
R_{T=100\mathrm{K}}\approx 15.4 \sqrt{\frac{L}{L_\odot}} \mathrm{AU},
\label{eq:RT100}
\end{align}
with $L$ the source luminosity, implies an $R_{T=100\mathrm{K}}$ of {37.4}~AU, 35.8~AU, and {11.6}~AU for B1-c, S68N, and B1-bS, respectively. Our Band~6 beam size of $\sim$0.45" (corresponding to a radius of $\sim72$~AU and $\sim98$~AU for Perseus and Serpens, respectively) is larger than this radius and thus agrees with all \ac{COM} emission being unresolved, except for any CH$_3$OH emission related to an outflow.

\begin{figure*}
\centering
\includegraphics[width=1\linewidth]{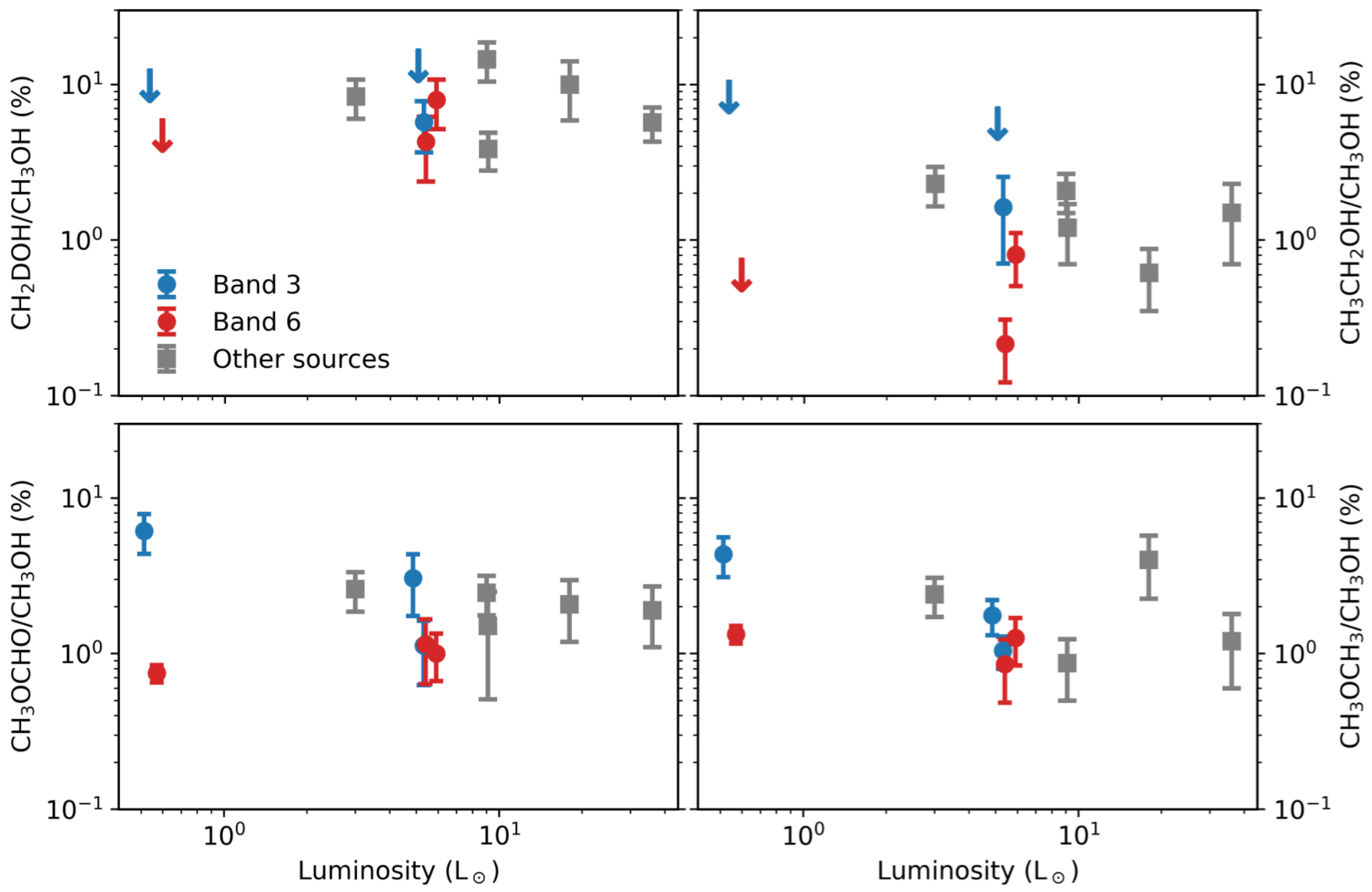}
\caption{Abundance of CH$_2$DOH (top left), CH$_3$CH$_2$OH (top right), CH$_3$OCHO (bottom left), and CH$_3$OCH$_3$ (bottom right) with respect to CH$_3$OH as function of bolometric source luminosity. Both the abundances in Band~3 (blue) and Band~6 (red) are plotted, where the Band~3 data-points are {slightly} shifted in luminosity for clarity. {Arrows denote upper limits}. In gray, the abundances of IRAS~2A and IRAS~4A \citep{Taquet2015,Taquet2019}, HH~212 \citep{Lee2019_HH212}, IRAS~16293A \citep{Manigand2020}, and IRAS~16293B \citep{Jorgensen2018} are presented. }
\label{fig:abun-L}
\end{figure*} 

Previous studies have found that a higher luminosity implies more emission of \ac{COM}s \citep[e.g.,][]{Jorgensen2002,Young2005,Visser2009}. Following Eq.~\ref{eq:RT100} we would indeed expect that the occurrence of \ac{COM} emission is correlated to the source luminosity. Here, three of the seven sources studied show emission of \ac{COM}s. However, our most luminous source, SMM3, does not exhibit any \ac{COM} emission. On the other end, the two sources with the lowest luminosity, B1-bN and B1-b, do not show any \ac{COM} emission either. Interestingly, B1-bS also has a rather low luminosity but does exhibit emission of \ac{COM}s. This suggests that, for the targets that we observed, no clear correlation exists between the occurrence of gaseous \ac{COM}s and source luminosity.   

{Not finding such a} correlation is surprising. The relation for $R_{T=100\mathrm{K}}$ in Eq.~\eqref{eq:RT100}, however, assumes a spherically symmetric infalling envelope \citep{Bisschop2007_hotcores}. In reality, our Class~0 sources may be surrounded by an accretion disk \citep[e.g.,][]{Tobin2012,Murillo2013}. The presence of a disk or disk-like structure may shift the $R_{T=100\mathrm{K}}$ inwards as the temperature in the mid-plane of a disk-like structure is lower than in the surface layers \citep[e.g.,][]{Jorgensen2005,Visser2009}. However, even for a fixed $R_{T=100\mathrm{K}}$, the amount of material with $T_\mathrm{dust}>100\mathrm{K}$ (e.g., high enough to release the \ac{COM}s from the dust grains) can vary between different sources due to, for example, differences in the disk mass and radius \citep{Persson2016}.

The absence of \ac{COM} emission {can} also originate from other source properties. B1-bN has been suggested as a first hydrostatic core candidate \citep[e.g.,][]{Pezzuto2012,Hirano2014,Gerin2017}; the temperature in the inner regions may not have reached the threshold for ice \ac{COM} sublimation. B1-b has a high bolometric temperature, and {could} therefore be more evolved than the other sources. Here only a weak continuum is detected, indicating that B1-b may have a cleared inner region; all gaseous \ac{COM}s have either already been destroyed or are incorporated as ices in the cold outer part. The strong ice absorption observed toward this source {might} originate in a region that is more extended than the inner B1-b envelope \citep{Oberg2011}. SMM3 shows strong continuum, which may imply that increased dust opacity hides any \ac{COM} emission.

Despite the absence of a clear correlation between the occurrence of \ac{COM} emission and source luminosity, an increase in luminosity should result in {higher temperatures and an increased \ac{UV} radiation field} \citep{Visser2009,Drozdovskaya2016}. The abundance of \ac{COM}s with respect to CH$_3$OH may therefore be related to the source luminosity. However, \cite{Taquet2015} and \cite{Belloche2020} have shown that for most \ac{COM} abundances no clear correlation exists with luminosity, similar to what is found here; see Fig.~\ref{fig:abun-L}. Our range of source luminosities is significantly smaller \citep[up to $\sim$~20~L$_\odot$ compared to $10^6$~L$_\odot$;][]{Taquet2015}, and a similar scatter of abundances can be seen for their lower luminosities. 

\subsection{Dependence on source size}
\label{subsec:sourcesize}
In modeling the \ac{COM} emission, a source size equal to the beam size (0.45") was assumed for all sources. However, since the emission is spatially unresolved, the source size in reality will be smaller. A smaller emitting region will increase the derived column densities, and may result in spectral lines actually being optically thick. We can test this scenario by assuming the 100~K {diameter} as a source size. This gives an emitting region of {0.23"}, 0.16" and {0.07}" for B1-c, S68N, and B1-bS, respectively. Assuming the lines remain optically thin, the column densities derived for a 0.45" source size can be scaled to a smaller size by computing the difference in beam dilution (see Eq.~\eqref{eq:beamdil}):
\begin{align}\label{eq:beamdilcomparison}
N_1 \frac{\theta^2_\mathrm{s_1}}{\theta^2_\mathrm{b} + \theta^2_\mathrm{s_1}} = N_2 \frac{\theta^2_\mathrm{s_2}}{\theta^2_\mathrm{b} + \theta^2_\mathrm{s_2}},
\end{align}
where $N_{1,2}$ are the column densities for respective source sizes $\theta_\mathrm{s_{1,2}}$, and $\theta_\mathrm{b}$ is the beam size. 

For B1-c and S68N, the resulting column densities still give optically thin ($\tau<0.1$) lines for all species except CH$_3$OH. However, the lines of CH$_3$OH are already optically thick for a 0.45" source size and the column density is derived from the $^{13}$C and $^{18}$O isotopologues, whose lines remain optically thin. This means that the abundances derived previously in this work for these two sources remain the same. For B1-bS most lines do become optically thick (i.e., \mbox{$\tau>0.1$}), indicating that the abundances derived {above in Table~\ref{tab:cassis_abundances}} for B1-bS should be used with care.  

\subsection{From cold (Band~3) to hot (Band~6) COMs}
The temperature in the envelopes of Class~0 sources is expected to have an onion shell layered structure, with increasing temperature when moving closer to the star \citep[e.g.,][]{Jorgensen2002}. 
In warmer layers the \ac{COM}s {could be} be further processed to form more complex molecules (on the dust grains) and, more importantly, will all be released from the dust grains when the temperature reaches $T_\mathrm{dust}\approx100-300$~K. 
With our multi-Band observations, we can check for hints of this onion-layered temperature structure by comparing the excitation temperature {of \ac{COM}s in Band~3, which is more sensitive to extended emission, to the \ac{COM}s in Band~6, which is more sensitive to compact emission}. Indeed, the average excitation temperature for the few species for which an excitation temperature could be derived in Band~3 (i.e., CH$_3$CH$_2$OH, CH$_3$OCHO, and CH$_3$OCH$_3$) is about 30~K lower than in Band~6. However, as mentioned in Section~\ref{subsec:colTex}, no significant changes in the column densities and abundances arise if the Band~3 excitation temperature of all species for which $T_\mathrm{ex}$ was fixed is decreased by 30~K. The outcome of the Band~6 model does not change when all lines with a low $E_\mathrm{up}<100$~K are excluded. 

\cite{Jorgensen2018} modeled the abundances of O-bearing \ac{COM}s {in IRAS~16293B} and found that they can be divided in two classes with temperatures of $\sim125$~K (warm) and $\sim300$~K (hot). Most \ac{COM}s in our sample have excitation temperatures of $\sim200$~K, which falls in the middle between these two classes. Only CH$_3$OCH$_3$ consistently shows $T_\mathrm{ex}\sim100$~K, both in the Band~3 and Band~6 observations for all sources. This agrees with \cite{Jorgensen2018}, who classified CH$_3$OCH$_3$ as a warm species. However, some ambiguity remains, since CH$_3$OCH$_3$ is also observed to have hot components in, for instance, higher mass cores \citep[][]{Belloche2013,Isokoski2013}.

\subsubsection{Multi-component model}
\label{subsubsec:multicomp}
The emission of \ac{COM}s can be a superposition of a warm/hot inner component, and a second colder and more extended component. Earlier observational studies have shown that indeed a cold ($T<100$~K, i.e., below the desorption temperature), more extended component may be present \citep[e.g.,][]{Bisschop2007_hotcores,Isokoski2013,Fayolle2015,Marcelino2018}. Using our data, a possible cold component can be probed by fitting a multi-component model to the Band~3 data. This is only done for B1-c and S68N since the \ac{SN} of B1-bS in Band~3 is insufficient to carry out such analysis. H$_2$CCO, CH$_3$COCH$_3$, (CH$_2$OH)$_2$, and CH$_2$OHCHO are excluded from this analysis due to the lack of clear detections in our Band~3 data. 

For the warm/hot component, the best-fit model of Band~6 is adopted. This component is kept fixed. As a second component, a colder model is introduced. Here, an excitation temperature of 60~K was assumed and only the column density was fitted for. 
{The \ac{FWHM} was initially set as a free parameter, but fixed to that of the warm/hot component in the cases that the fit results were unconstrained.}
 A source size of 2.0" (equal to Band~3 beam) is assumed for the cold component, compared with the source size of the warm component of 0.45". 

\begin{sidewaystable}
\caption{Column densities resulting from the multi-component analysis.}
\begin{tabular}{@{\extracolsep{3pt}}lcccccccc@{}}
\hline \hline
& \multicolumn{4}{c}{B1-c} & \multicolumn{4}{c}{S68N} \\
\cline{2-5} \cline{6-9} 
& \multicolumn{2}{c}{Component 1: warm/hot} & \multicolumn{2}{c}{Component 2: cold} & \multicolumn{2}{c}{Component 1: warm/hot} & \multicolumn{2}{c}{Component 2: cold} \\
\cline{2-3} \cline{4-5} \cline{6-7} \cline{8-9} 
Species & $N$ (cm$^{-2}$) & X/CH$_3$OH (\%) & $N$ (cm$^{-2}$) & X/CH$_3$OH (\%) & $N$ (cm$^{-2}$) & X/CH$_3$OH (\%) & $N$ (cm$^{-2}$) & X/CH$_3$OH (\%) \\
\hline
CH$_3$OH & $(1.9 \pm 0.6) \times 10^{18}$ & $\equiv 100$ & $(1.2 \pm 0.7) \times 10^{16}$ & $\equiv 100$ & $(1.4 \pm 0.6) \times 10^{18}$ & $\equiv 100$ & $<4.8 \times 10^{14}$ & $\equiv 100$ \\ 
$^{13}$CH$_3$OH & $(1.8 \pm 0.2) \times 10^{16}$ & $1.0 \pm 0.3$ & $(1.7 \pm 1.0) \times 10^{14}$ & $\equiv 1.43$ & $(1.0 \pm 0.1) \times 10^{16}$ & $0.7 \pm 0.3$ & $<1.0 \times 10^{14}$ & $\equiv 1.43$ \\ 
CH$_3^{18}$OH & $(3.4 \pm 1.1) \times 10^{15}$ & $\equiv 0.18$ & $<7.0 \times 10^{15}$ & $<62$ & $(2.5 \pm 1.1) \times 10^{15}$ & $\equiv 0.18$ & $<1.6 \times 10^{15}$ & --\tablefootmark{1} \\
CH$_2$DOH & $(1.6 \pm 0.1) \times 10^{17}$ & $8.4 \pm 2.8$ & $<1.6 \times 10^{14}$ & $<15$ & $(6.0 \pm 0.7) \times 10^{16}$ & $4.3 \pm 1.9$ & $<1.0 \times 10^{14}$ & --\tablefootmark{1} \\
CH$_3$CH$_2$OH & $(1.5 \pm 0.3) \times 10^{16}$ & $0.8 \pm 0.3$ & $(1.2 \pm 0.2) \times 10^{15}$ & $9.8 \pm 6.1$ & $(3.0 \pm 0.2) \times 10^{15}$ & $0.22 \pm 0.09$ & $<1.0 \times 10^{14}$ & --\tablefootmark{1} \\ 
CH$_3$CHO & $(4.6 \pm 1.0) \times 10^{15}$ & $0.24 \pm 0.10$ & $(1.8 \pm 0.2) \times 10^{14}$ & $1.5 \pm 0.9$ & $(1.0 \pm 0.1) \times 10^{15}$ & $0.07 \pm 0.03$ & $<1.0 \times 10^{14}$ & --\tablefootmark{1} \\ 
CH$_3$OCHO & $(1.9 \pm 0.1) \times 10^{16}$ & $1.0 \pm 0.3$ & $(4.7 \pm 0.3) \times 10^{14}$ & $3.8 \pm 2.3$ & $(1.6 \pm 0.2) \times 10^{16}$ & $1.1 \pm 0.5$ & $(1.0 \pm 0.5) \times 10^{14}$ & $> 1$\tablefootmark{2} \\ 
CH$_3$OCH$_3$ & $(2.4 \pm 0.1) \times 10^{16}$ & $1.3 \pm 0.4$ & $<1.0 \times 10^{14}$ & $<0.22$ & $(1.2 \pm 0.1) \times 10^{16}$ & $0.9 \pm 0.4$ & $<1.0 \times 10^{14}$ & --\tablefootmark{1} \\
t-HCOOH & $(7.0 \pm 2.0) \times 10^{14}$ & $0.04 \pm 0.02$ & $(2.2 \pm 0.2) \times 10^{14}$ & $1.8 \pm 1.1$ & $(9.1 \pm 2.1) \times 10^{14}$ & $0.07 \pm 0.03$ & $<1.0 \times 10^{14}$ & --\tablefootmark{1} \\
\hline
\end{tabular}
\tablefoot{The values of the warm/hot Component 1 are the same as derived for Band~6 in a 0.45" beam and are fixed during the fitting (typical $T_\mathrm{ex}\sim200$~K). Component 2 represents cold (typical $T_\mathrm{ex}=60$~K), more extended material in a 2.0" beam. The column density of CH$_3$OH is determined by scaling from $^{13}$CH$_3$OH for Band~3 and CH$_3^{18}$OH for Band~6 using $^{12}\mathrm{C}/^{13}\mathrm{C}=70$ and $^{16}\mathrm{O}/^{18}\mathrm{O}=560$. 
Abundance ratio's which are set to their respective isotope ratio are indicated with a $\equiv$ symbol.
\\
\tablefoottext{1}{No abundance is computed since the column densities of both this species and CH$_3$OH are upper limits.}
\tablefoottext{2}{Lower limit to the abundance due to an upper limit of the CH$_3$OH column density.}
}
\label{tab:multicomp}
\end{sidewaystable}

\begin{figure}[t]
\centering
\includegraphics[width=1.0\linewidth]{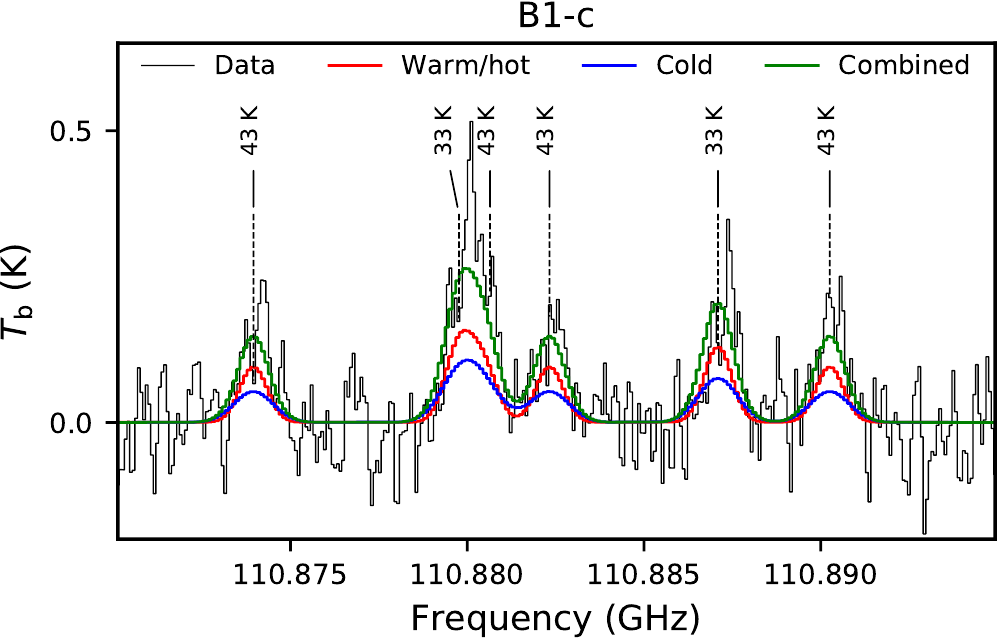}
\caption{Part of the Band~3 spectrum of the warm/hot ($T_\mathrm{ex}\sim100-300$~K) compact (red), cold ($T_\mathrm{ex}=60$~K) extended (blue), and combined (green) model overlayed on the data (black). The spectral features are from CH$_3$OCHO, with the upper energy indicated on top. }
\label{fig:B1c_multicompmodel}
\end{figure}

\begin{figure*}[t]
\centering
\includegraphics[width=1.0\linewidth]{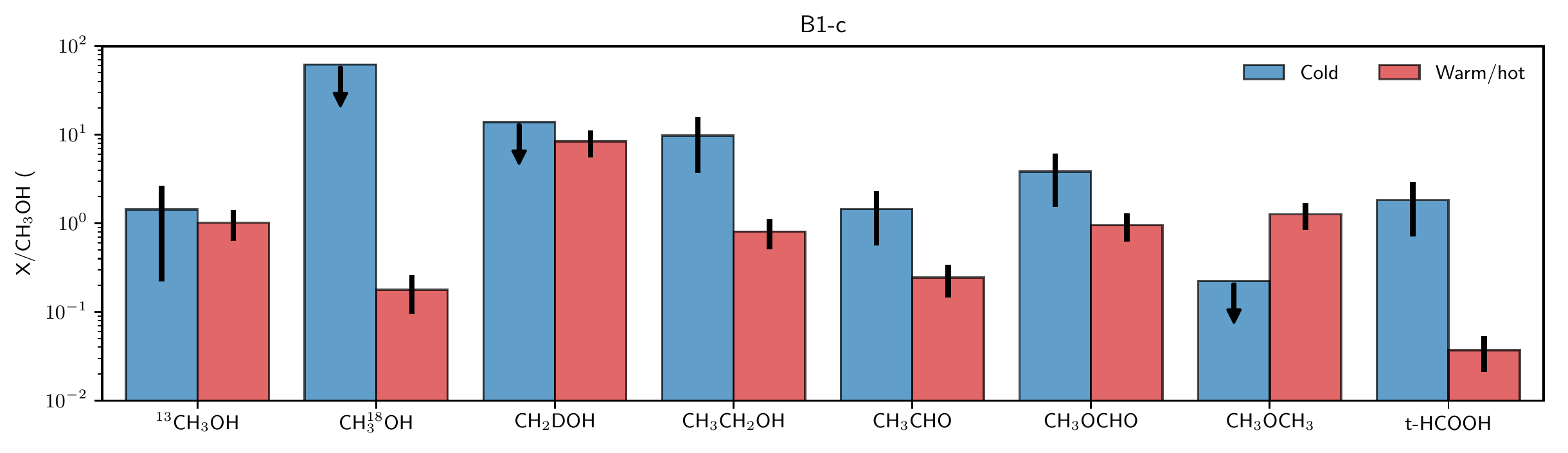}
\caption{Abundance of several species with respect to CH$_3$OH for our multi-component analysis. The warm ($T_\mathrm{ex}\sim200$~K) component is fixed to the Band~6 model with a 0.45" source size, whereas the cold ($T_\mathrm{ex}=60$~K) component is fitted with a more extended 2.0" source size. The 2$\sigma$ (95\%) errors are shown in black, with arrows denoting upper limits. The CH$_3$OH column density is determined by scaling from $^{13}$CH$_3$OH using $^{12}\mathrm{C}/^{13}\mathrm{C}=70$ and $^{16}\mathrm{O}/^{18}\mathrm{O}=560$, respectively. {The higher abundances in the cold component are likely due to an underestimate of the column density of $^{13}$CH$_3$OH}.}
\label{fig:multicomp}
\end{figure*}

The results are presented in Table~\ref{tab:multicomp}. In Fig.~\ref{fig:B1c_multicompmodel} part of the Band~3 spectrum of B1-c centered around several CH$_3$OCHO lines is shown. The abundances of Table~\ref{tab:multicomp} are presented in a bar plot in Fig.~\ref{fig:multicomp}. In B1-c, a cold ($T_\mathrm{ex}=60$~K) and more extended component is detected for most species. The column density of this cold component is typically a few \% but can rise up to 10\% of the warm/hot inner component. {The abundances of more complex species in the cold component are higher than those in the warm/hot component, which contradicts the idea that \ac{COM}s are more effectively processed and released from icy grains in warmer layers. However, one should also note that the column density of $^{13}$CH$_3$OH, and therefore CH$_3$OH, is likely underestimated. }

{Only for CH$_3$CH$_2$OH and CH$_3$OCHO, a \ac{FWHM} of $3.3\pm0.8$ could be derived for the cold component which agrees with the average \ac{FWHM} of the warm/hot component of $3.3\pm0.2$. The line width of the cold component can be expected to be smaller since the emitting material is located further out in the envelope. However, \cite{Murillo2018} find that lower resolution data can also result in a larger line width, especially if the molecule is associated with the outflow or its cavity.}

{The presence of a cold component of \ac{COM} emission} indicates that some gas-phase formation routes are already efficient in the cold envelope as most \ac{COM}s formed through grain-surface chemistry are still frozen out at these temperatures. Or alternatively, that other mechanisms are capable of transferring the \ac{COM}s from the solid state to the gas phase. {Possible mechanisms include photodesorption \citep{Fayolle2011,MunozCaro2016}, desorption of fragments following photo-dissociation \citep{Bertin2016}, low temperature co-desorption \citep{Fayolle2013,Ligterink2018_codesorpion}, or reactive desorption via a thermal hot-spot \citep{Minissale2016,Chuang2018}}. Only for CH$_3$OCH$_3$ and the deuterated and $^{18}$O isotopologues of CH$_3$OH, an extended component seems to be absent.

\begin{figure*}[t]
\centering
\includegraphics[width=1.0\linewidth]{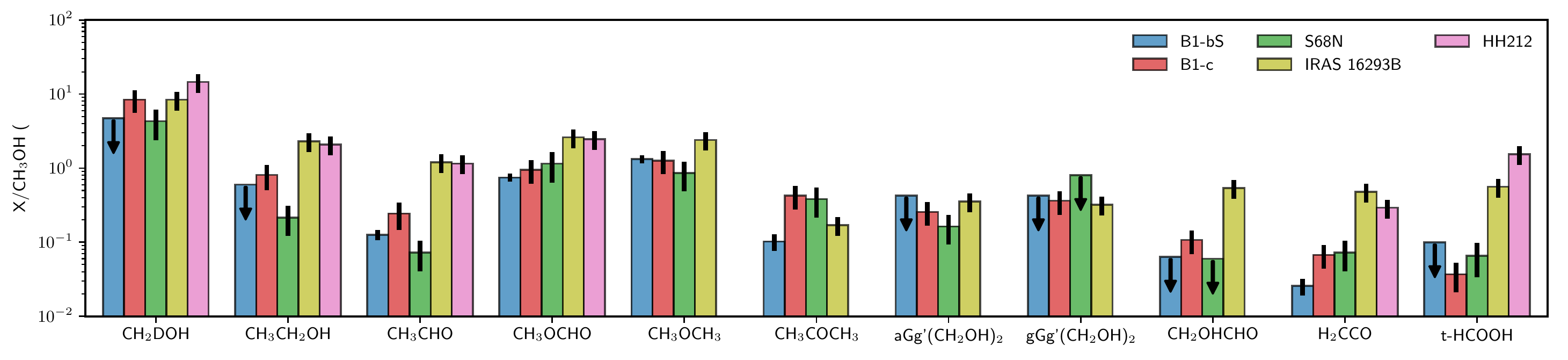}
\caption{Bar plot comparing our derived Band~6 abundances to other sources. The IRAS~16293B abundances for all species except (CH$_2$OH)$_2$, CH$_2$OHCHO, {and CH$_3$COCH$_3$} are from \cite{Jorgensen2018}; the (CH$_2$OH)$_2$ and CH$_2$OHCHO abundances are from \cite{Jorgensen2016}, {and the CH$_3$COCH$_3$ abundances from \cite{Lykke2017}}. The HH~212 abundances are from \cite{Lee2019_HH212}. The 2$\sigma$ (95\%) errors are shown in black, with arrows denoting upper limits. We note that {the abundances of our targets are remarkably similar and show comparable values to those derived for} IRAS 16293B and HH~212. 
}
\label{fig:IRAS16293barplot}
\end{figure*}

Formic acid is very abundant in the cold, extended component. This disagrees with the hot identification in IRAS~16293B \citep{Jorgensen2018}, but is more consistent with the cold identification in high-mass star-forming regions \citep{Bisschop2007_hotcores,Belloche2013} and it likely being an abundant ice species \citep{Oberg2011,Boogert2015}. However, both in Band~3 and Band~6, only 1--3 lines could be fitted, hence the identification of \mbox{t-HCOOH} {should be considered} rather tentative. 

In S68N, a cold component is less evident. Only for CH$_3$OCHO a cold column density is derived; for all other species only upper limits are found. These results suggest that a possible cold component would have a column density of $\lesssim1$\% of the warm component.

\subsection{Comparison to other sources}
Our Band~6 abundances are compared to the \ac{ALMA} Band~7 abundances of IRAS~16293B \citep{Jorgensen2018} and HH~212 \citep{Lee2019_HH212}, see Fig.~\ref{fig:IRAS16293barplot}. The abundances of several \ac{COM}s (i.e., CH$_2$DOH, CH$_3$OCHO, CH$_3$OCH$_3$, and CH$_3$COCH$_3$) from this work are very similar to IRAS~16293B and HH~212; the distribution of abundances seems to span only a factor of a few. The abundances {are also within a factor 3} of IRAS~2A and IRAS~4A \citep{Taquet2015}. This similarity in abundances is rather interesting given that the source properties (e.g., $L_\mathrm{bol}$, $T_\mathrm{bol}$, etc.) are different for all sources. For example, {IRAS~2A} has a luminosity of {36~L$_\odot$}, whereas the luminosity of B1-bS is two orders of magnitude lower ({0.57}~L$_\odot$). A similarly large range is seen for the bolometric temperature and envelope mass. Moreover, these sources originate from four different star-forming regions: Perseus (\mbox{B1-bS}, \mbox{B1-c}, IRAS~2A, IRAS~4A), Serpens (S68N), Ophiuchus (IRAS~16293B), and Orion (HH~212). Apparently, despite the differences in source properties and local environment, the internal distribution of these \ac{COM} abundances is roughly similar. This indicates that their formation schemes must follow a rather universal recipe, for instance following the atom addition and abstraction reaction networks proposed to govern the surface chemistry in cold interstellar clouds \citep[e.g.,][]{Chuang2016}. This implies that solid-state processes are already enriching the complex organic inventory of the interstellar medium in the prestellar phase.

Interestingly, our sources are under-abundant {up to more than one order of magnitude} in CH$_3$CH$_2$OH, CH$_3$CHO, and CH$_2$OHCHO compared with IRAS~16293B, HH~212, IRAS~2A and IRAS~4A. 
Our abundances of these \ac{COM}s {do fall within a factor of a few with} IRAS~16293A \citep{Manigand2020}.
This suggests that for some \ac{COM}s local source conditions do become important for the formation or the destruction. The column density ratio of CH$_3$OCHO to CH$_2$OHCHO of B1-c is consistent with the lower part of the bimodal distribution found by \cite{ElAbd2019}.
For H$_2$CCO and \mbox{t-HCOOH}, our sources also have lower abundances. The abundance of \mbox{t-HCOOH} in the cold component of the multi component analysis (see Section~\ref{subsubsec:multicomp}) is consistent with IRAS~16293B and HH~212. However, the detections of these species are rather tentative since only $<3$ clear unblended lines (not taking into account (hyper)fine structured lines) are detected.

In more massive star-forming regions, a larger scatter in \ac{COM} abundances seems to be present.
AFGL~4176 has abundances very similar to ours for most \ac{COM}s \citep{Bogelund2019}, whereas those in Orion~KL show deviations: the abundance of CH$_3$CH$_2$OH is similar \citep{Tercero2018}, but CH$_3$OCHO, CH$_3$OCH$_3$ have lower abundances in our sources, and CH$_3$COCH$_3$ higher {abundances} \citep[see Table~6 in][]{Bogelund2019}. Even larger deviations are found toward Sagittarius~B2, where the abundances of all \ac{COM}s are higher by up to an order of magnitude \citep{Belloche2016}. 

The abundances derived here are higher than those in chemical models of a collapsing cloud for most \ac{COM}s \citep[e.g.,][]{Garrod2013}. Especially for CH$_3$COCH$_3$ and (CH$_2$OH)$_2$, these models underestimate the abundances by up to an order of magnitude. {This may be a direct result for a still not fully characterized solid-state and gas-phase chemistry and underlying gas-grain interactions} \citep[see discussion by][]{Bogelund2019}.

\subsection{Comparison with ices}
No clear \ac{COM} ice features have (yet) been observed toward any of our \ac{COM}-rich sources. However, toward B1-b (from the same cloud as B1-c) hints for CH$_3$CHO, CH$_3$CH$_2$OH, and \mbox{t-HCOOH} ice have been detected with an abundance of $\sim10$\% with respect to CH$_3$OH \citep{Oberg2011,Boogert2015}. These are higher than those derived for the gaseous warm/hot component in this work ($\sim0.1-1$\%). 
A possible explanation for this is that, besides from typical grain-surface chemistry, additional gas-phase chemistry in the cold envelope can contribute in transforming CH$_3$OH into more complex species such as CH$_3$CHO and CH$_3$CH$_2$OH and thus lowering the abundance of these species with respect to CH$_3$OH {\citep[e.g.,][]{Balucani2015}}. Another explanation is that {other non-thermal desorption mechanisms, such as desorption of fragments following photo-dissociation \citep{Bertin2016}, are active.}

To better compare the abundances of \ac{COM}s in the ice and gas phase, it is essential to directly observe the ices. However, the above mentioned detections are still very uncertain, with only CH$_3$OH being securely identified in interstellar ices. Measuring the column densities of interstellar \ac{COM} ices remains difficult since current infrared telescopes lack the sensitivity or wavelength coverage, and because the spectral features originating from ices are broad and depend on the ice composition. Moreover, the number of {\ac{COM}s} for which extensive {high resolution infrared laboratory spectra} have become available is limited \citep{Hudson2018,TerwisschavanScheltinga2018}. 
Future \acl{MIR} facilities such as \ac{JWST}/MIRI will allow for directly observing \ac{COM}s in the ice, which will be essential for studying their formation and evolution from the cold envelope toward the warm inner regions. 

\subsection{Temperature dependence of deuterated methanol}
The deuteration fraction of methanol is an indication of the temperature during the {solid-state} formation of this species. At low temperatures, deuterium is more efficient than regular hydrogen in the hydrogenation process on interstellar dust grains \citep[e.g.,][]{Nagaoka2005}. This leads to a higher deuteration fraction than the elemental ratio \citep[$\mathrm{D}/\mathrm{H}\sim10^{-5}$;][]{Caselli2012}. \cite{Bogelund2018} investigated the effect of temperature during the formation of methanol with the astrochemical GRAINOBLE model \citep{Taquet2012}. In order to compare our methanol D/H ratios to their model results, our {gas-phase} CH$_2$DOH abundances have to be corrected for statistical weights; a D atom has a 3 times higher probability to stick into the CH$_3$ group than into the OH group. 
In Band~3, this gives a D/H ratio of $1.9\pm0.7\%$, $<4.3\%$, and $<3.3\%$ for B1-c, S68N, and B1-bS respectively.
In Band~6, this is $2.7\pm0.9\%$, $1.4\pm0.6\%$, and $<1.6\%$. These deuteration fractions correspond to a temperature during the formation of methanol of $\sim15$~K, where a density of $n_\mathrm{H}=10^{5-6}$~cm$^{-3}$ was assumed \citep[cf. Fig.~8 of ][]{Bogelund2018}.

Our derived methanol D/H fractions agree well with other well-studied low-mass protostars such as IRAS16293B \citep[$\sim2\%$;][]{Jorgensen2018}, IRAS~2A and IRAS~4A \citep[$\sim2\%$ and $\sim1\%$, respectively;][]{Taquet2019}, and HH~212 \citep[$\sim~1\%$ after taking into account statistical weights;][]{Bianchi2017,Taquet2019}. Our fractions are up to an order of magnitude higher than for high-mass star-forming regions such as NGC~6334I \citep[$\sim0.1\%$ on average;][]{Bogelund2018}, and Sagittarius~B2 \citep[$0.04\%$ after statistical correction;][]{Belloche2016}. The lower D/H ratios in more massive star-forming regions hint at a higher temperature (i.e., $>20$~K) already during the earliest phase of high-mass star formation. 

\section{Summary}
\label{sec:summary}
In this paper, the content of oxygen bearing \ac{COM}s is studied in seven young protostellar systems. High resolution ALMA observations in both Band~3 and Band~6 were used to determine the column densities and excitation temperatures. The main conclusions are as follows:
\begin{itemize}
\item Three out of the seven sources exhibit warm \ac{COM} emission features with typical $T_\mathrm{ex}\sim200$~K. There seems to be no correlation between the bolometric luminosity (and temperature) and occurrence of \ac{COM}s of these Class~0 sources. {Other effects, such as disk formation and clearing of an inner region, may play a role as well.}

\item A multi-component model was introduced to check whether a cold ($T_\mathrm{ex}=60$~K), more extended component of \ac{COM} emission is present. Only B1-c shows clear signs of such a cold component, with typically a few \% of the column density of the warm component. {This indicates that in the cold envelope already gas-phase formation may take place and that this could be a direct consequence of the grain-gas transitions of COMs or (radical) COM fragments.} \mbox{t-HCOOH} is significantly more abundant in the cold component, hinting at the cold {origin} of this species. 

\item The abundance of several \ac{COM}s (i.e., CH$_2$DOH, CH$_3$OCHO, CH$_3$OCH$_3$, CH$_3$COCH$_3$, and (CH$_2$OH)$_2$) are remarkably similar in comparison to other young Class 0 protostars in different star-forming regions such as IRAS~16293-2422B and HH~212. This similarity suggests that the distribution of these \ac{COM} abundances is roughly fixed at an early cold, i.e, prestellar stage. However, the abundances of some \ac{COM}s (i.e., CH$_3$CH$_2$OH, CH$_3$CHO, and CH$_2$OHCHO) in our sources differ significantly compared to IRAS~16293B and HH~212, possibly because local source conditions become important {and affect formation and destruction pathways in different ways. Astrochemical modeling will be needed to extend on these findings.}

\item The D/H ratio of deuterated CH$_3$OH is around a few \% for all \ac{COM}-rich sources, suggesting a dust temperature of $\sim 15$~K during its solid-state formation. The derived formation temperature is similar to other low-mass protostars, but lower than in high-mass star-forming regions ($> 20$~K).
\end{itemize}

\noindent Our sample of \ac{COM}-rich protostars studied on solar system scales remains small. It is still unknown how much of the prestellar cloud material is reprocessed during the evolution toward protostellar systems and more evolved sources. The abundance of most \ac{COM}s in the ice phase is still highly uncertain. Future mid-infrared facilities, most notably the {\it James Webb Space Telescope}, will provide vital information on \ac{COM} ices. A larger sample of \ac{COM}-rich sources with ALMA, both in the Class 0 phase and in the more evolved Class~I phase where most \ac{COM} emission has already disappeared, would provide better constraints on the chemical complexity during these earliest phases of star formation. The present work already shows that such studies will be valuable, given the rich distribution in COMs discussed here.

\begin{acknowledgements}
This paper makes use of the following ALMA data: ADS/JAO.ALMA\#2017.1.01174.S and ADS/JAO.ALMA\#2017.1.01350.S. ALMA is a partnership of ESO (representing its member states), NSF (USA) and NINS (Japan), together with NRC (Canada), MOST and ASIAA (Taiwan), and KASI (Republic of Korea), in cooperation with the Republic of Chile. The Joint ALMA Observatory is operated by ESO, AUI/NRAO and NAOJ.  

Astrochemistry in Leiden is supported by the Netherlands Research School for Astronomy (NOVA).

H.B. acknowledges support from the European Research Council under the Horizon 2020 Framework Program via the ERC Consolidator Grant CSF-648505. H.B. also acknowledges support from the Deutsche Forschungsgemeinschaft (DFG) via Sonderforschungsbereich (SFB) 881 “The Milky Way System” (sub-project B1).
\end{acknowledgements}


\bibliographystyle{aa}
\bibliography{refs}

\clearpage

\begin{appendix}

\section{Laboratory spectroscopic data}
\label{app:lab}
{All laboratory spectroscopic data are acquired using the CDMS \citep{Muller2001,Muller2005,Endres2016} and JPL \citep{Pickett1998} catalogs. In some of the entries, vibrational or torsional states are not taken into account in the calculation of the partition function. At low temperatures, the contribution of these states are negligible. At higher temperatures (i.e., $T>100~K$), however, a so-called vibrational correction factor must be applied if the torsional and vibrational states are not accounted for in the partition functions. Below, we detail the laboratory spectroscopy of each individual species discussed in this paper and indicate if a vibrational correction factor was applied.

The data of CH$_3$OH are taken from the CDMS database and are based on the work of \cite{Xu2008}. The ground state and first three torsional states are taken into account in the partition function. The entry of the $^{13}$C isotopologue takes the first two torsional states into account \citep{Xu1997}. They assumed that the dipole moment is the same as the main isotopologue and calculated the partition function taking only the permanent dipole moment into account. The CH$_3^{18}$OH entry is based mostly on the data of \cite{Fisher2007}. The spectroscopic data of CH$_2$DOH are taken from the JPL catalog, where the entry is based on the work of \cite{Pearson2012}. The vibrational correction factor at 200~K is calculated to be 1.181 by summing up over torsional substates extrapolated from \citet{Lauvergnat2009}

CH$_3$CH$_2$OH exists as two conformers: {\it trans} and {\it gauche}, which originates from the torsion in the OH group. The spectroscopic data are taken from the CDMS catalog, where the entry is based on \citet{Pearson2008} and updated by \citet{Muller2016}, who noticed that the old catalog entry did not predict the line intensities correctly around 3~mm in Sagittarius~B2. This applies also to other frequency regions, but not to all transitions. Both conformers are included in the entry. 

The data of CH$_3$CHO are taken from the JPL catalog. It is based on laboratory spectroscopy of \cite{Kleiner1996} and includes the first and second torsional state in its partition function.

The spectroscopic data of CH$_3$OCHO are taken from the JPL database. The entry is based on the work of \cite{Ilyushin2009} and \cite{Brown1975}. The partition function includes states up to the first torsional state.

CH$_3$OCH$_3$ data are taken from the CDMS entry, which includes mostly the work of \cite{Endres2009}. The partition function includes the ground vibrational state and excited states up to 500~cm$^{-1}$. However, for the excitation temperatures derived in this work ($\sim$100~K), the contribution of higher order vibrational states to the partition function is negligible. 

The CH$_3$COCH$_3$ data are taken form the JPL catalog. The rotational transitions were calculated by \citet{Groner2002} and include only the ground state. Here, we use an updated version of the entry. It is based on the study of \cite{Ordu2019} with the majority of the lines from \citet{Groner2002}. 

The (CH$_2$OH)$_2$ structure consists of 3 coupled rotors, which give rise to ten different stable configurations \citep{Christen2001}. Here, only the two lowest energy conformers of (CH$_2$OH)$_2$ are studied. The aGg' state is the lowest energy state whereas the gGg' state lies about 290~K above that \citep{Muller2004}. The spectroscopic data are taken from the CDMS entries, which are based on the work of \cite{Christen1995} and \cite{Christen2003} for the aGg' conformer and \cite{Christen2001} and \cite{Muller2004} for the gGg' conformer. For gGg'(CH$_2$OH)$_2$, the partition function and line intensities were calculated assuming as if it is the lowest energy state conformer. Since for these calculations only the ground state is considered, we approximate the effect of vibrational states by using a vibrational correction factor of 2.143 for both conformers at 200~K.

The CH$_2$OHCHO data are taken from the CDMS catalog. The entry is based on the work of \cite{Butler2001} and only includes data on the ground vibrational state. A vibrational correction factor of 1.5901 is calculated at 200~K  in the harmonic approximation.

The spectroscopic data of H$_2$CCO are acquired from the CDMS catalog. The entry is calculated mostly by \cite{Johnson1952}, \cite{Fabricant1977}, and \cite{Brown1990}. The contribution of vibrational states is assumed to be negligible. 

Formic acid, HCOOH, exists both in a $trans$ and $cis$ state, where the lowest energy state of \mbox{c-HCOOH} is about 2000~K higher than for \mbox{t-HCOOH}. In this study we only detect \mbox{t-HCOOH}, for which we use the entry of the JPL catalog. This is based on the work of \cite{Bellet1971}. The vibrational correction is $<1.1$ at 200~K and is therefore neglected.
}

\clearpage

\section{CASSIS modeling results}
\label{app:cassisresults}
\renewcommand{\arraystretch}{1.1}
\noindent\begin{minipage}{\textwidth}
\begin{center}
\captionof{table}{Computed Band~3 and Band~6 column densities and excitation temperatures of our sources.}
\vspace{-4.5mm}
\begin{tabular}{@{\extracolsep{3pt}}llcccccc@{}}
\multicolumn{8}{c}{\textbf{B1-c}} \\
\hline \hline
& & \multicolumn{3}{c}{Band 3} & \multicolumn{3}{c}{Band 6} \\
\cline{3-5} \cline{6-8} 
Species  & Catalog & $T_\mathrm{ex}$ (K) & $N$\tablefootmark{1} (cm$^{-2}$) & X/CH$_3$OH (\%) & $T_\mathrm{ex}$ (K) & $N$\tablefootmark{1} (cm$^{-2}$) & X/CH$_3$OH (\%) \\
\hline
CH$_3$OH & CDMS & -- & $(1.8 \pm 0.4) \times 10^{18}$ & $\equiv 100$ & -- & $(1.9 \pm 0.6) \times 10^{18}$ & $\equiv 100$ \\ 
$^{13}$CH$_3$OH & CDMS & [190] & $(2.6 \pm 0.6) \times 10^{16}$ & $\equiv 1.43$ & $190 \pm 30$ & $(1.8 \pm 0.2) \times 10^{16}$ & $1.0 \pm 0.3$ \\ 
CH$_3^{18}$OH & CDMS & [140] & $<1.8 \times 10^{16}$ & $<1.0$ & $140 \pm 60$ & $(3.4 \pm 1.1) \times 10^{15}$ & $\equiv 0.18$ \\ 
CH$_2$DOH & JPL & $210 \pm 60$ & $(1.0 \pm 0.3) \times 10^{17}$ & $5.7 \pm 2.1$ & $180 \pm 20$ & $(1.6 \pm 0.1) \times 10^{17}$ & $8.4 \pm 2.8$ \\ 
CH$_3$CH$_2$OH & CDMS & $160 \pm 60$ & $(2.9 \pm 1.5) \times 10^{16}$ & $1.6 \pm 0.9$ & $250 \pm 60$ & $(1.5 \pm 0.3) \times 10^{16}$ & $0.8 \pm 0.3$ \\ 
CH$_3$CHO & JPL & $260 \pm 50$ & $(2.4 \pm 0.8) \times 10^{16}$ & $1.3 \pm 0.5$ & $260 \pm 60$ & $(4.6 \pm 1.0) \times 10^{15}$ & $0.24 \pm 0.10$ \\ 
CH$_3$OCHO & JPL & $120 \pm 40$ & $(2.0 \pm 0.8) \times 10^{16}$ & $1.1 \pm 0.5$ & $180 \pm 20$ & $(1.9 \pm 0.1) \times 10^{16}$ & $1.0 \pm 0.3$ \\ 
CH$_3$OCH$_3$ & CDMS & $100 \pm 10$ & $(1.9 \pm 0.1) \times 10^{16}$ & $1.0 \pm 0.2$ & $120 \pm 10$ & $(2.4 \pm 0.1) \times 10^{16}$ & $1.3 \pm 0.4$ \\ 
CH$_3$COCH$_3$ & JPL & [200] & $<5.0 \times 10^{16}$ & $<2.8$ & [200] & $(8.0 \pm 0.9) \times 10^{15}$ & $0.4 \pm 0.1$ \\ 
aGg'(CH$_2$OH)$_2$ & CDMS & [180] & $<4.2 \times 10^{16}$ & $<2.4$ & $180 \pm 100$ & $(4.8 \pm 0.6) \times 10^{15}$ & $0.3 \pm 0.1$ \\  
gGg'(CH$_2$OH)$_2$ & CDMS & [200] & $<2.2 \times 10^{16}$ & $<1.2$ & [200] & $(6.8 \pm 0.8) \times 10^{15}$ & $0.4 \pm 0.1$ \\ 
CH$_2$OHCHO & CDMS & [200] & $<1.6 \times 10^{16}$ & $<0.89$ & [200] & $(2.0 \pm 0.2) \times 10^{15}$ & $0.11 \pm 0.04$ \\ 
H$_2$CCO & CDMS & -- & -- & -- & [200] & $(1.3 \pm 0.1) \times 10^{15}$ & $0.07 \pm 0.02$ \\
t-HCOOH & JPL & [200] & $(9.0 \pm 1.0) \times 10^{15}$ & $0.5 \pm 0.1$ & [200] & $(7.0 \pm 2.0) \times 10^{14}$ & $0.04 \pm 0.02$ \\  
\hline
\\
\multicolumn{8}{c}{\textbf{S68N}} \\
\hline \hline
CH$_3$OH & CDMS & -- & $(3.6 \pm 0.8) \times 10^{17}$ & $\equiv 100$ & -- & $(1.4 \pm 0.6) \times 10^{18}$ & $\equiv 100$ \\ 
$^{13}$CH$_3$OH & CDMS & [160] & $(5.1 \pm 1.2) \times 10^{15}$ & $\equiv 1.43$ & $160 \pm 20$ & $(1.0 \pm 0.1) \times 10^{16}$ & $0.7 \pm 0.3$ \\ 
CH$_3^{18}$OH & CDMS & [200] & $<1.0 \times 10^{16}$ & $<2.8$ & $200 \pm 80$ & $(2.5 \pm 1.1) \times 10^{15}$ & $\equiv 0.18$ \\ 
CH$_2$DOH & JPL & [200] & $<4.7 \times 10^{16}$ & $<13$ & [200] & $(6.0 \pm 0.7) \times 10^{16}$ & $4.3 \pm 1.9$ \\ 
CH$_3$CH$_2$OH & CDMS & [200] & $<2.0 \times 10^{16}$ & $<5.6$ & [200] & $(3.0 \pm 0.2) \times 10^{15}$ & $0.22 \pm 0.09$ \\
CH$_3$CHO & JPL & [200] & $<5.0 \times 10^{15}$ & $<1.4$ & [200] & $(1.0 \pm 0.1) \times 10^{15}$ & $0.07 \pm 0.03$ \\  
CH$_3$OCHO & JPL & $220 \pm 80$ & $(1.6 \pm 0.9) \times 10^{16}$ & $3.9 \pm 2.5$ & $290 \pm 30$ & $(1.6 \pm 0.2) \times 10^{16}$ & $1.1 \pm 0.5$ \\ 
CH$_3$OCH$_3$ & CDMS & $110 \pm 10$ & $(6.4 \pm 0.7) \times 10^{15}$ & $1.8 \pm 0.4$ & $90 \pm 10$ & $(1.2 \pm 0.1) \times 10^{16}$ & $0.9 \pm 0.4$ \\ 
CH$_3$COCH$_3$ & JPL & [200] & $<5.0 \times 10^{16}$ & $<14$ & [200] & $(5.3 \pm 0.3) \times 10^{15}$ & $0.4 \pm 0.2$ \\ 
aGg'(CH$_2$OH)$_2$ & CDMS & [200] & $<4.2 \times 10^{16}$ & $<12$ & [200] & $(2.3 \pm 0.1) \times 10^{15}$ & $0.16 \pm 0.07$ \\ 
gGg'(CH$_2$OH)$_2$ & CDMS & [200] & $<2.2 \times 10^{16}$ & $<6$ & [200] & $<1.1 \times 10^{16}$ & $<0.81$ \\ 
CH$_2$OHCHO & CDMS & [200] & $<8.0 \times 10^{15}$ & $<2.2$ & [200] & $<8.0 \times 10^{14}$ & $<0.06$ \\ 
H$_2$CCO & CDMS & -- & -- & -- & [200] & $(1.0 \pm 0.1) \times 10^{15}$ & $0.07 \pm 0.03$ \\ 
t-HCOOH & JPL & [200] & $<5.0 \times 10^{15}$ & $<1.4$ & [200] & $(9.1 \pm 2.1) \times 10^{14}$ & $0.07 \pm 0.03$ \\ 
\hline
\\
\multicolumn{8}{c}{\textbf{B1-bS}} \\
\hline \hline
\hline
CH$_3$OH & CDMS & -- & $(2.4 \pm 0.7) \times 10^{17}$ & $\equiv 100$ & -- & $(5.0 \pm 0.6) \times 10^{17}$ & $\equiv 100$ \\ 
$^{13}$CH$_3$OH & CDMS & [160] & $(3.5 \pm 1.0) \times 10^{15}$ & $\equiv 1.43$ & $160 \pm 40$ & $(1.5 \pm 0.5) \times 10^{15}$ & $0.3 \pm 0.1$ \\ 
CH$_3^{18}$OH & CDMS & [200] & $<1.0 \times 10^{16}$ & $<4.2$ & [200] & $(9.0 \pm 1.0) \times 10^{14}$ & $\equiv 0.18$ \\ 
CH$_2$DOH & JPL & [200] & $<2.4 \times 10^{16}$ & $<9.9$ & [200] & $<2.4 \times 10^{16}$ & $<4.7$ \\ 
CH$_3$CH$_2$OH & CDMS & [200] & $<2.0 \times 10^{16}$ & $<8.3$ & [200] & $<3.0 \times 10^{15}$ & $<0.6$ \\ 
CH$_3$CHO & JPL & [200] & $<5.0 \times 10^{15}$ & $<2.1$ & [200] & $(6.4 \pm 0.7) \times 10^{14}$ & $0.13 \pm 0.02$ \\ 
CH$_3$OCHO & JPL & [200] & $(1.5 \pm 0.1) \times 10^{16}$ & $6.1 \pm 1.8$ & [200] & $(3.8 \pm 0.2) \times 10^{15}$ & $0.75 \pm 0.10$ \\ 
CH$_3$OCH$_3$ & CDMS & $90 \pm 10$ & $(1.1 \pm 0.1) \times 10^{16}$ & $4.3 \pm 1.2$ & $110 \pm 10$ & $(6.7 \pm 0.4) \times 10^{15}$ & $1.3 \pm 0.2$ \\ 
CH$_3$COCH$_3$ & JPL & [200] & $<5.0 \times 10^{16}$ & $<21$ & [200] & $(5.1 \pm 1.2) \times 10^{14}$ & $0.10 \pm 0.03$ \\ 
aGg'(CH$_2$OH)$_2$ & CDMS & [200] & $<4.2 \times 10^{16}$ & $<18$ & [200] & $<2.2 \times 10^{15}$ & $<0.43$ \\ 
gGg'(CH$_2$OH)$_2$ & CDMS & [200] & $<2.2 \times 10^{16}$ & $<8.9$ & [200] & $<2.2 \times 10^{15}$ & $<0.43$ \\ 
CH$_2$OHCHO & CDMS & [200] & $<8.0 \times 10^{15}$ & $<3.3$ & [200] & $<3.2 \times 10^{14}$ & $<0.064$ \\ 
H$_2$CCO & CDMS & -- & -- & -- & [200] & $(1.3 \pm 0.3) \times 10^{14}$ & $0.03 \pm 0.01$ \\ 
t-HCOOH & JPL & [200] & $<1.0 \times 10^{16}$ & $<4.2$ & [200] & $<5.0 \times 10^{14}$ & $<0.1$ \\ 
\hline
\end{tabular}
\label{tab:cassisresults_B1bS}
\tablefoot{The column density of CH$_3$OH is determined by scaling $^{13}$CH$_3$OH for Band~3 and CH$_3^{18}$OH for Band~6 using $^{12}\mathrm{C}/^{13}\mathrm{C}=70$ and $^{16}\mathrm{O}/^{18}\mathrm{O}=560$, respectively. A value between squared brackets means the parameter was fixed to that value during the computation. 
Abundance ratio's which are set to their respective isotope ratio are indicated with a $\equiv$ symbol.\\
\tablefoottext{1}{{The presented column densities are derived for a 0.45" source size. However, in reality the source sizes will be smaller, and the presented column densities thus represent lower limits.} }
}
\end{center}
\end{minipage}

\clearpage

\begin{table}
\begin{center}
\caption{Number of transitions and lines fitted for all sources.}
\begin{tabular}{@{\extracolsep{3pt}}lcccccc@{}}
\multicolumn{7}{c}{\textbf{B1-c}} \\
\hline \hline
& \multicolumn{3}{c}{Band 3} & \multicolumn{3}{c}{Band 6}  \\
\cline{2-4} \cline{5-7}
Species & T & L & F & T & L & F \\
\hline
CH$_3$OH & 1 & 1 & 0 & 3 & 3 & 0\\
$^{13}$CH$_3$OH & 2 & 1 & 1 & 8 & 7 & 4\\
CH$_3^{18}$OH & 4 & 0 & 0 & 20 & 16 & 8\\
CH$_2$DOH & 5 & 2 & 2 & 16 & 9 & 6 \\
CH$_3$CH$_2$OH & 12 & 3 & 3 & 30 & 14 & 8 \\
CH$_3$CHO & 17 & 2 & 2 & 27 & 6 & 5 \\
CH$_3$OCHO & 34 & 16 & 13 & 111 & 12 & 7\\
CH$_3$OCH$_3$ & 16 & 8 & 8 & 112 & 27 & 20\\
CH$_3$COCH$_3$ & 59 & 0 & 0 & 128 & 6 & 5\\
aGg'(CH$_2$OH)$_2$ & 25 & 0 & 0 & 105 & 18 & 8\\
gGg'(CH$_2$OH)$_2$ & 43 & 0 & 0 & 115 & 4 & 2 \\
CH$_2$OHCHO & 15 & 0 & 0 & 27 & 5 & 3\\
H$_2$CCO & 0 & 0 & 0 & 6 & 2 & 2 \\
t-HCOOH & 6 & 3 & 3 & 3 & 1 & 1\\
\hline
\\
\multicolumn{7}{c}{\textbf{S68N}} \\
\hline \hline
& \multicolumn{3}{c}{Band 3} & \multicolumn{3}{c}{Band 6}  \\
\cline{2-4} \cline{5-7}
Species & T & L & F & T & L & F \\
\hline
CH$_3$OH & 1 & 1 & 0 & 3 & 3 & 0\\
$^{13}$CH$_3$OH & 2 & 1 & 1 & 8 & 5 & 5\\
CH$_3^{18}$OH & 4 & 0 & 0 & 20 & 2 & 2 \\
CH$_2$DOH & 5 & 0 & 0 & 16 & 2 & 2\\
CH$_3$CH$_2$OH & 12 & 0 & 0 & 30 & 3 & 2\\
CH$_3$CHO & 17 & 0 & 0 & 27 & 2 & 2\\
CH$_3$OCHO & 34 & 7 & 7 & 111 & 10 & 7 \\
CH$_3$OCH$_3$ & 16 & 8 & 8 & 112 & 27 & 20\\
CH$_3$COCH$_3$ & 59 & 0 & 0 & 128 & 2 & 1\\
aGg'(CH$_2$OH)$_2$ & 25 & 0 & 0 & 105 & 1 & 1\\
gGg'(CH$_2$OH)$_2$ & 43 & 0 & 0 & 115 & 0 & 0 \\
CH$_2$OHCHO & 15 & 0 & 0 & 27 & 0 & 0 \\
H$_2$CCO & 0 & 0 & 0 & 6 & 2 & 1 \\
t-HCOOH & 6 & 0 & 0 & 3 & 1 & 1 \\
\hline
\\
\multicolumn{7}{c}{\textbf{B1-bS}} \\
\hline \hline
& \multicolumn{3}{c}{Band 3} & \multicolumn{3}{c}{Band 6}  \\
\cline{2-4} \cline{5-7}
Species & T & L & F & T & L & F \\
\hline
CH$_3$OH & 1 & 1 & 0 & 3 & 3 & 0 \\
$^{13}$CH$_3$OH & 2 & 1 & 1 & 8 & 4 & 3 \\
CH$_3^{18}$OH & 4 & 0 & 0 & 20 & 13 & 5 \\
CH$_2$DOH & 3 & 0 & 0 & 16 & 0 & 0 \\
CH$_3$CH$_2$OH & 12 & 0 & 0 & 30 & 0 & 0\\
CH$_3$CHO & 17 & 2 & 2 & 27 & 3 & 2 \\
CH$_3$OCHO & 34 & 8 & 8 & 111 & 10 & 7 \\
CH$_3$OCH$_3$ & 16 & 6 & 6 & 112 & 63 & 60\\
CH$_3$COCH$_3$ & 59 & 0 & 0 & 128 & 1 & 1 \\
aGg'(CH$_2$OH)$_2$ & 25 & 0 & 0 & 0 & 0 & 0 \\
gGg'(CH$_2$OH)$_2$ & 43 & 0 & 0 & 0 & 0 & 0 \\
CH$_2$OHCHO & 15 & 0 & 0 & 0 & 0 & 0 \\
H$_2$CCO & 0 & 0 & 0 & 6 & 1 & 1\\
t-HCOOH & 6 & 0 & 0 & 3 & 0 & 0 \\
\hline
\end{tabular}
\label{tab:Nlines}
\tablefoot{T is the number of transitions, L is the number of lines in the model with a 3$\sigma$ detection, and F is the number of unblended lines (i.e., no other line within one FWHM) that are included in the fit. }
\end{center}
\end{table} 

\begin{table}
\begin{center}
\caption{Computed column densities and abundance ratios in Band~3, assuming $T_\mathrm{ex}=100$~K.}
\begin{tabular}{@{\extracolsep{0pt}}lccc@{}}
\multicolumn{4}{c}{\textbf{B1-c}} \\
\hline \hline
& \multicolumn{3}{c}{Band 3}  \\
\cline{2-4}
Species & $T_\mathrm{ex}$ (K) & $N$ (cm$^{-2}$) & X/CH$_3$OH (\%) \\
\hline
CH$_3$OH & -- & $(6.7 \pm 1.1) \times 10^{17}$ & $\equiv 100$ \\ 
$^{13}$CH$_3$OH & [100] & $(9.6 \pm 1.6) \times 10^{15}$ & $\equiv 1.43$ \\ 
CH$_3^{18}$OH & [100] & $<2.2 \times 10^{16}$ & $<3.4$ \\ 
CH$_2$DOH & [100] & $(5.7 \pm 0.7) \times 10^{16}$ & $8.4 \pm 1.7$ \\ 
CH$_3$CH$_2$OH & [100] & $(1.8 \pm 0.2) \times 10^{16}$ & $2.7 \pm 0.5$ \\ 
CH$_3$CHO & [100] & $(4.7 \pm 1.6) \times 10^{15}$ & $0.7 \pm 0.3$ \\ 
CH$_3$OCHO & [100] & $(1.6 \pm 0.2) \times 10^{16}$ & $2.4 \pm 0.5$ \\ 
CH$_3$OCH$_3$ & [100] & $(1.9 \pm 0.1) \times 10^{16}$ & $2.8 \pm 0.5$ \\ 
CH$_3$COCH$_3$ & [100] & $<5.0 \times 10^{16}$ & $< 7.5$ \\ 
aGg'(CH$_2$OH)$_2$ & [100] & $<2.1 \times 10^{16}$ & $<3.2$ \\ 
gGg'(CH$_2$OH)$_2$ & [100] & $<2.1 \times 10^{16}$ & $<3.2$ \\ 
CH$_2$OHCHO & [100] & $<8.0 \times 10^{15}$ & $<1.2$ \\ 
t-HCOOH & [100] & $(4.3 \pm 0.7) \times 10^{15}$ & $0.6 \pm 0.2$\\
\hline
\\
\multicolumn{4}{c}{\textbf{S68N}} \\
\hline \hline
& \multicolumn{3}{c}{Band 3}  \\
\cline{2-4}
Species & $T_\mathrm{ex}$ (K) & $N$ (cm$^{-2}$) & X/CH$_3$OH (\%) \\
\hline
CH$_3$OH & -- & $(2.1 \pm 0.7) \times 10^{17}$ & $\equiv 100$ \\ 
$^{13}$CH$_3$OH & [100] & $(3.0 \pm 1.0) \times 10^{15}$ & $\equiv 1.43$ \\ 
CH$_3^{18}$OH & [100] & $<2.0 \times 10^{16}$ & $<9.8$ \\ 
CH$_2$DOH & [100] & $<1.0 \times 10^{17}$ & $<49$ \\ 
CH$_3$CH$_2$OH & [100] & $<2.0 \times 10^{16}$ & $<9.8$ \\ 
CH$_3$CHO & [100] & $<2.9 \times 10^{14}$ & $<0.14$ \\ 
CH$_3$OCHO & [100] & $(4.7 \pm 0.3) \times 10^{15}$ & $2.3 \pm 0.8$ \\ 
CH$_3$OCH$_3$ & [100] & $(6.0 \pm 0.3) \times 10^{15}$ & $2.9 \pm 1.0$ \\ 
CH$_3$COCH$_3$ & [100] & $<5.0 \times 10^{16}$ & $<25$ \\ 
aGg'(CH$_2$OH)$_2$ & [100] & $<4.3 \times 10^{16}$ & $<21$ \\ 
gGg'(CH$_2$OH)$_2$ & [100] & $<4.3 \times 10^{16}$ & $<21$ \\ 
CH$_2$OHCHO & [100] & $<1.6 \times 10^{16}$ & $<7.8$ \\ 
t-HCOOH & [100] & $<5.0 \times 10^{15}$ & $<2.5$ \\
\hline
\\
\multicolumn{4}{c}{\textbf{B1-bS}} \\
\hline \hline
& \multicolumn{3}{c}{Band 3}  \\
\cline{2-4}
Species & $T_\mathrm{ex}$ (K) & $N$ (cm$^{-2}$) & X/CH$_3$OH (\%) \\ 
\hline
CH$_3$OH & -- & $(1.1 \pm 0.3) \times 10^{17}$ & $\equiv 100$ \\ 
$^{13}$CH$_3$OH &[100] & $(1.6 \pm 0.4) \times 10^{15}$ & $\equiv 1.43$ \\ 
CH$_3^{18}$OH & [100] & $<1.0 \times 10^{16}$ & $<8.9$ \\ 
CH$_2$DOH & [100] & $<2.0 \times 10^{16}$ & $<18$ \\ 
CH$_3$CH$_2$OH & [100] &  $<2.0 \times 10^{16}$ & $<18$ \\ 
CH$_3$CHO & [100] & $(1.4 \pm 0.2) \times 10^{15}$ & $1.2 \pm 0.3$ \\ 
CH$_3$OCHO & [100] & $(5.7 \pm 0.7) \times 10^{15}$ & $5.0 \pm 1.3$ \\ 
CH$_3$OCH$_3$ & [100] & $(1.2 \pm 0.1) \times 10^{16}$ & $11 \pm 3$ \\ 
CH$_3$COCH$_3$ & [100] & $<1.0 \times 10^{16}$ & $<8.9$ \\ 
aGg'(CH$_2$OH)$_2$ & [100] & $<1.1 \times 10^{16}$ & $<9.6$ \\ 
gGg'(CH$_2$OH)$_2$ & [100] & $<2.1 \times 10^{16}$ & $<19$ \\ 
CH$_2$OHCHO & [100] & $<8.0 \times 10^{15}$ & $<7.2$ \\ 
t-HCOOH & [100] & $<1.0 \times 10^{16}$ & $<8.9$ \\ 
\hline
\end{tabular}
\label{tab:B3100K}
\tablefoot{The column density of CH$_3$OH is determined by scaling $^{13}$CH$_3$OH using $^{12}\mathrm{C}/^{13}\mathrm{C}=70$. A value between squared brackets means the parameter was fixed to that value during the computation. 
Abundance ratio's which are set to their respective isotope ratio are indicated with a $\equiv$ symbol.
}
\end{center}
\end{table} 

\clearpage

\section{Full ALMA Band~6 spectra}
\label{app:full_spectra}
\noindent\begin{minipage}{\textwidth}
\begin{center}
\centering
\includegraphics[width=0.95\hsize]{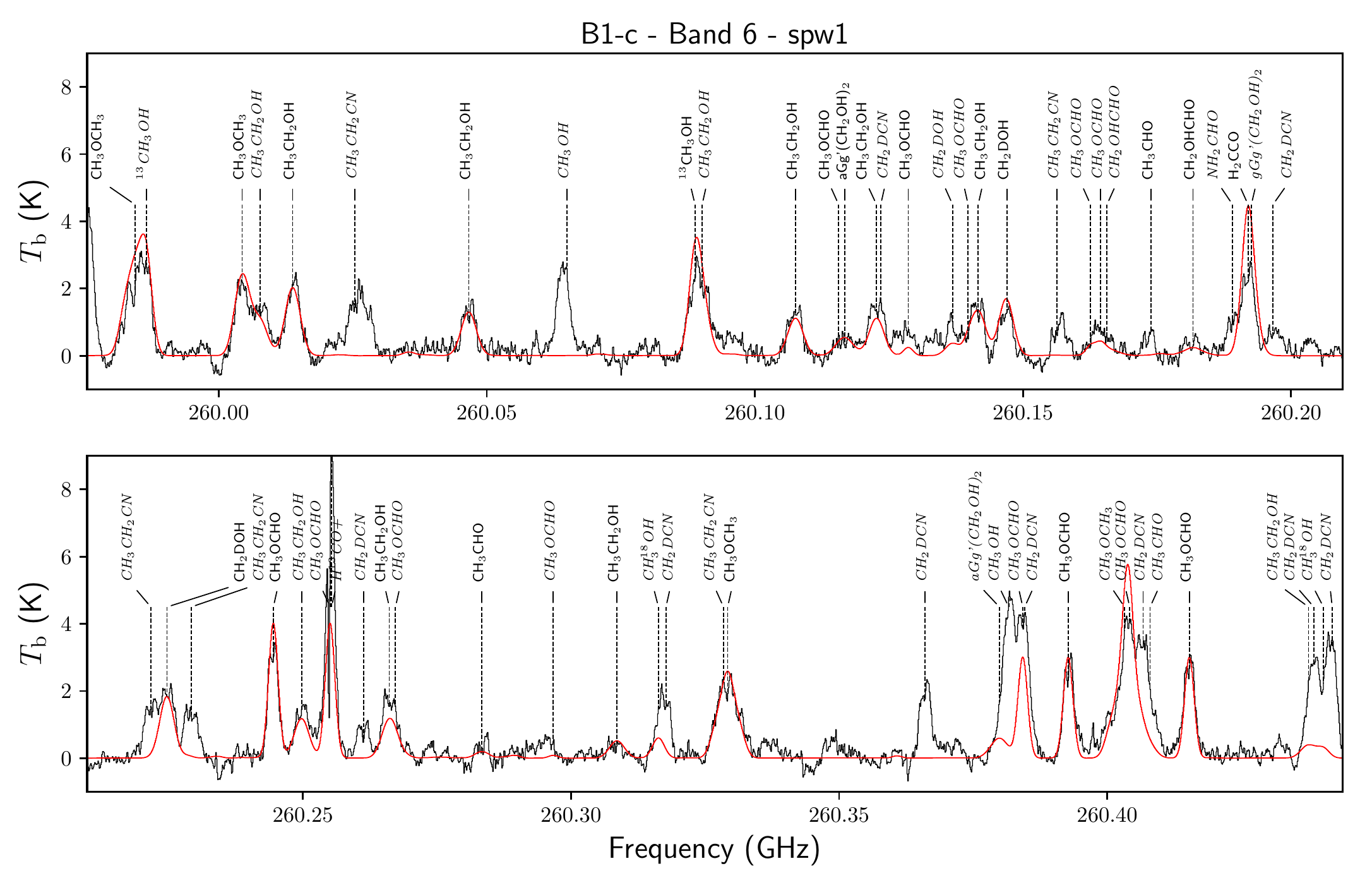}
\includegraphics[width=0.95\hsize]{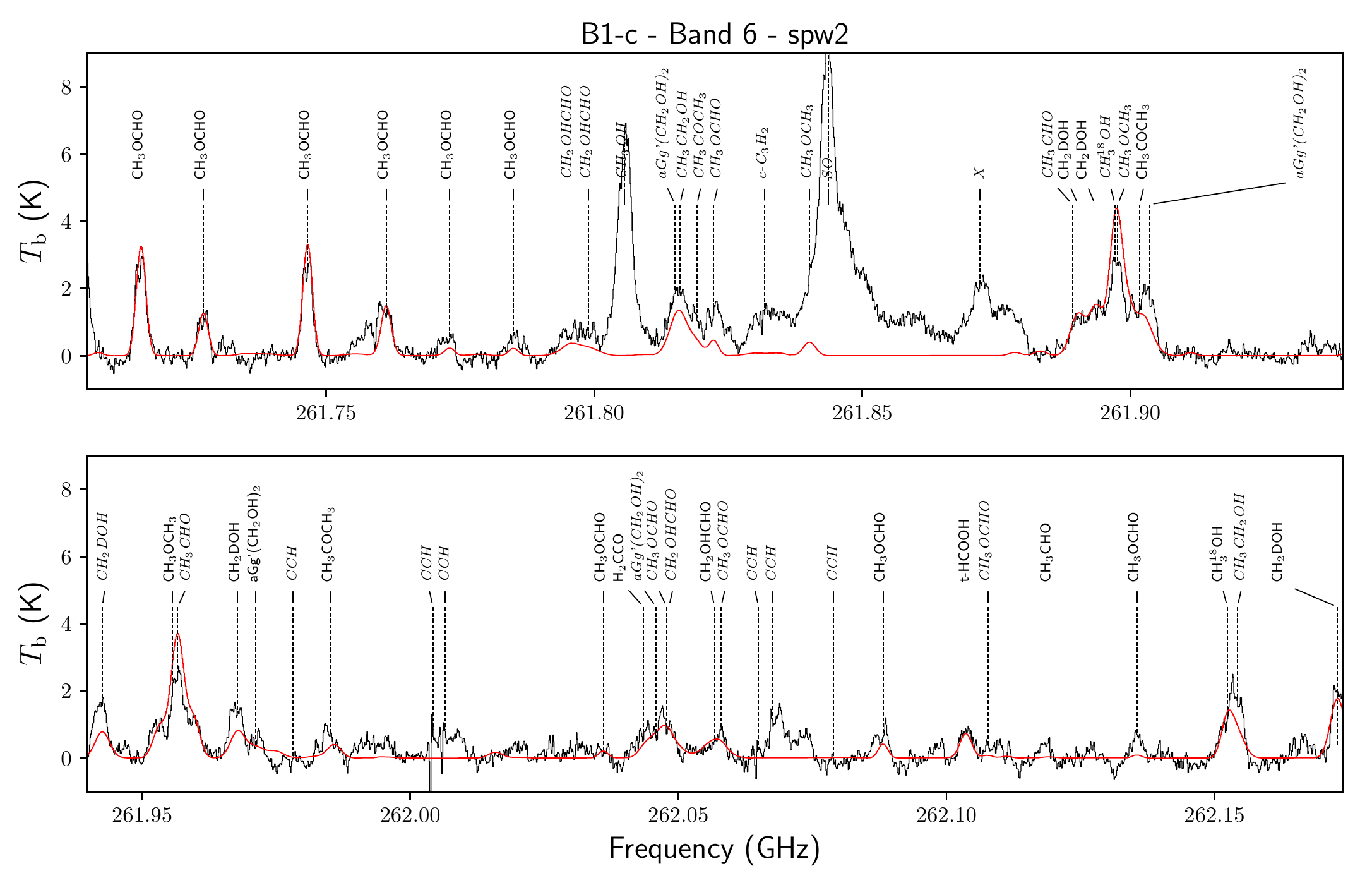}
\captionof{figure}{Full Band~6 spectrum of B1-c (black) with best fitting CASSIS model overplotted (red). {We indicate the positions of species, where lines in italic are excluded in the fitting}. Lines annotated with an 'X' are unidentified.}
\label{fig:B1c_fullspectrum}
\end{center}
\end{minipage}

\begin{figure*}[p]
\includegraphics[width=1.0\linewidth]{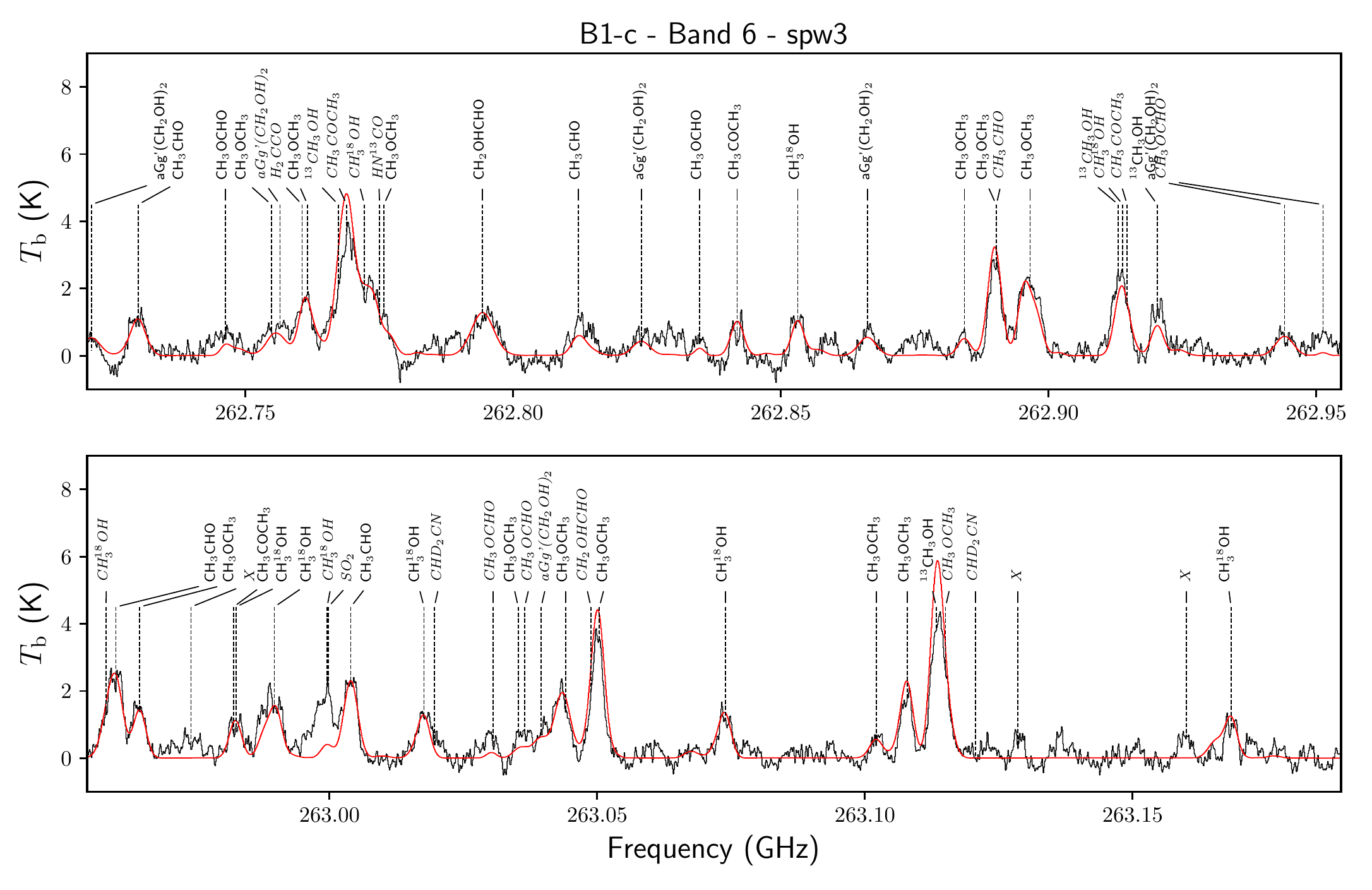}
\includegraphics[width=1.0\linewidth]{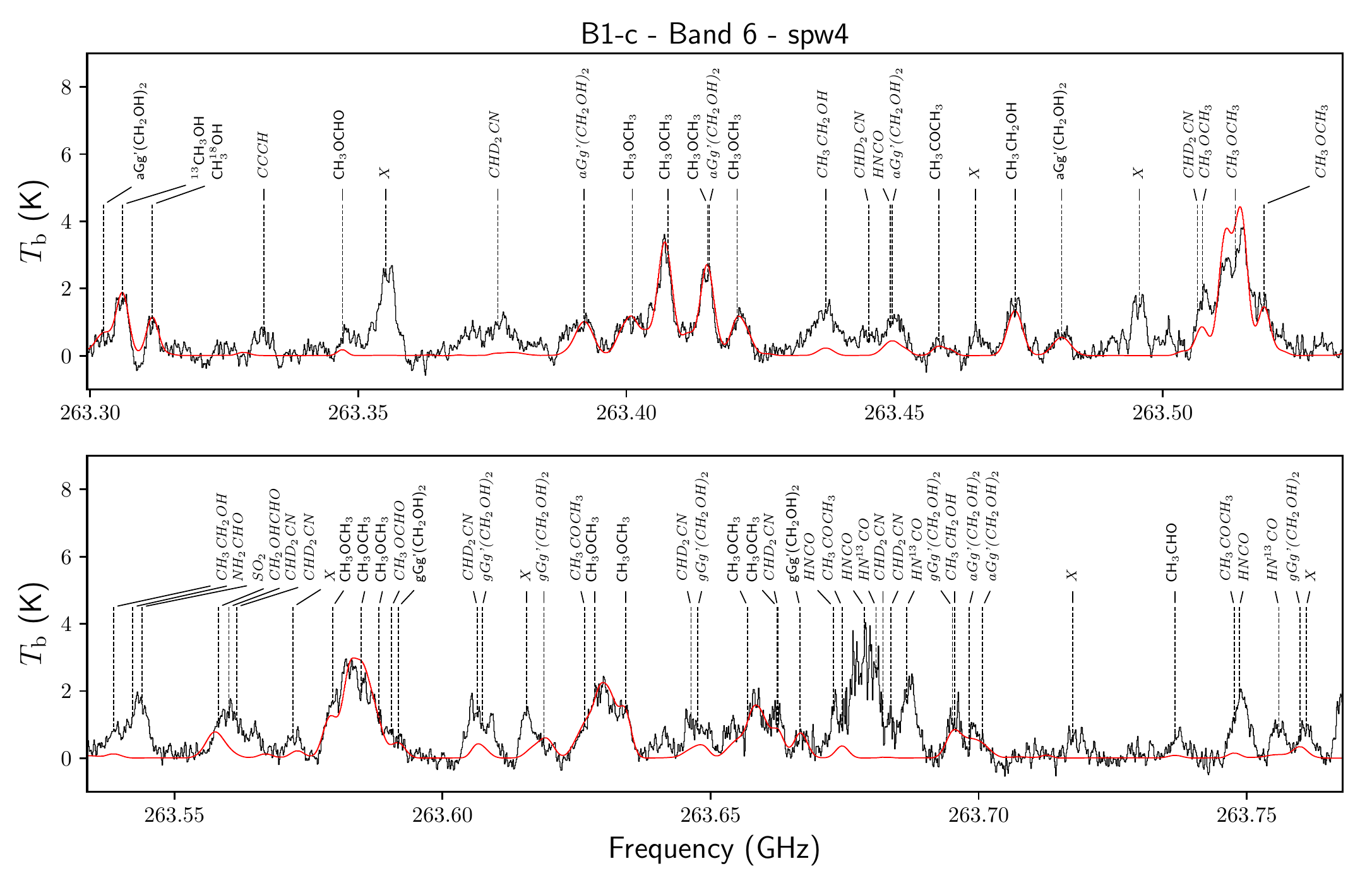}
\caption{(Continued)}
\end{figure*}

\begin{figure*}[p]
\centering
\includegraphics[width=1.0\linewidth]{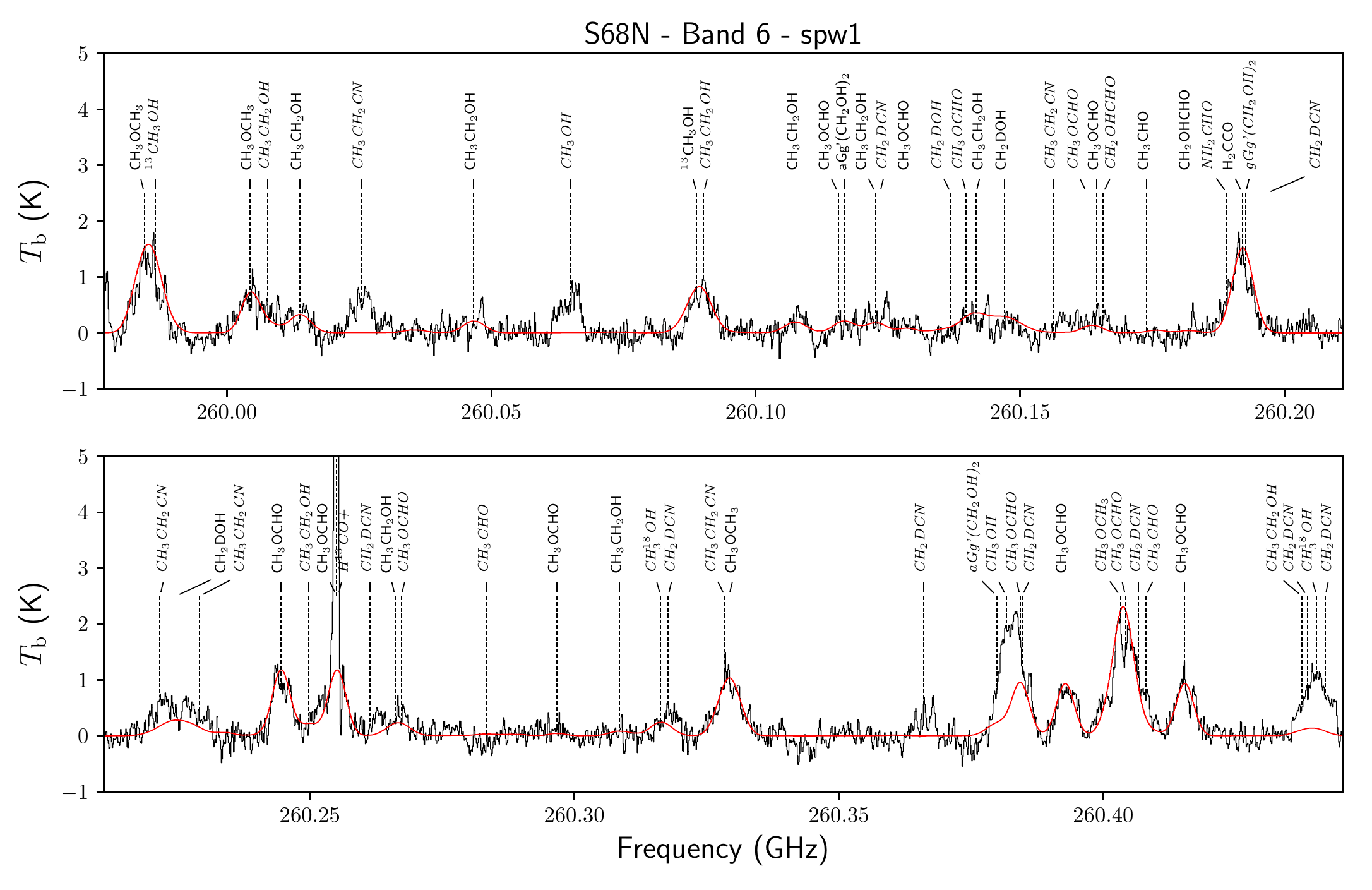}
\includegraphics[width=1.0\linewidth]{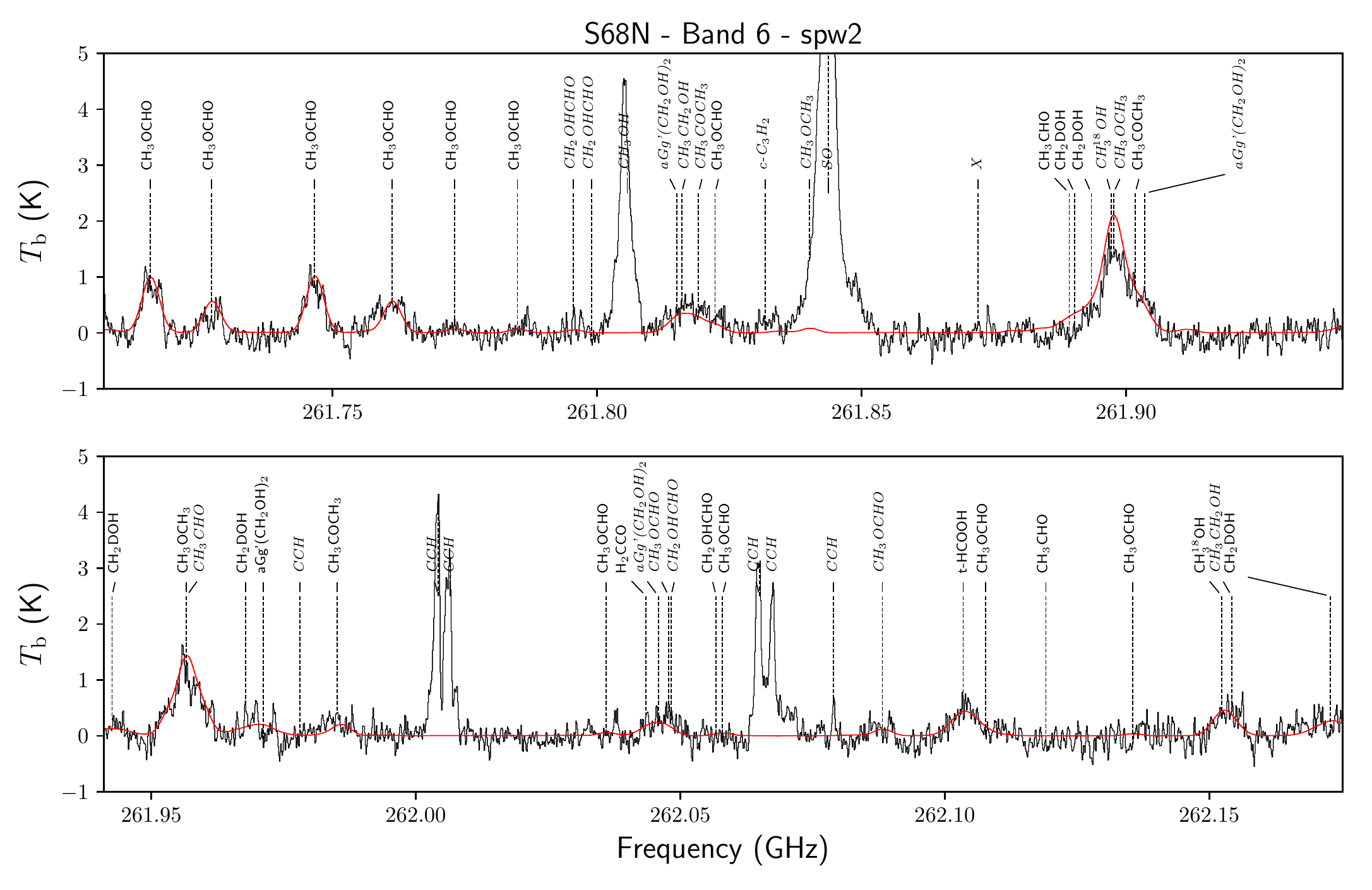}
\caption{Same as Fig.~\ref{fig:B1c_fullspectrum} but now for S68N.}
\end{figure*}
\begin{figure*}[p]
\includegraphics[width=1.0\linewidth]{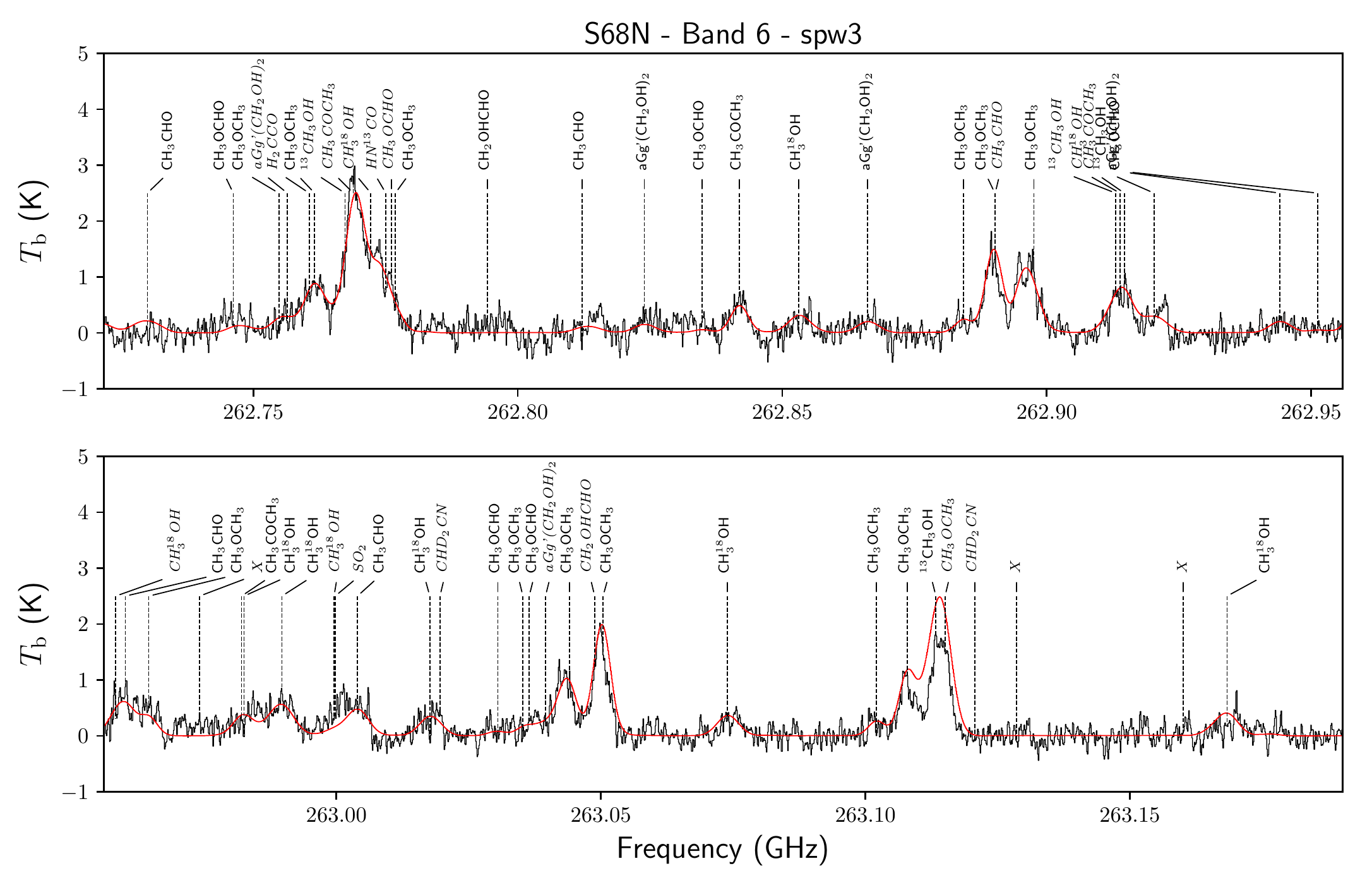}
\includegraphics[width=1.0\linewidth]{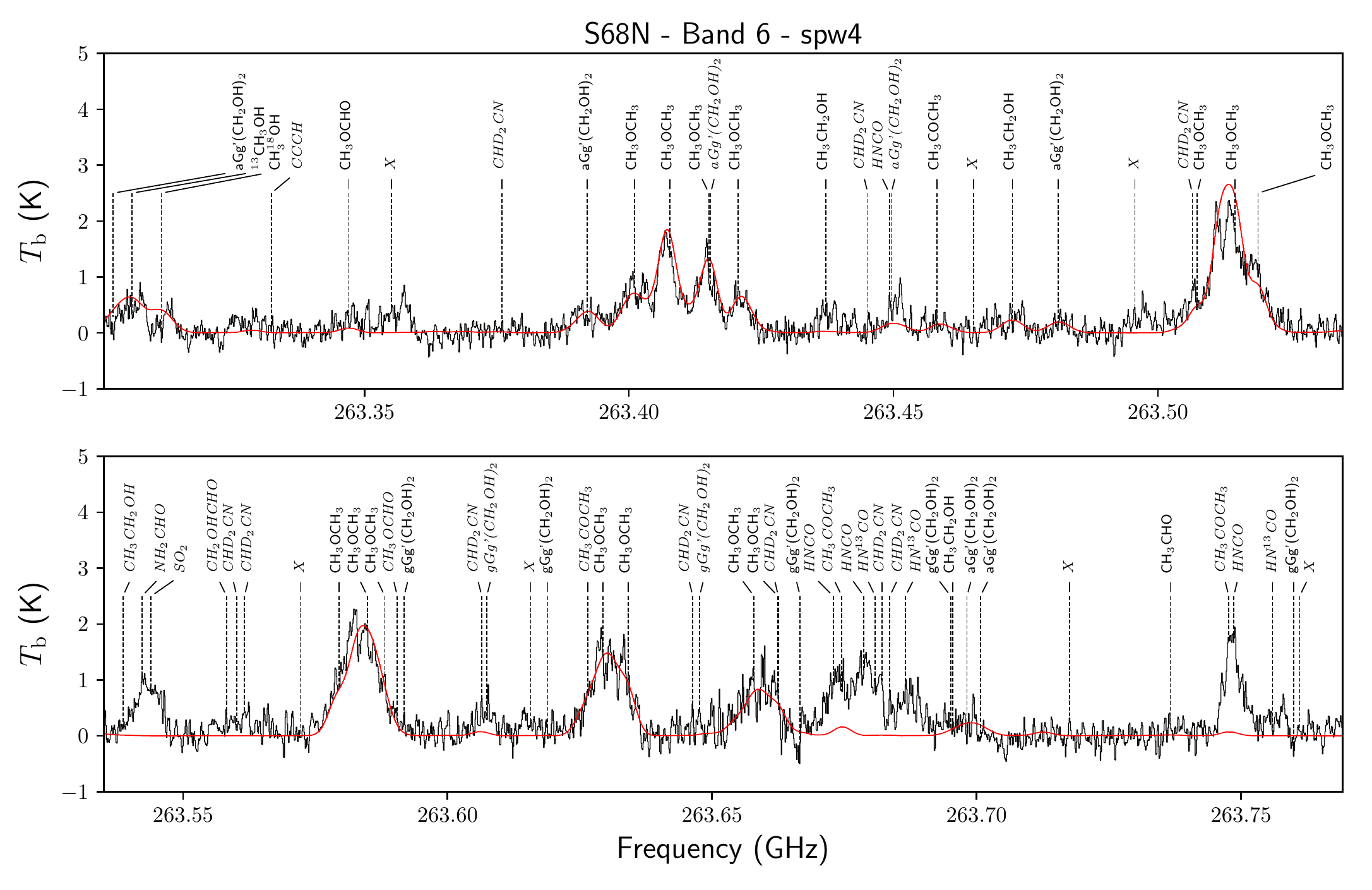}
\caption{(Continued)}
\label{fig:S68N_fullspectrum}
\end{figure*}

\begin{figure*}[p]
\centering
\includegraphics[width=1.0\linewidth]{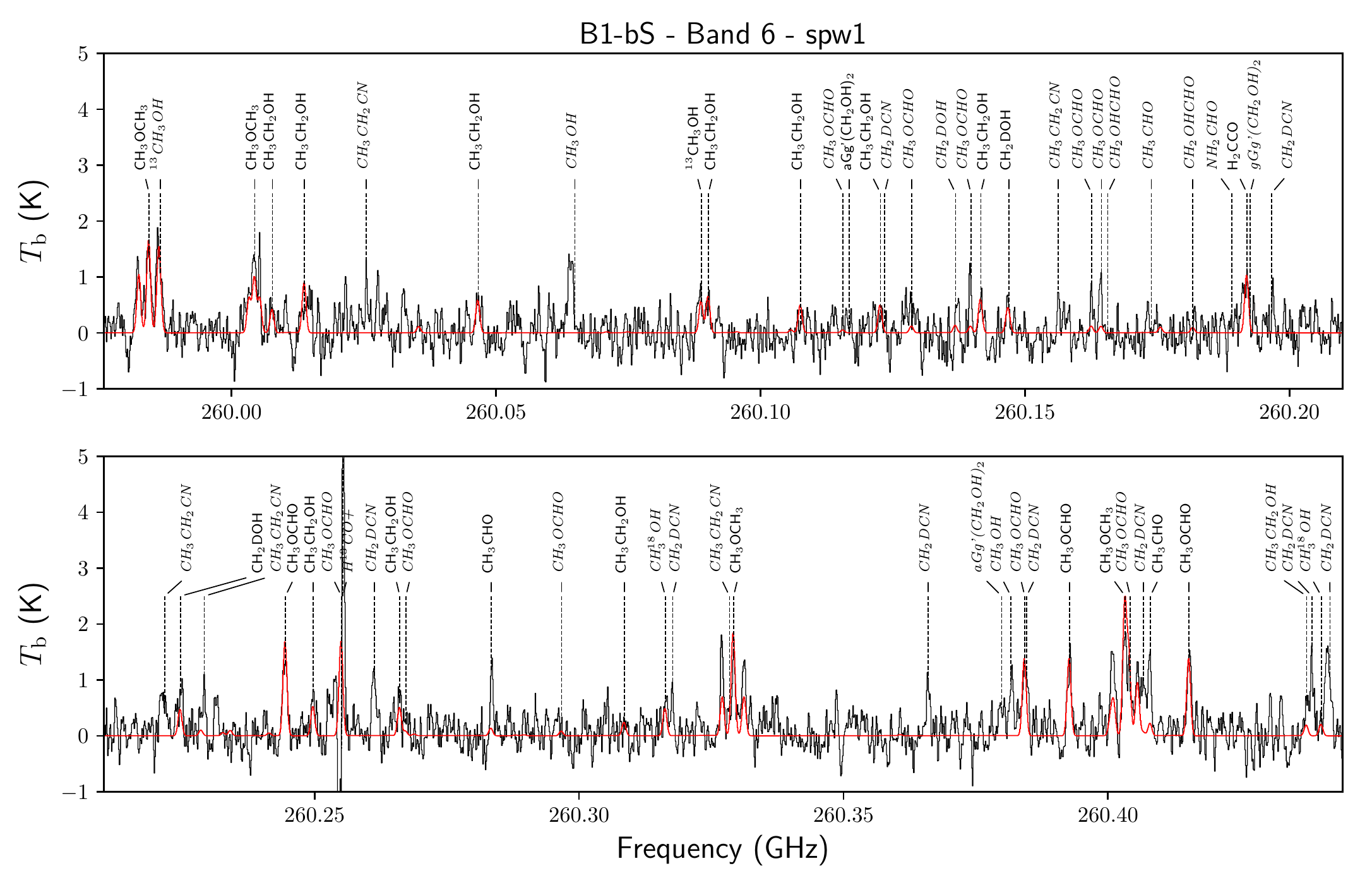}
\includegraphics[width=1.0\linewidth]{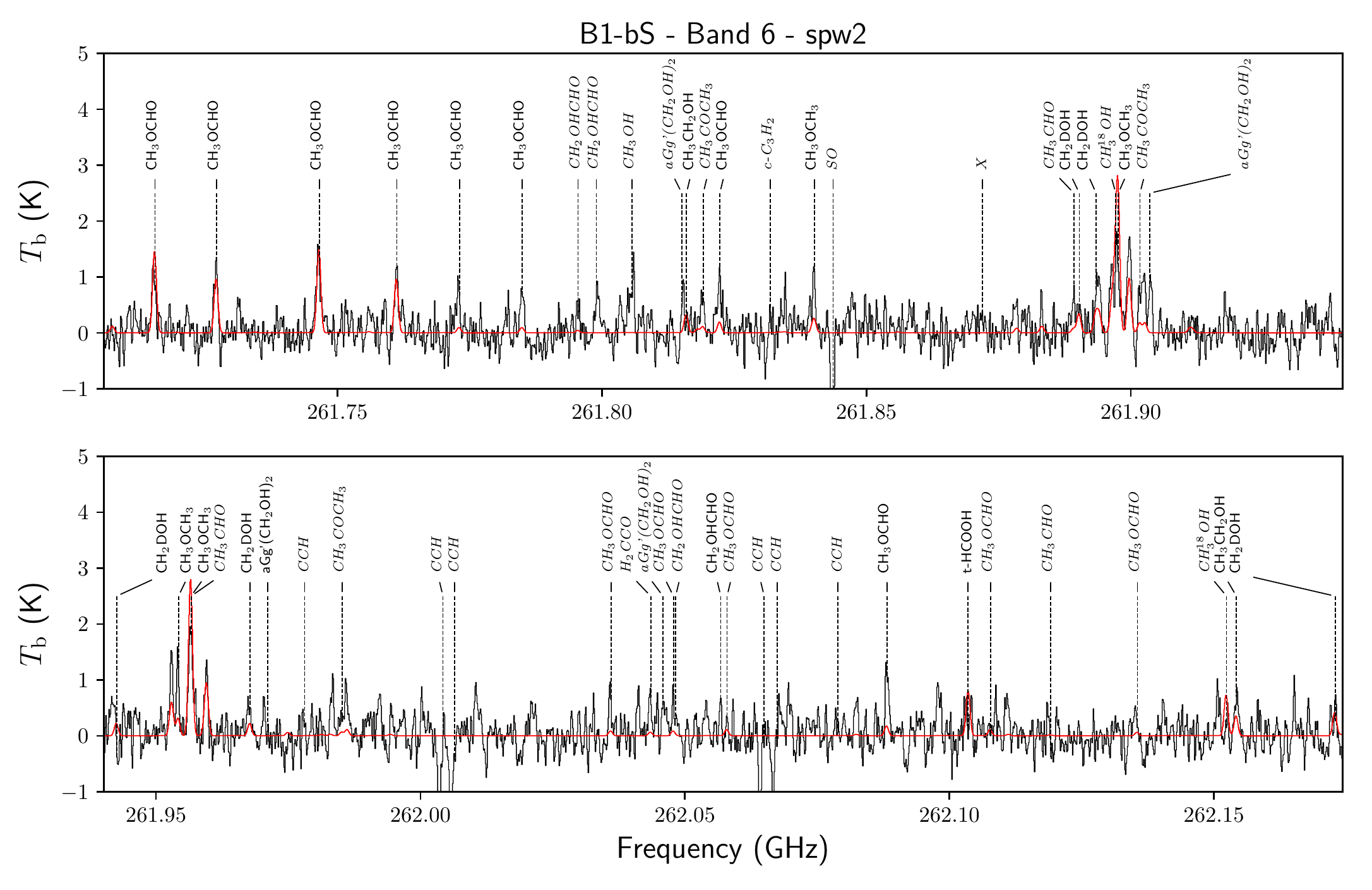}
\caption{Same as Fig.~\ref{fig:B1c_fullspectrum} but now for B1-bS.}
\end{figure*}
\begin{figure*}[p]
\includegraphics[width=1.0\linewidth]{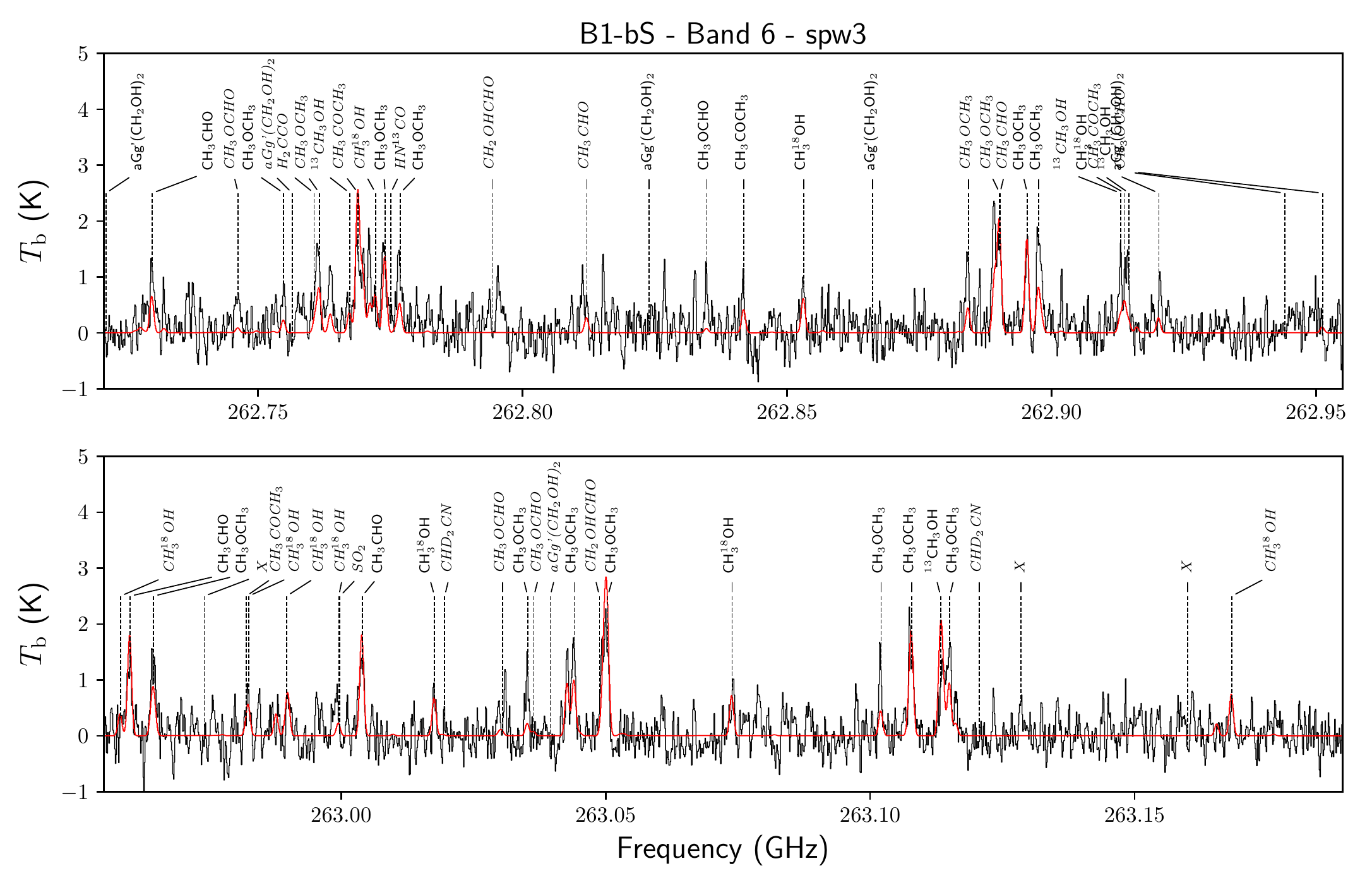}
\includegraphics[width=1.0\linewidth]{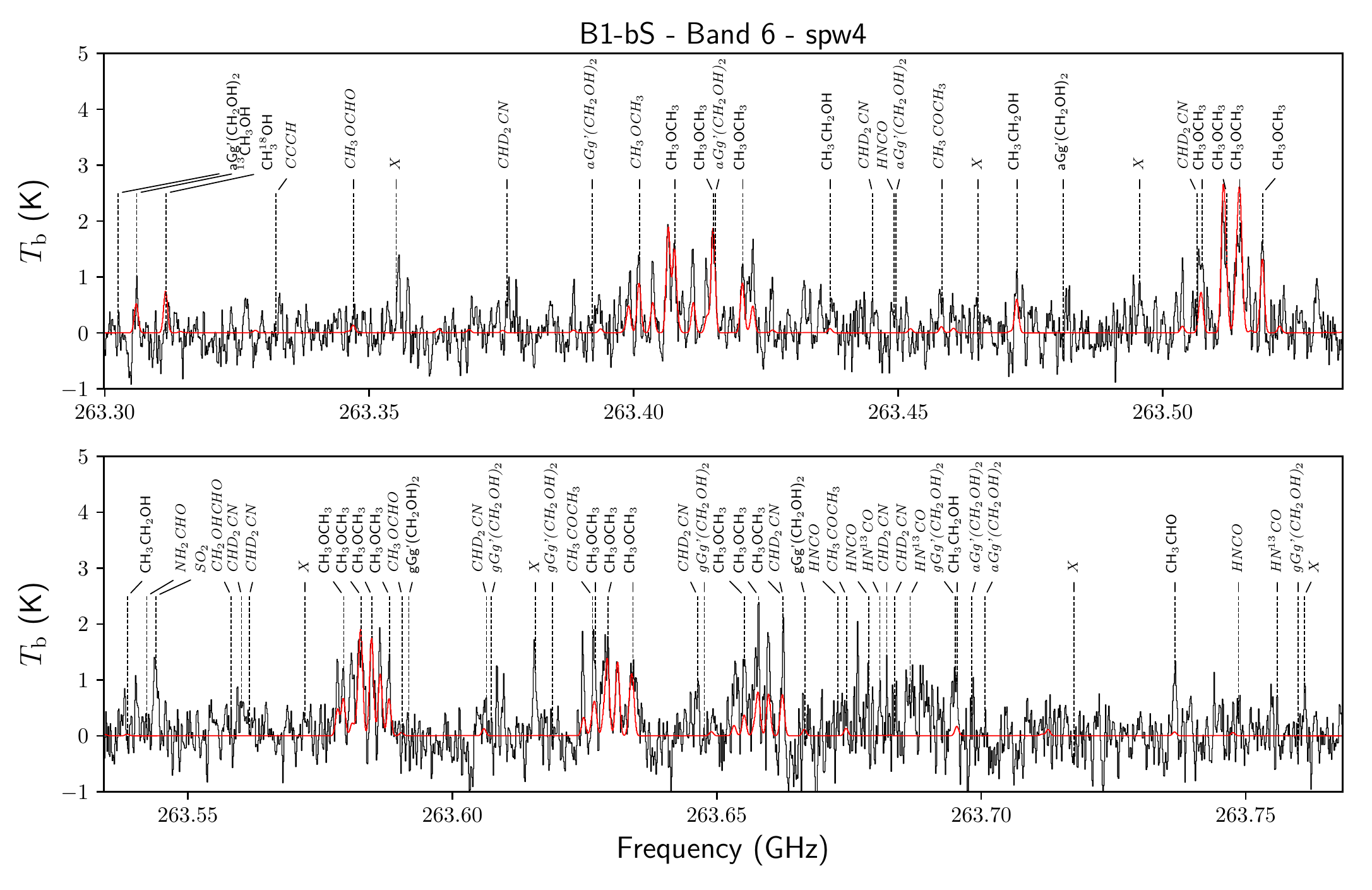}
\caption{(Continued)}
\label{fig:B1bS_fullspectrum}
\end{figure*}

\clearpage

\section{B1-c Band~3 spectrum}
\label{app:full_B3spectrum}
\noindent\begin{minipage}{\textwidth}
\begin{center}
\centering
\includegraphics[width=0.95\hsize]{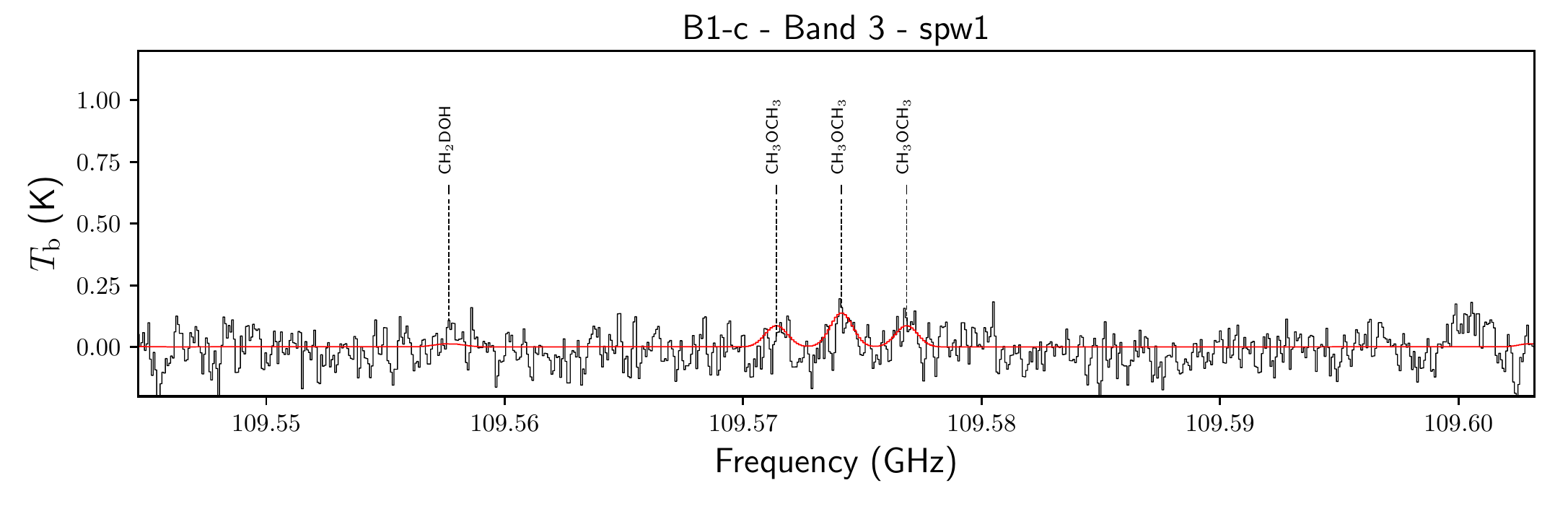}
\includegraphics[width=0.95\hsize]{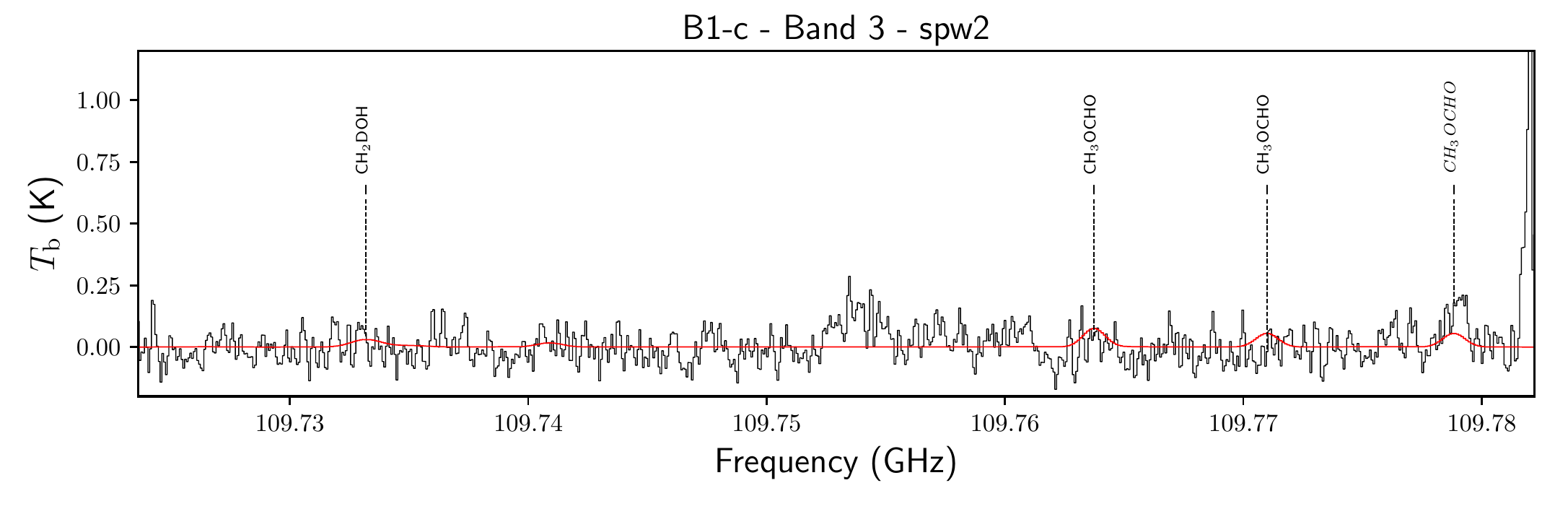}
\includegraphics[width=0.95\hsize]{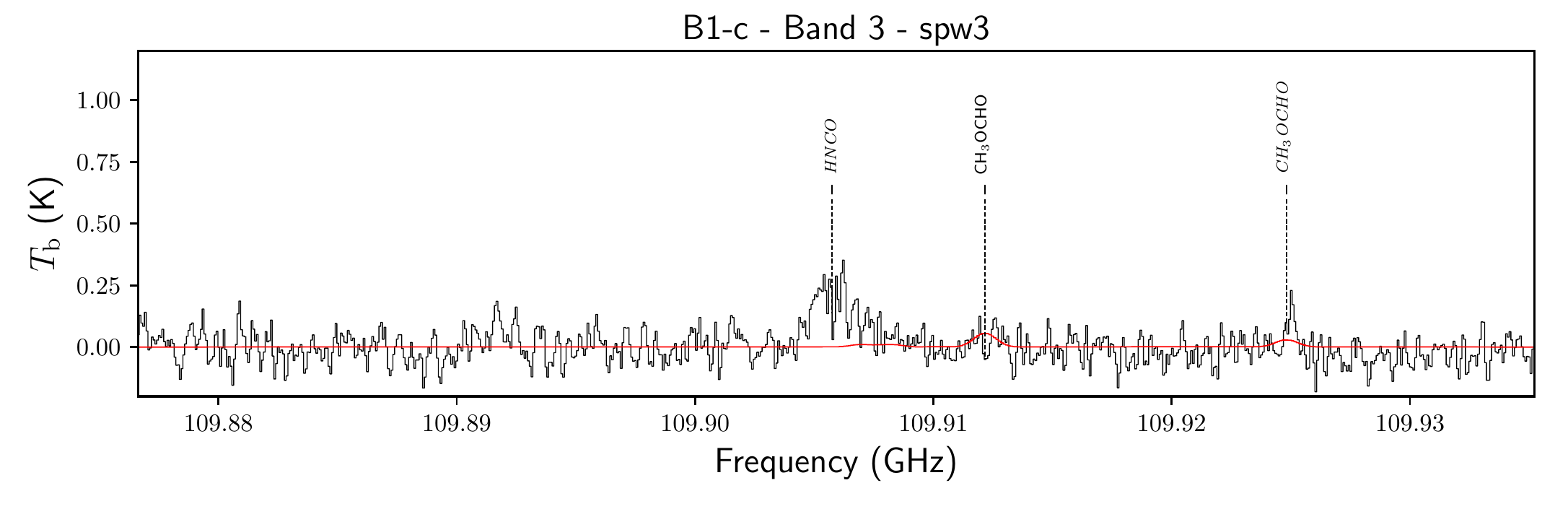}
\includegraphics[width=0.95\hsize]{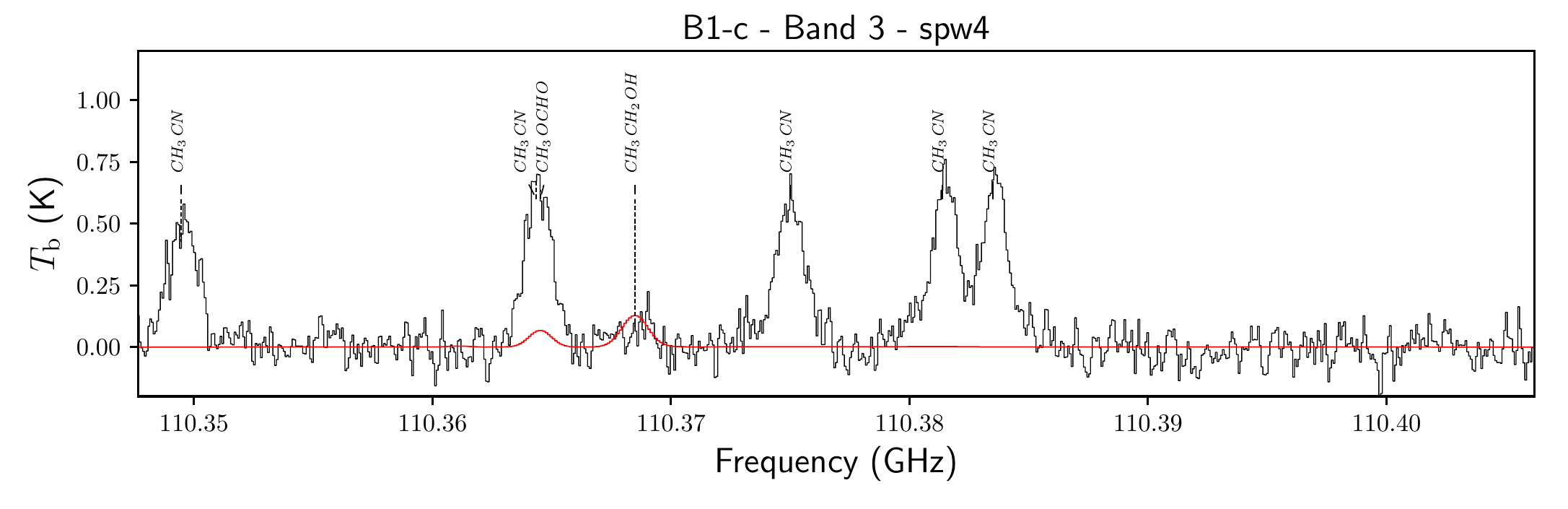}
\captionof{figure}{{Full Band~3 spectrum of B1-c (black) with best fitting CASSIS model overplotted (red). We indicate the positions of species, where lines in italic are excluded in the fitting.}}
\label{fig:B1c_B3_fullspectrum}
\end{center}
\end{minipage}

\begin{figure*}[p]
\includegraphics[width=1.0\linewidth]{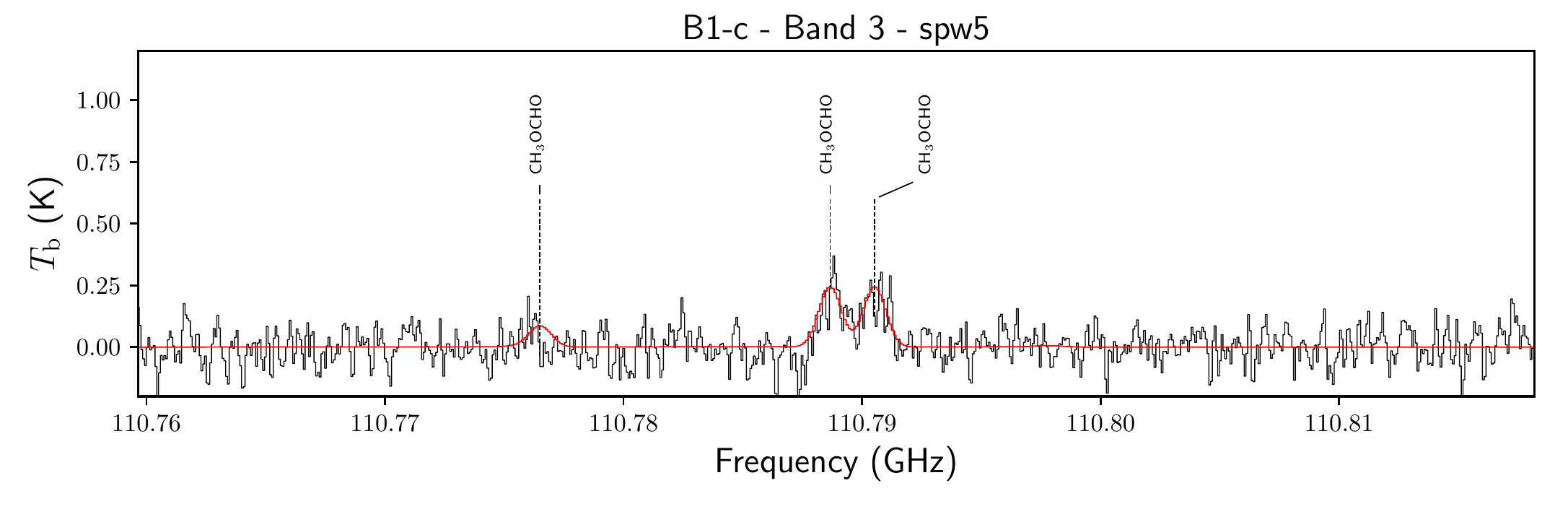}
\includegraphics[width=1.0\linewidth]{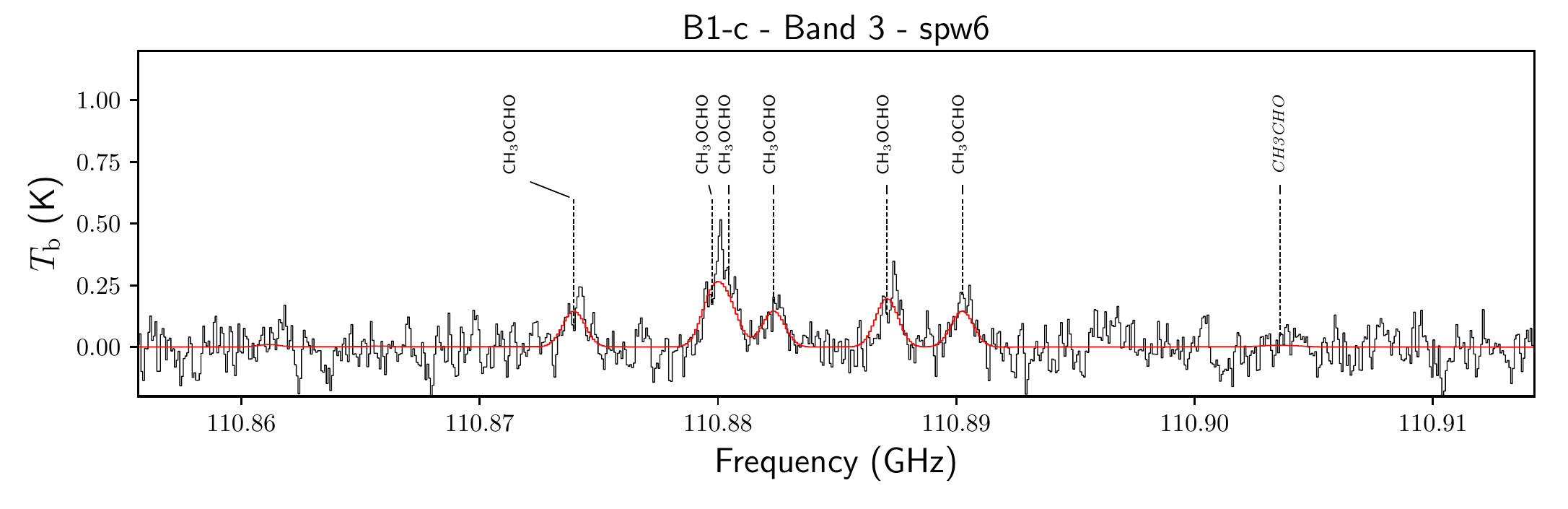}
\includegraphics[width=1.0\linewidth]{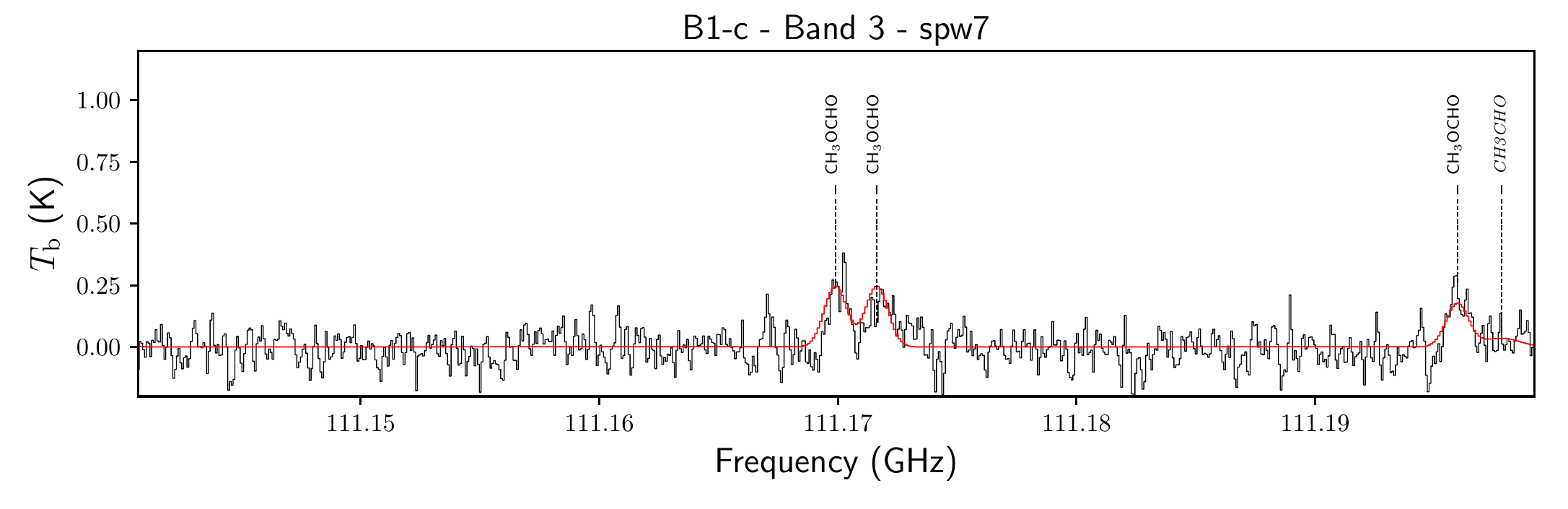}
\includegraphics[width=1.0\linewidth]{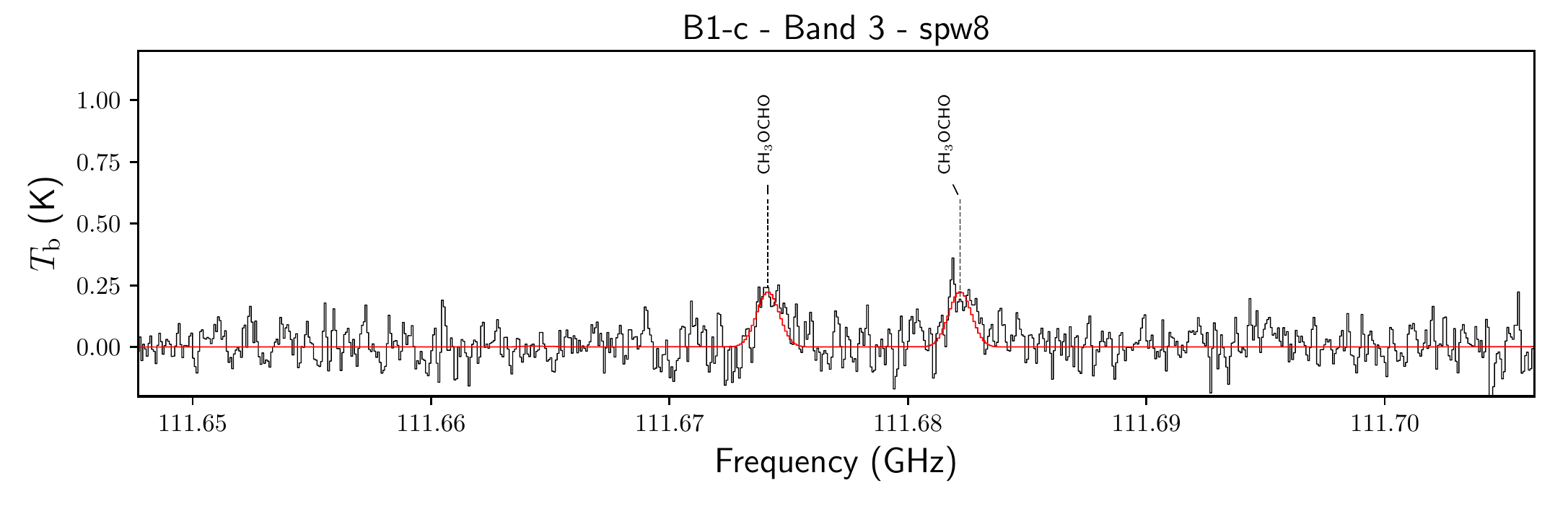}
\caption{(Continued)}
\end{figure*}

\begin{figure*}[p]
\includegraphics[width=1.0\linewidth]{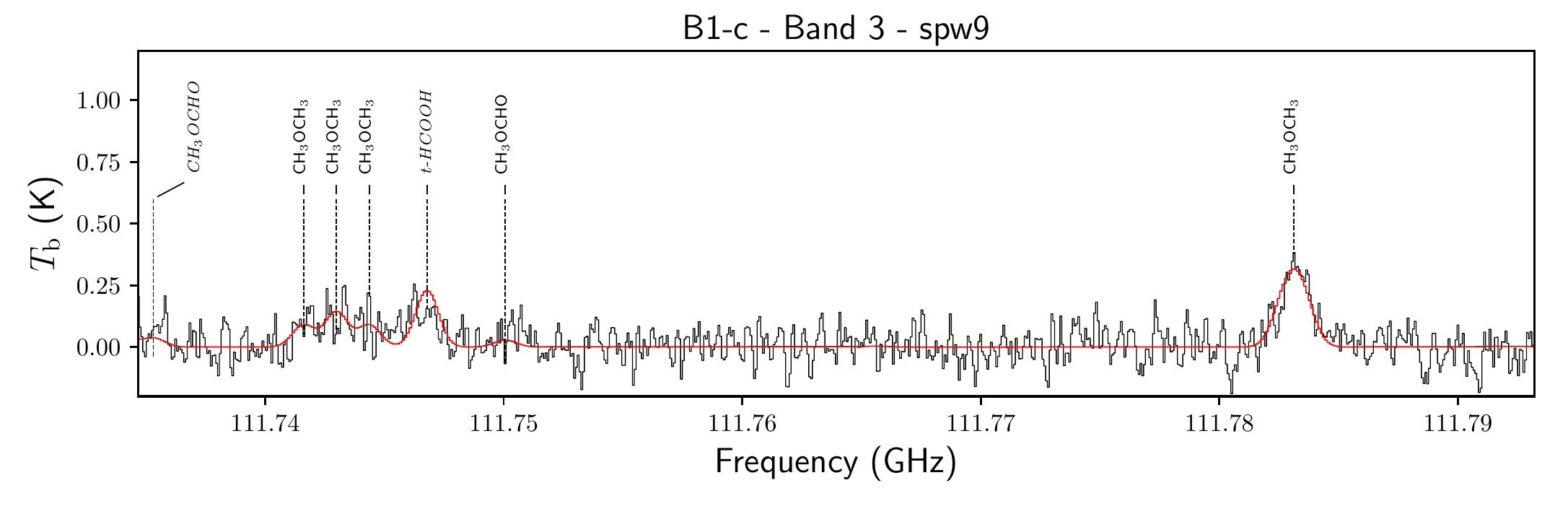}
\includegraphics[width=1.0\linewidth]{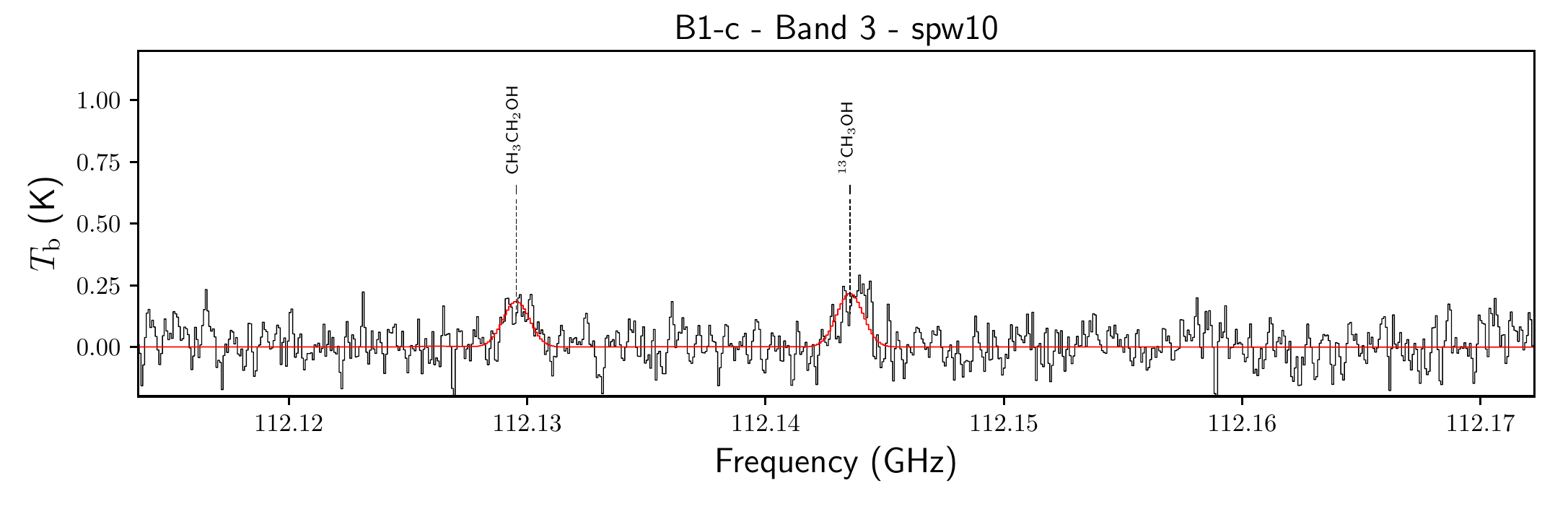}
\includegraphics[width=1.0\linewidth]{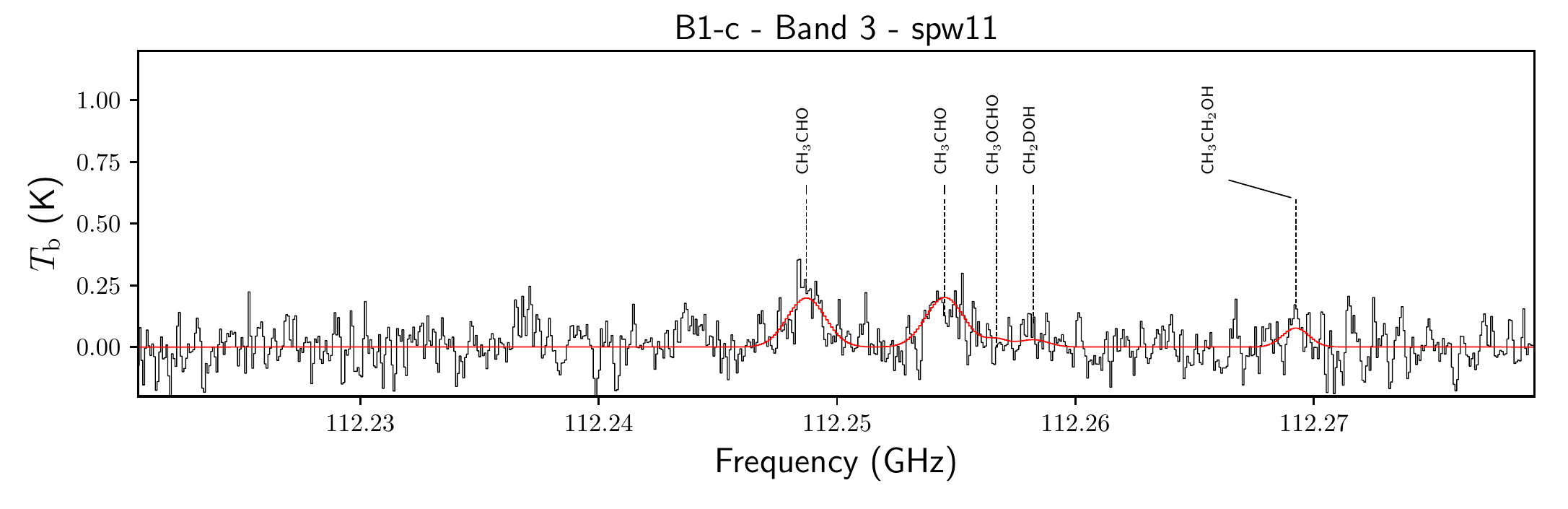}
\includegraphics[width=1.0\linewidth]{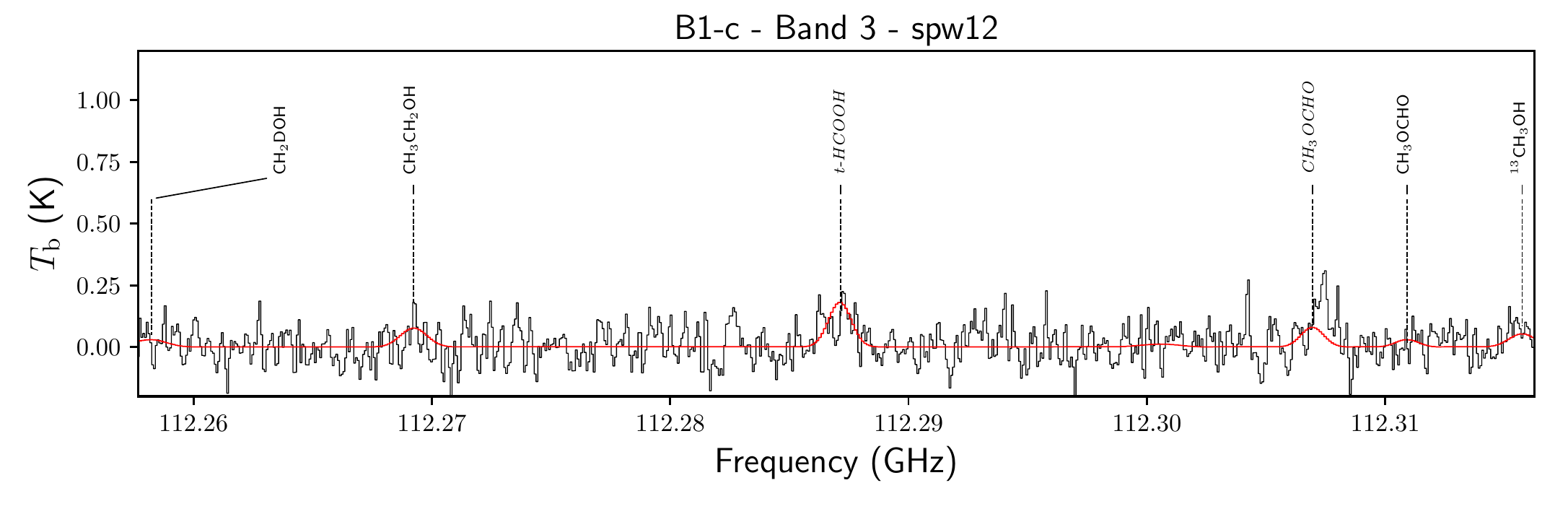}
\caption{(Continued)}
\end{figure*}

\begin{figure*}[p]
\includegraphics[width=1.0\linewidth]{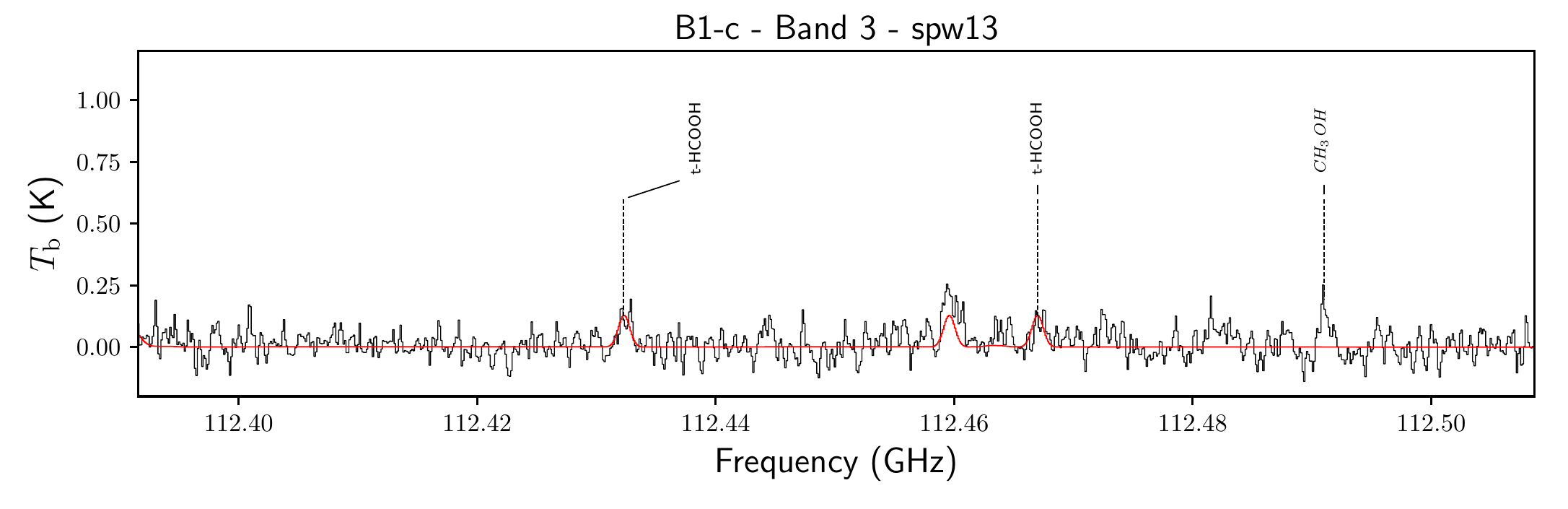}
\includegraphics[width=1.0\linewidth]{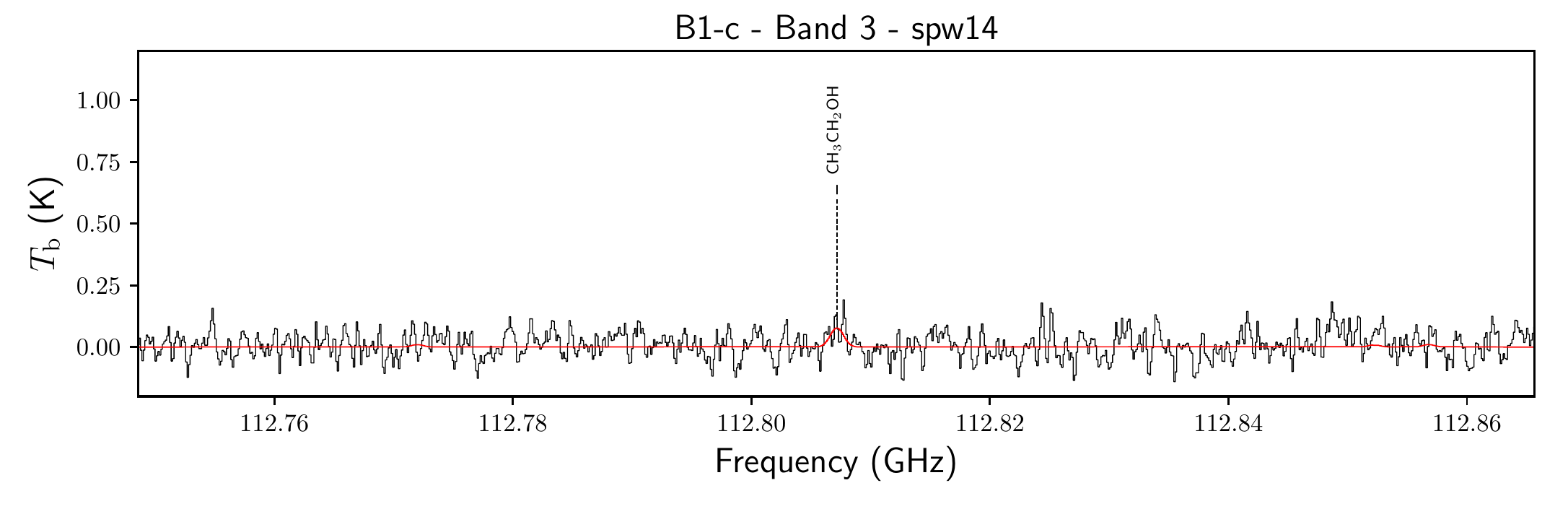}
\includegraphics[width=1.0\linewidth]{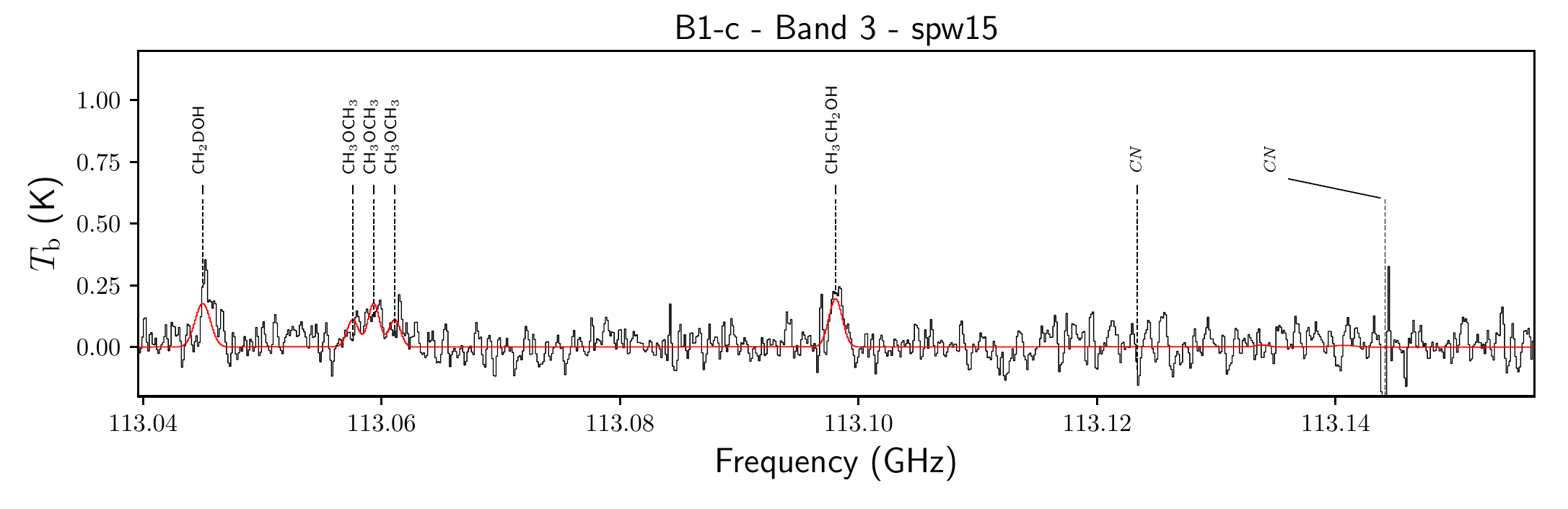}
\includegraphics[width=1.0\linewidth]{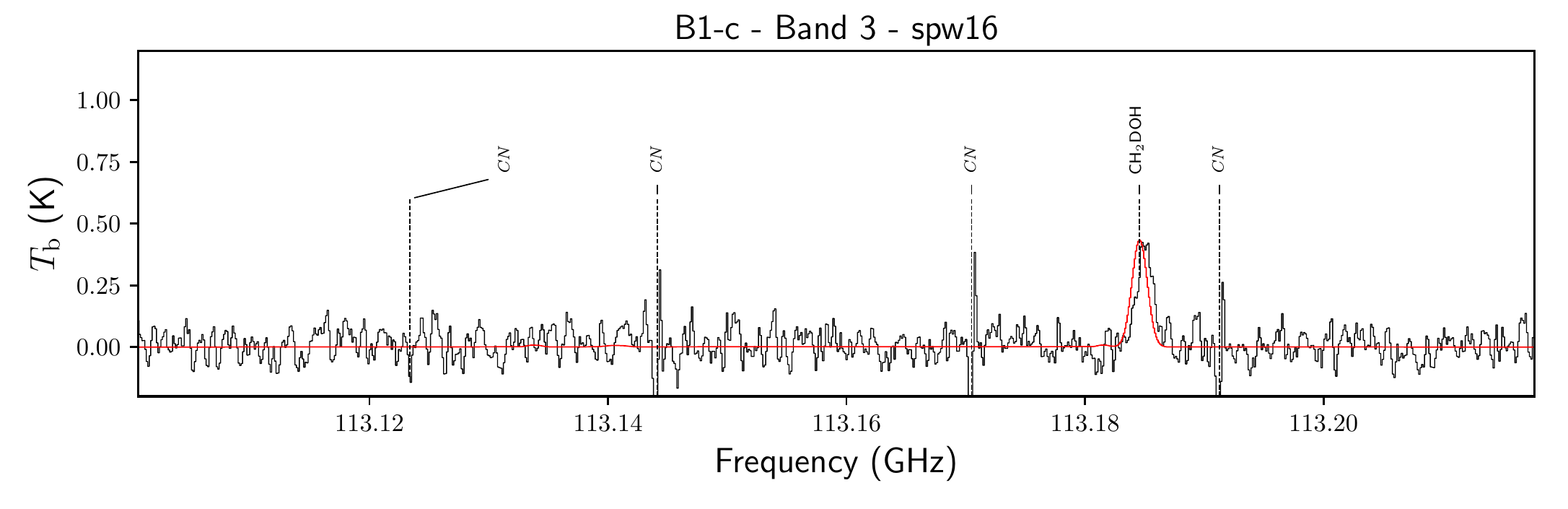}
\caption{(Continued)}
\end{figure*}

\clearpage

\section{Additional tables}
\renewcommand{\arraystretch}{1.12}
\noindent\begin{minipage}{\textwidth}
\begin{center}
\captionof{table}{List of ALMA image properties from program 2017.1.01774.S.}
\begin{tabular}{@{\extracolsep{3pt}}lcccccccc@{}}
\hline\hline
 & \multicolumn{2}{c}{B1-b\tablefootmark{1}} & \multicolumn{2}{c}{B1-c} & \multicolumn{2}{c}{S68N\tablefootmark{2}} & \multicolumn{2}{c}{SMM3} \\
 \cline{2-3}\cline{4-5}\cline{6-7}\cline{8-9}
 & Band~3 & Band~6 & Band~3 & Band~6 & Band~3 & Band~6 & Band~3 & Band~6 \\
\hline
Configuration & C43-3 & C43-4 & C43-3 & C43-4 & C43-2 & C43-4 & C43-2 & --\\
Beam (") & 2.1$\times$1.4 & 0.58$\times$0.39 & 2.1$\times$1.4 & 0.58$\times$0.39 & 2.8$\times$1.8 & 0.46$\times$0.42 & 2.8$\times$1.8 & --\\
LAS (") & 16 & 6 & 16 & 6 & 22 & 6 & 22 & --\\
$\Delta$V (\kms) & 0.2 / 0.4 & 0.14 & 0.2 / 0.4 & 0.14 & 0.17 / 0.32 & 0.14 & 0.17 / 0.32 & --  \\
Continuum rms (mJy) & 8.5 & 5.4 & 14.4 & 24.4 & 14.6 & 9.2 & 5.1 & -- \\
Line rms (K) & 0.05 & 0.16 & 0.06 & 0.14 & 0.04 & 0.15 & 0.06 & -- \\
Flux uncertainty (\%) & 3 & 12 & 3 & 12 & 5 & 9 & 5 & -- \\
\hline
\end{tabular}
\label{tab:obsprop}
\tablefoot{\tablefoottext{1}{Including B1-bN and B1-bS.} \tablefoottext{2}{Including Ser-emb~8~(N).}}
\end{center}
\end{minipage}

\longtab[2]{
\begin{longtable}{lrclccl}
\caption{Lines per species used for fitting the Band~3 data.} \\
\hline \hline
Species & \multicolumn{3}{c}{Transition} & Frequency & $A_{ij}$ & $E_\mathrm{up}$ \\
 & (J K L M & - & J K L M) & (GHz) & (s$^{-1}$) & (K) \\
\hline
\endfirsthead
\caption{continued.}\\
\hline\hline
Species & \multicolumn{3}{c}{Transition} & Frequency & $A_{ij}$ & $E_\mathrm{up}$ \\
 & (J K L M & - & J K L M) & (GHz) & (s$^{-1}$) & (K) \\
\hline
\endhead
\hline
\endfoot
$^{13}$CH$_3$OH & 3 1 3 +0 & - & 4 0 4 +0 & 112.1435 & $7.1 \times 10^{-6}$ & 28.0 \\
 & 17 -2 16 0 & - & 17 1 16 0 & 112.3157 & $2.1 \times 10^{-6}$ & 373.0 \\
\hline
CH$_3^{18}$OH & 21 3 18 4 & - & 20 2 18 4 & 110.9133 & $5.0 \times 10^{-6}$ & 834.9 \\
 & 18 1 17 4 & - & 19 2 17 4 & 111.6514 & $1.1 \times 10^{-6}$ & 790.6 \\
 & 13 3 11 1 & - & 14 2 13 1 & 112.3039 & $2.6 \times 10^{-6}$ & 265.0 \\
 & 25 0 25 1 & - & 25 1 25 1 & 113.0484 & $1.2 \times 10^{-6}$ & 730.9 \\
\hline
CH$_2$DOH & 28 0 28 2 & - & 28 1 28 1 & 109.7332 & $5.3 \times 10^{-6}$ & 878.3 \\
 & 12 1 11 1 & - & 12 0 12 1 & 113.0450 & $2.5 \times 10^{-6}$ & 185.1 \\
\end{longtable}
}

%
\longtab[3]{
\begin{longtable}{lrclccl}
\caption{Lines per species used for fitting the Band~6 data.} \\
\hline \hline
Species & \multicolumn{3}{c}{Transition} & Frequency & $A_{ij}$ & $E_\mathrm{up}$ \\
 & (J K L M & - & J K L M) & (GHz) & (s$^{-1}$) & (K) \\
\hline
\endfirsthead
\caption{continued.}\\
\hline\hline
Species & \multicolumn{3}{c}{Transition} & Frequency & $A_{ij}$ & $E_\mathrm{up}$ \\
 & (J K L M & - & J K L M) & (GHz) & (s$^{-1}$) & (K) \\
\hline
\endhead
\hline
\endfoot
CH$_3$OH & 20 8 13 2 & - & 21 7 14 2 & 260.0643 & $2.1 \times 10^{-5}$ & 808.2 \\
 & 20 3 18 0 & - & 20 2 19 0 & 260.3815 & $9.1 \times 10^{-5}$ & 536.9 \\
 & 2 1 1 1 & - & 1 0 1 1 & 261.8057 & $5.6 \times 10^{-5}$ & 28.0 \\
\hline
$^{13}$CH$_3$OH & 2 1 1 0 & - & 1 0 1 0 & 259.9865 & $5.5 \times 10^{-5}$ & 27.9 \\
 & 18 3 16 +0 & - & 18 2 17 -0 & 260.0888 & $9.2 \times 10^{-5}$ & 437.2 \\
 & 20 3 18 +0 & - & 20 2 19 -0 & 262.7673 & $9.4 \times 10^{-5}$ & 525.4 \\
 & 7 4 4 -0 & - & 8 3 5 -0 & 262.9130 & $1.7 \times 10^{-5}$ & 144.2 \\
 & 7 4 3 +0 & - & 8 3 6 +0 & 262.9203 & $1.7 \times 10^{-5}$ & 144.2 \\
\end{longtable}
}

\end{appendix}

\end{document}